\documentclass[12pt]{article}
\pdfoutput=1

\usepackage{fullpage,mathrsfs}

\usepackage{amsfonts}
\usepackage{amsmath,amssymb,amsthm}
\usepackage{graphicx}
\usepackage{color}
\usepackage{setspace}

\def\no{\nonumber}
\newcommand{\half}{\frac{1}{2}}
\newcommand{\p}{\partial}
\newcommand{\dd}{\delta}
\newcommand{\N}{\mathcal{N}}
\newcommand{\ti}{\widetilde}
\newcommand{\wat}{\widehat}

\def\pq#1#2{{\left[ \begin{array}{c}
#1 \\
#2
\end{array}
\right] }}

\def\abcd#1#2#3#4{{\left( \begin{array}{cc}
#1 & #2 \\
#3 & #4
\end{array}
\right) }}

\newcommand{\ba}{\begin{align}}
\newcommand{\ea}{\end{align}}

\newcommand{\lp}{\left(}
\newcommand{\rp}{\right)}
\newcommand{\tr}{\textrm{Tr}}
\newcommand{\ch}{\textrm{ch}}
\newcommand{\sh}{\textrm{sh}}
\newcommand{\thh}{\textrm{th}}

\newcommand{\bC}{\ensuremath{\mathbb{C}}}

\newcommand{\bN}{\ensuremath{\mathbb{N}}}

\newcommand{\bR}{\ensuremath{\mathbb{R}}}

\newcommand{\bZ}{\ensuremath{\mathbb{Z}}}

\newcommand{\scC}{\ensuremath{\mathcal{C}}}

\newcommand{\scI}{\ensuremath{\mathcal{I}}}

\newcommand{\scS}{\ensuremath{\mathcal{S}}}

\newcommand{\scW}{\ensuremath{\mathcal{W}}}

\begin{document}

\numberwithin{equation}{section}

\title{{\bf Hanany-Witten effect and $SL(2,\mathbb{Z})$ dualities\\
 in matrix models}}

\author{ \\
{\large Benjamin~Assel,$^{\natural}$}}

\maketitle

\vskip -8.5cm
\rightline{KCL-MTH-14-11}
\vskip 8.5cm

\centerline{$^\natural$ Department of Mathematics, King's College London, }
\centerline{The Strand, London WC2R 2LS, United Kingdom}
\centerline{{\em benjamin.assel@kcl.ac.uk}}

 \vskip 1cm

\begin{abstract}
We provide tests of dualities for three-dimensional $\N=4$ quiver SCFTs with brane realizations in IIB string theory, by matching their exact partition functions on $S^3$. The dualities are generated by $SL(2,\bZ)$ transformations and Hanany-Witten 5-brane moves. These contain mirror symmetry as well as dualities identifiying fixed points of Yang-Mills quivers and Chern-Simons theories. The partition function is given by a matrix model, that can be nicely rearranged into a sequence of factors mimicking the brane realization. Identities obeyed by these elementary factors can be used to match the partition functions of dual theories, providing tests for the full web of dualities. In particular we are able to check mirror symmetry for linear and circular quivers with gauge nodes of arbitrary ranks. Our analysis also leads to a proof of a conjectured formula evaluating the matrix models of linear quiver SCFTs.
\end{abstract}

\newpage

\tableofcontents

\newpage

\section{Introduction}

$\N=4$ Yang-Mills theories in three dimensions admit infrared strongly coupled fixed points, subject to the mirror symmetry duality \cite{Intriligator:1996ex}. Their moduli spaces split into a Coulomb branch and a Higgs branch that are exchanged under this duality. A prominent feature of these moduli spaces is that the Higgs branch does not receive quantum corrections and thus captures the quantum corrections to the Coulomb branch of the mirror theory.   The $\N=4$ theories can be deformed by mass and Fayet-Iliopoulos terms, whose parameters are also exchanged under the duality. Mirror symmetry was extended to quiver theories in \cite{deBoer:1996mp, deBoer:1996ck}, using the brane realizations of the theories in type IIB string theory  \cite{Hanany:1996ie}. The fixed point of a three-dimensional quiver theory was obtained as the low-energy theory of D3-branes suspended between NS5-branes and D5-branes. Mirror symmetry was then identified with the action of S-duality in IIB string theory, exchanging NS5-branes and D5-branes.

In \cite{Kapustin:2009kz} it was shown that the partition function of three-dimensional SCFTs deformed by mass and FI terms and defined on the round 3-sphere $S^3$ can be computed exactly using localization techniques (see \cite{Marino:2011nm} for a review) and reduces to a simple matrix model. 
The exact partition function on $S^3$ was then used in \cite{Kapustin:2010xq} to test mirror symmetry for IR fixed points of $\N=4$ circular Yang-Mills quiver theories with nodes of equal rank.

More tests of mirror symmetry by matching exact partition functions were given in \cite{Dey:2011pt,Dey:2013nf,Dey:2014dwa} for A- and D-type quiver SCFTs.
Other remarkable tests have been performed in \cite{Hanany:2011db,Cremonesi:2013lqa,Cremonesi:2014vla,Cremonesi:2014kwa,Dey:2014tka}, where the Hibert series of the Coulomb branch were shown to coincide with the Hilbert series of the Higgs branch of the mirror dual theories. Several cases were investigated there, involving mirror symmetry for A, D quivers and various star-shaped quivers. One of the interesting ideas developed in \cite{Dey:2014tka} is the possibility to generate new mirror pairs from the known dualities by gauging global symmetries. 

Despite the significant advancements in testing mirror symmetry, very few results are known for the cases of quiver theories with nodes of different ranks. This gave the initial motivation for the present work, whose first ambition was to complete the tests of mirror symmetry for quiver theories with nodes of arbitrary ranks.
The second idea was to understand the full $SL(2,\bZ)$ duality group of the three-dimensional quiver theories, inherited from type IIB string theory \cite{Witten:2003ya}. It was shown in \cite{Jafferis:2008em} that these more general dualities can map Yang-Mills fixed points to Chern-Simons theories coupled to matter. These are very interesting dualities since the Chern-Simons theories are superconformal, so that they would be indeed the infrared fixed point of the ``dual" Yang-Mills theories. Three-dimensional Chern-Simons couplings have a priori only up to $\N=3$ supersymmetry, however with specific matter content and superpotentials the amount of supersymmetry can be enhanced to $\N=4$ \cite{Gaiotto:2008sd, Hosomichi:2008jd, Imamura:2008nn}, and up to $\N=8$ \cite{Aharony:2008ug,Aharony:2008gk}.

In this paper we investigate these dualities by comparing the exact partition functions on $S^3$ of the $\N=4$ Yang-Mills and Chern-Simons SCFTs deformed by mass and FI parameters, for linear and circular quivers with unitary nodes of arbitrary ranks. All the theories we consider admit brane realizations in IIB string theory as arrays of D3-branes suspended between two types of 5-branes. Yang-Mills quivers are realizated with NS5 and D5-branes; Chern-Simons quivers are realized with NS5 and $(1,k)$-5branes. Other choices of 5-branes lead to theories that were described in \cite{Gaiotto:2008ak} as Chern-Simons theories with ``interpolating $T(U(N))$" couplings.
A web of dualities is generated by $SL(2,\bZ)$ actions combined with Hanany-Witten 5-brane moves (HW moves). The HW moves exchange two 5-branes of different type in the sequence of 5-branes realizing a quiver theory. Mirror symmetry corresponds to the action of $S$-duality combined with HW moves transforming the brane realization of a quiver theory into the brane realization of its mirror-dual. Other interesting dualities are generated, one being a level-rank duality for $\N=4$ Chern-Simons theories, corresponding to HW moves exchanging NS5 and $(1,k)$-5branes. Another duality relates Yang-Mills quivers to Chern-Simons quivers with Chern-Simons levels at each node being $\pm1$ or $0$. A node with vanishing Chern-Simons level has an ``auxiliary vector multiplet" \cite{Gaiotto:2008sd, Imamura:2008dt}. A class of Yang-Mills fixed points admit Chern-Simons duals with non-vanishing levels at all nodes, which correspond to microscopic descriptions of the infrared fixed point. Each mass or FI deformation parameter is associated to a the position of a 5-brane in the brane realization and the map between parameters is obtained by following the 5-branes through the duality transformations.

To test these dualities we decompose the matrix model which computes the partition function into a sequence of elementary factors, that can be associated to the 5-branes in the brane realization. This approach was already developed in \cite{Gulotta:2011si} for quiver with nodes of equal ranks and we generalize it to quivers of arbitrary ranks and including mass and FI deformation terms. We then provide an action of $SL(2,\bZ)$ dualities on the 5-brane factors, based on the analysis of \cite{Gaiotto:2008ak}, again generalizing \cite{Gulotta:2011si}. This action on the matrix factors can be understood as a local $SL(2,\bZ)$ duality on the brane configuration, introducing duality walls, that are associated to their own factors in the matrix model.
The $SL(2,\bZ)$ action on the 5-brane factors is used to prove the equality of the partition functions of any pair of $SL(2,\bZ)$-dual theories, in a straightforward way. In practice, only two of the $SL(2,\bZ)$ dualities will be relevant for us: $S$ duality which is related to mirror symmetry and the duality changing Yang-Mills theories into Chern-Simons theories.
To include HW move dualities, we prove an identity for products of two 5-brane factors, mimicking the HW 5-brane exchange. Combining the $SL(2,\bZ)$ action and this HW move-identity we are able to match the partition functions of any pair of quiver SCFTs related by this web of dualities. In particular
we match the exact partition function of mirror dual YM quiver SCFTs with nodes of arbitrary ranks, deformed by parameters. 

The main technical challenge in our computations was to find a matrix factor for a D5-brane with different numbers of D3-branes ending on its left and right sides. We found that it must be defined in terms of an unusual distribution $\hat\delta$. Although they look very exotic, these distributions proved very useful to derive the HW move identity. 
As a by-product our analysis leads to a proof of the formula conjectured in \cite{Nishioka:2011dq} (NTY formula) evaluating the matrix models of arbitrary deformed YM linear quiver SCFTs.

The map between partition functions that we found is only exact up to phases that depends on the mass and FI parameters. We argue, following \cite{Closset:2012vg, Closset:2012vp}, that these phases affect contact terms in global current correlation functions and are unphysical in the sense that they arise from different UV regularizations (they can be removed by local counter-terms).

One lesson we learned from our analysis is that there is a close link between the matrix models and the brane realizations of the theories, and that what happens on the brane side always has a counterpart on the matrix model side, so that the brane picture is (as usual) a powerful guide to understand the properties of the gauge theories.

We also point out that the $AdS_4$ gravity duals of the $\N=4$ quiver SCFTs studied in this paper (at vanishing deformation parameters) where constructed in \cite{Assel:2011xz,Assel:2012cj} in type IIB ten-dimensional supergravity. The counterparts in supergravity of the dualities we study are standard $SL(2,\bZ)$ transformations of the gravity backgrounds.

The rest of the paper is organized as follows. In section \ref{sec:3dTheories} we give a brief description of the $\N=4$ three-dimensional theories that admit brane realizations. In section \ref{sec:SL2Z} we introduce the matrix models computing the partition functions on $S^3$, we explain the decomposition into 5-brane factors, we derive the $SL(2,\bZ)$ action on the factors and prove the equality of partition functions for $SL(2,\bZ)$ dual theories. In section \ref{sec:MirrorSym} we prove the equality of the partition functions of theories related by HW move duality. We emphasize the cases of mirror symmetry and Yang-Mills/Chern-Simons dualities. In section \ref{sec:ExplicitResults} we give a proof of the NTY formula. Section \ref{sec:Discussion} contains our conclusions and perspectives for future work. We also included three appendices: appendix \ref{app:deltahat} provides more details about the $\hat\delta$ distributions, appendix \ref{app:formulas} gathers a few useful formulas and appendix \ref{app:computations} contains details of computations.

\section{$\N=4$ theories with brane realizations}
\label{sec:3dTheories}

In this section we describe the $\N=4$ quiver theories and their type IIB brane realizations. We explain how to summerize the brane configuration, and thus the content of the quiver theory, in a graph with two types of nodes (or dots) and various labels.

\bigskip

Three-dimensional $\N=4$ gauge theories contain an  $\N=4$ vector multiplet transforming in the adjoint representation of a gauge group $G$ and hypermultiplets transforming in arbitrary representations $R$ of $G$. The $\N=4$ vector multiplet is made out of an  $\N=2$ vector multiplet $V$ (in superfield notations) whose bosonic fields are a vector field $A_{\mu}$, a real scalar $\sigma$ and a real auxiliary field $D$, and a chiral multiplet $\Phi$ containing a complex scalar $\phi$ and a complex auxiliary field $F$. A hypermultiplet contains two chiral multiplets $Q, \ti Q$ transforming in conjugate representations $R$ and $\bar R$ of the gauge group. The R-symmetry group is $SU(2)_V \times SU(2)_H$, with  the real scalars in the vector multiplet forming a $SU(2)_V$ triplet and the  scalars in the hypermultiplet forming two $SU(2)_H$ doublets. $\N=4$ Yang-Mills theories are naturally endowed with canonical vector- and hyper-multiplet actions, given in terms of $\N=2$ superfields by:
\begin{align}
S^{\N=4}_{\rm vec} &=  \frac {1}{g_{YM}^2} \int d^3 x d^2\theta d^2\bar\theta \ \tr \lp \frac 1 4 \Sigma^2 + \Phi^{\dagger} e^{V} \Phi \rp \\
S^{\N=4}_{\rm hyper} &=  \int d^3 x d^2\theta d^2\bar\theta \lp Q^{\dagger} e^{V} Q + \ti Q e^{-V} \ti Q^{\dagger} \rp \ ,
\end{align}
with $\Sigma=\bar D^{\alpha} D_{\alpha} V$ (linear multiplet).
In addition there is a superpotential coupling the adjoint chiral $\Phi$ and the matter chirals $Q_i,\ti Q_i$ of the form:
\begin{align}
S_{\rm sp} =  \int d^3 x d^2\theta \lp \sum_i \ti Q_i \Phi Q_i \rp + c.c. \ . 
\end{align}

In 3d the Yang-Mills coupling is dimensionful, $g_{YM}^2$ has the dimension of a mass, implying that the UV limit is free, while the IR limit is strongly interacting. For this reason, the above microscopic Lagrangian description is reliable only at high energies.

The theories can be deformed by supersymmetric mass terms for the hypermultiplets, obtained by coupling $Q, \ti Q$ to a background abelian $\N=4$ vector multiplet $(V_m,\Phi_m)$, with $V_m = m \, \bar\theta \theta$ and $\Phi_m= \phi_m \bar\theta\theta$. $(m,$Re$(\phi_m)$,Im$(\phi_m))$ form a triplet under $SU(2)_V$.

Supersymmetric FI deformation terms can aslo be added through a BF coupling to a background abelian $\N=4$ twisted vector multiplet $(V_{\eta},\Phi_{\eta})$:
\begin{align}
S_{\eta} &= \int d^3 x d^2\theta d^2\bar\theta \ \tr \lp \Sigma V_{\eta} + \Phi \Phi_{\eta}  \rp
\label{FIterm}
\end{align}
with $V_{\eta}  = \eta \, \bar\theta \theta$ and $\Phi_{\eta} = \phi_{\eta}  \bar\theta\theta$. $(\eta,$Re$(\phi_{\eta})$,Im$(\phi_{\eta}))$ form a triplet under $SU(2)_H$.

In this paper we focus on $\N=4$ theories which can be engineered on type IIB brane configurations involving D3-branes stretched between two types of 5-branes. The SCFTs arise in the infrared limit and the superconformal algebra is $OSp(4|4)$.

\bigskip

\subsection{Yang-Mills quiver SCFTs}
\label{ssec:YMquivers}

We consider infrared fixed points of three-dimensional $\N=4$ theories that admit a realization as the low-energy limit of brane configurations in type-IIB string theory  \cite{Hanany:1996ie}. The brane configuration realizing Yang-Mills quivers consists of an array of D3, D5 and NS5 branes oriented as shown in the table.
The D3 branes span a finite interval along the $x^3$ direction and terminate on the five-branes. In the low-energy limit the world-volume SCFT on the D3-branes is effectively three-dimensionnal.

\begin{table}[h]
\label{tab:probeconfig}
\begin{center}
\begin{tabular}{|c||c|c|c|c|c|c|c|c|c|c||}
  \hline
      & 0 & 1 & 2 & 3 & 4 & 5 & 6 & 7 & 8 & 9 \\ \hline
  D3  & X & X & X & X &   &   &   &   &   &   \\
  D5  & X & X & X &   & X & X & X &   &   &   \\
  NS5 & X & X & X &   &   &   &   & X & X & X \\ \hline
\end{tabular}
\caption{\footnotesize Brane array for three-dimensional quiver gauge theories}
\end{center}
\end{table}
\smallskip

A stack of $N$ coincident D3-branes stretched between two NS5-branes and intersecting $M$ D5-branes give rise to a $U(N)$ gauge group with Yang-Mills $\N=4$ vector multiplet, plus $M$ hypermultiplets in the fundamental representation of $U(N)$. Moreover each NS5-brane introduces an additionnal hypermulitplet transforming in the bifundamental representation of $U(N_i)\times U(N_{i+1})$, where $N_i$ and $N_{i+1}$ are the numbers of D3-branes ending on the left and right of the NS5-brane respectively.

\begin{figure}[th]
\centering
\includegraphics[scale=0.5]{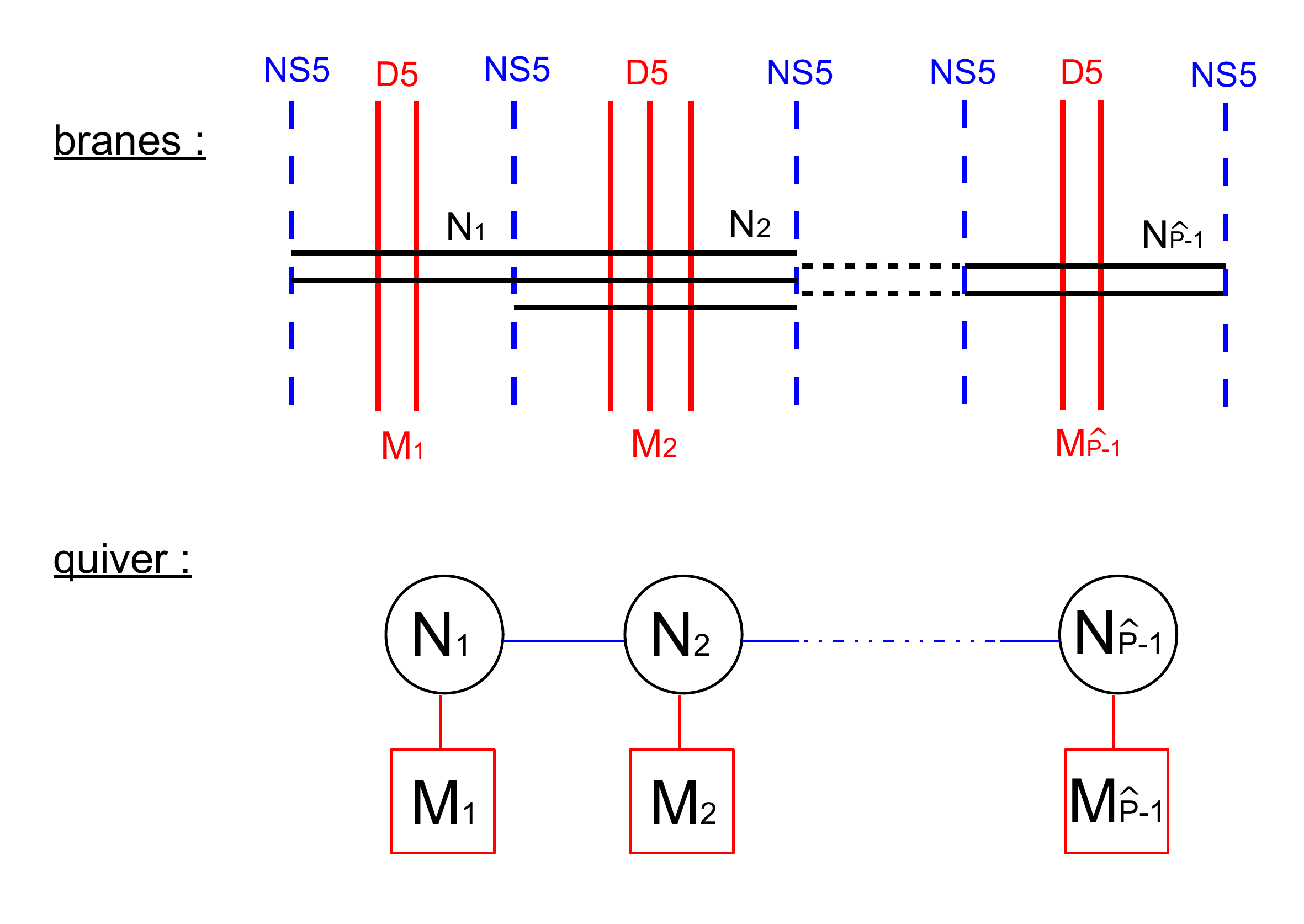}
\vspace{-0.5cm}
\caption{\footnotesize Brane realization of linear quivers. $N_j$ refers to the number of coincident D3-branes (parallel black lines) in a D3-segment and $M_j$ to the number of D5-branes crossing it. The horizontal direction can be thought as $x^3$. The vertical direction can be $x^{456}$ for the D5s and $x^{789}$ for the NS5s.}
\label{linquiv}
\end{figure}

\noindent The brane configuration depicted in figure \ref{linquiv} is mapped to the linear quiver gauge theory summarized in the quiver diagram shown in the same figure. It contains $\hat P$ NS5-branes with $N_j$ coincident D3-branes stretched between the $j$-th and $(j+1)$-th NS5-brane and $M_j$ D5-branes intersecting the $N_j$ D3-branes. In the quiver diagram a $U(N_j)$ gauge node, with $M_j$ fundamental hypermultiplets, is indicated by a circle with $N_j$ and an attached square with $M_j$. Bifundamental hypermultiplets are symbolized by lines joining two nodes.
 In total there are $\hat P -1$ D3-segments (stacks of D3-branes), whose low energy dynamics is the same as the infrared fixed point of the quiver theory with $\hat P -1$ gauge nodes. We also denote $P \equiv \sum_{j=1}^{\hat P -1} M_j$ the total number of D5-branes.\\

\smallskip

\begin{figure}[th]
\centering
\includegraphics[scale=0.5]{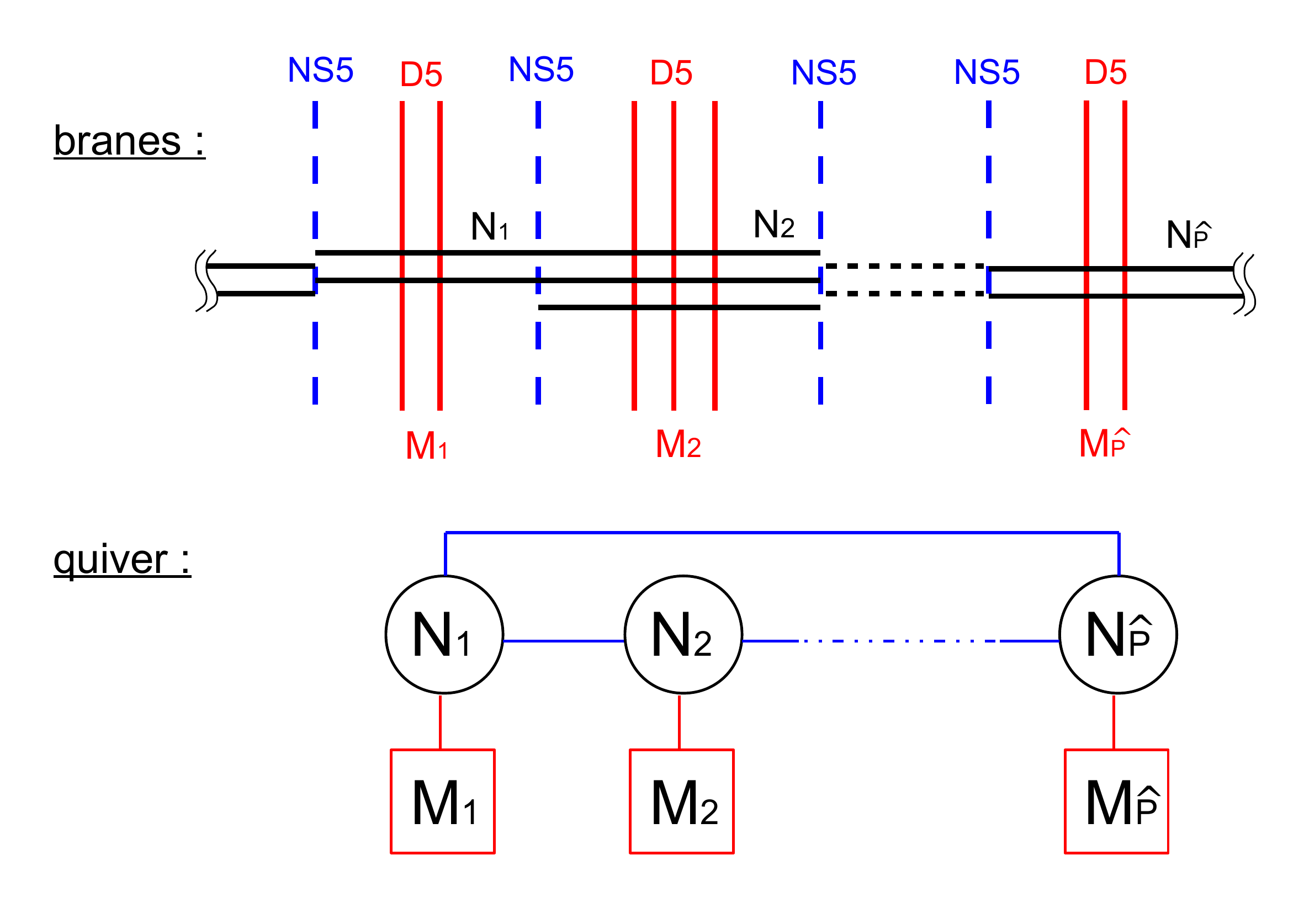}
\vspace{-0.5cm}
\caption{\footnotesize Brane realization of circular quivers. The D3-segments on the left and on the right are identified.}
\label{circquiv}
\end{figure}

One can also compactify the $x^3$ direction to a circle. The brane configuration is depicted in figure \ref{circquiv} and its low-energy theory corresponds to the low-energy limit of a circular quiver. The difference with respect to the linear quiver case is that the bifundamentals connecting the nodes form a closed loop.

\bigskip

The deformations of the quiver SCFT by mass terms and FI terms can be associated to displacements of the 5-branes in transverse space. For instance we can take the convention that, if $t_j$ denotes the position of the $j$th NS5-brane along $x^9$, then the FI parameter of the $j$th node is $\eta_j=t_j - t_{j+1}$, and if $m_j$ denotes the position of the $j$th D5-brane along $x^6$, then the corresponding hypermultiplet aquires a real mass $m_j$. Displacements of 5-branes along the other directions can be similarly mapped to the other FI and mass deformation parameters, but we will set them to zero in this work since they do not appear in the partition function on $S^3$ computed by supersymmetric localization.

\bigskip

{\bf Graphs:}

\bigskip

\begin{figure}[th]
\centering
\includegraphics[scale=0.8]{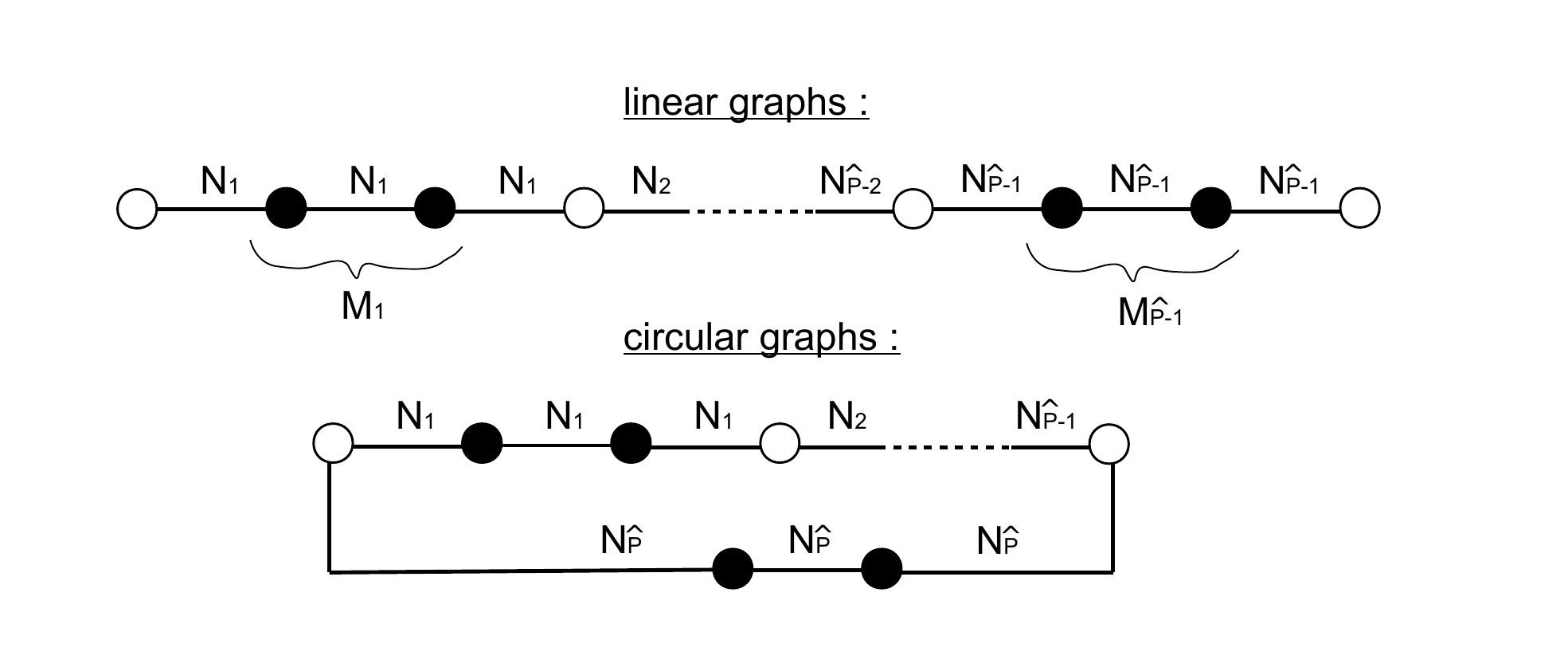}
\vspace{-0.5cm}
\caption{\footnotesize Graphs associated to linear and circular quivers.}
\label{graphquiv}
\end{figure}

The brane pictures themselves can be recast into graphs, where a line labelled by $N$ denotes $N$ coincident D3-branes extended between two consecutive 5-branes, a white dot denotes a NS5-brane and a black dot denotes a D5-brane. This provides graphs associated to the the linear and circular quivers of figures \ref{linquiv}, \ref{circquiv} as shown in figure \ref{graphquiv}.

\subsubsection{$T(U(N))$ theory}
\label{sssec:TUN}

\begin{figure}[th]
\centering
\includegraphics[scale=0.4]{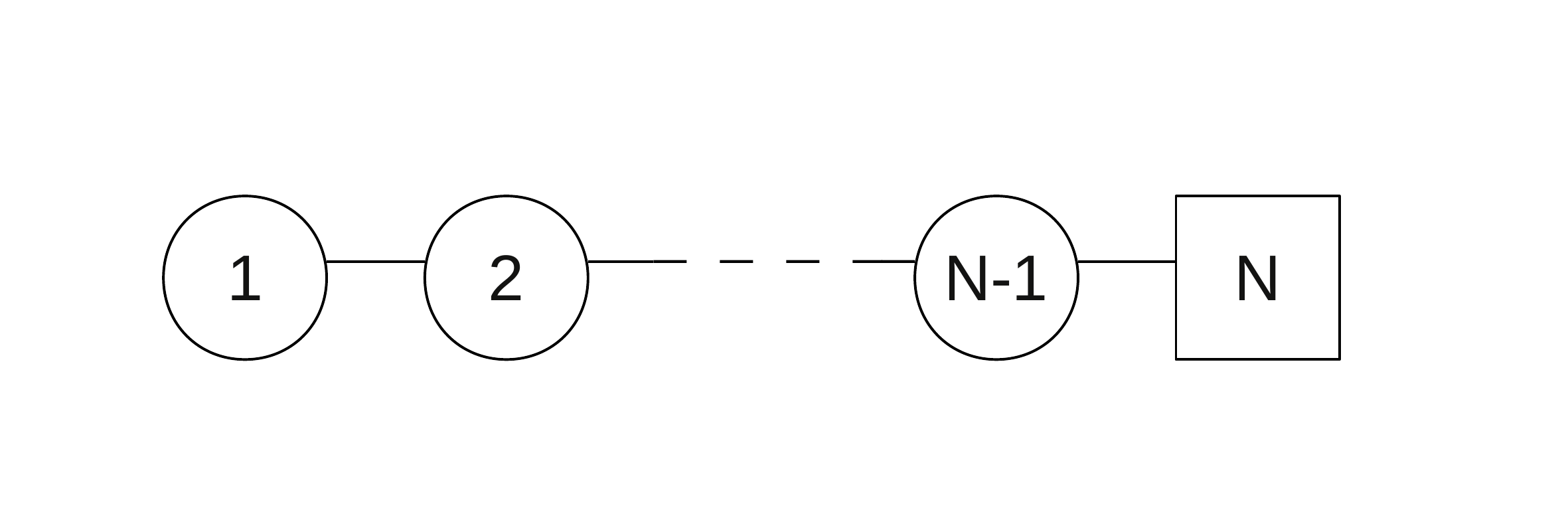}
\vspace{-1cm}
\caption{\footnotesize $T(SU(N))$ quiver diagram.}
\label{TSUNquiv}
\end{figure}

There is a Yang-Mills linear quiver SCFT that will play a distinguished role in our story: the $T(U(N))$ theory. Let us first described the theory $T(SU(N))$. The UV quiver description has $U(1)\times U(2) \times ... \times U(N-1)$ gauge group, bifundamental hypermultiplets for each pair of adjacent nodes $U(p)\times U(p+1)$, plus $N$ hypermultiplets  in the fundamental representation of the $U(N-1)$ node. $T(SU(N))$ is the IR fixed point SCFT of this linear quiver (figure \ref{TSUNquiv}).\\
 The theory has a group of global symmetry $SU(N)_F\times SU(N)_J$ with $SU(N)_F$ rotating the $N$ fundamental hypermultiplets and the $SU(N)_J$ arising as an enhancement of the topological $U(1)^{N-1}$ symmetry at the IR fixed point. The deformation parameters of $T(SU(N))$, up to R-symmetry transformations, are $N$ real masses $m_j$ for the $N$ fundamental hypermultiplets and $N-1$ real FI parameters $\eta_j$ for the $N-1$ nodes.
 $T(SU(N))$ is known to be invariant under mirror symmetry, which exchanges the two $SU(N)$ global symmetries and the mass and FI parameters: $\eta_j \leftrightarrow m_j-m_{j+1}$.
 
 The diagonal $U(1)_F$ flavor symmetry, rotating the $N$ fundamental hypermultiplets by the same phase, is usually identified with the diagonal $U(1)$ of the gauge group.
 In the presence of generic supersymmetric FI deformations the diagonal $U(1)$ is coupled to a background twisted vector multiplet through a BF term and the  flavor $U(1)_F$ may be considered formally as an independent symmetry, completing the flavor symmetry to $U(N)_F$.
 
To define the theory $T(U(N))$ we add an extra BF coupling between the background $U(N)_F$ and a new background (twisted) abelian vector multiplet, defining a new symmetry that we call $U(1)_J$. This is the same as adding a FI term for the $U(N)_F$ flavor symmetry, with a FI parameter $t_N$. As we will see in section \ref{ssec:PartFunction}, the partition function of the $T(U(N))$ theory deformed by masses and FI terms is invariant under the exchange of the $2N$ parameters $m_j \leftrightarrow t_j$, with  $t_j$, for $1 \le j \le N-1$ defined by $\eta_j=t_j-t_{j+1}$. It is then natural to combine the topological symmetry $U(1)_J$ with $SU(N)_J$ into a $U(N)_J$ global symmetry.

A priori the $T(U(N))$ and $T(SU(N))$ theories have identical dynamical properties, but it will be important for us to consider $T(U(N))$ to be able to gauge the $U(N)_F \times U(N)_J$ global symmetries.

To conclude we notice that the $T(U(1))$ theory is rather trivial: it contains no dynamical field, but only a BF coupling between two abelian background vector multiplets.

\bigskip

\subsection{Chern-Simons SCFTs}
\label{ssec:CSquivers}

\bigskip

Three-dimensional gauge theories admit supersymmetric Chern-Simons terms, with quantized Chern-Simons level $k\in \bZ$ \footnote{In principle, depending on the number of massive chirals and their charges, the Chern-Simons level could be quantized to half-integer values \cite{Aharony:1997bx}. However with the matter content of an $\N=4$ theory (only hypermultiplets), the quantization condition is simply $k\in \bZ$.}, preserving $\N=3$ sypersymmetry \cite{Gaiotto:2007qi}. The action reads:
\begin{align}
S^{\N=3}_{\rm CS} &= \frac{k}{4\pi} \int d^3x \ \tr \Big[ \epsilon^{\mu\nu\rho} \lp A_{\mu} \p_{\nu} A_{\rho} + \frac 23 A_{\mu}A_{\nu}A_{\rho} \rp + \sigma D - \bar\lambda \lambda \Big]  - \frac{k}{4\pi} \int d^3x d^2\theta \ \tr \Big[ \Phi^2 + h.c. \Big] \ .
\end{align}
The Chern-Simons couping introduces a mass $\sim k g_{\rm YM}^2$ for the fields in the vector multiplet $V$ as well as for the adjoint chiral multiplet $\Phi$. In the infrared limit $g_{YM}$ becomes effectively very large, the Yang-Mills kinetic term is irrelevant and the adjoint chiral can be integrated out as an auxiliary field \cite{Gaiotto:2007qi,Aharony:2008ug}, leading to a pure Chern-Simons theory with a new superpotential:
\begin{align}
W &= \frac{4\pi}{k} \lp \sum_i \ti Q_i \, T^a_{R_i} \, Q_i \rp  \lp \sum_j \ti Q_j \, T^a_{R_j} \, Q_j \rp \ ,
\end{align}
with $T^a_{R_i}$ the generators of the gauge group in the representation $R_i$. 

The $\N=3$ Chern-Simons theories are exactly conformal \footnote{There is no relevant or marginal quantum correction to the classical action \cite{Gaiotto:2007qi}.}.
For specific choices of gauge group and matter content the Chern-Simons SCFTs can have enhanced $\N=4$ supersymmetry \cite{Gaiotto:2008sd, Hosomichi:2008jd, Imamura:2008nn,Imamura:2008dt}, up to $\N=8$ supersymmetry corresponding to the ABJ(M) theory at Chern-Simons level $k=1,2$ ($\N=6$ for $k>2$) \cite{Aharony:2008ug,Aharony:2008gk}.
\bigskip

The brane realization of $\N=4$ Chern-Simons SCFTs described in \cite{Aharony:2008ug} involves D3-branes, NS5-branes and $(1,k)$-5branes. The D3-branes are along the directions $0123$, with the direction $3$ compact, they intersect NS5-branes spanning $012789$ and $(1,k)$-5branes spanning $012[4,7]_{\theta}[5,8]_{\theta}[6,9]_{\theta}$, with $[a,b]_{\theta}$ meaning that the brane is along the direction $\cos \, \theta \, x_a  + \sin \, \theta \, x_b$ in the $(x_a,x_b)$ plane and the angle $\theta$ is fixed as a function of $k$ so that the brane configuration preserves $\N=3$ supersymmetry \cite{Kitao:1998mf}, with a vanishing axion field. The low-energy theory living on the D3-branes has both Yang-Mills and Chern-Simons couplings. The Chern-Simons coupling for a $U(N)$ node corresponding to $N$ D3-segments stretched between a $(1,k_1)$- and a $(1,k_2)$-5brane has CS level $k_1-k_2 \in \{ -k, 0, k\}$ \cite{Imamura:2008nn,Imamura:2008ji}.
In the infrared limit, for nodes with non-zero Chern-Simons coupling $\pm k$, massive fields in the vector multiplet can be integrated out, as explained above, leading to a pure Chern-Simons node with a specific superpotential. 

However, as mentioned in section 3.5 of \cite{Aharony:2008ug}, there exist alternative (or simpler) brane configurations realizing Chern-Simons-Matter SCFTs, that preserve $\N=4$ sypersymmetry. These are the same brane configurations as presented in section \ref{ssec:YMquivers} with D5-branes replaced by $(1,k)$-5branes, namely the D3-branes are along $0123$, the NS5-branes along $012789$ and the $(1,k)$-5branes along $012456$. This corresponds to the previous brane configurations with $\theta=\frac{\pi}{2}$, which preserves $\N=4$ (8 real supercharges) when allowing for non-vanishing axion field (\cite{Aharony:2008ug}). For our purposes it makes more sense to consider these brane realizations that preserve the correct amount of supersymmetry and for which the $SL(2,\bZ)$ action of type IIB string theory will be easily understood.

The theory living on D3-branes intersecting alternating NS5-branes and $(1,k)$-5branes  involves $U(N)$ Chern-Simons gauge nodes with alternating Chern-Simons levels $\pm k$ and bifundamental hypermultiplets (twisted and untwisted alternatively) \cite{Hosomichi:2008jd}. Moreover if the sequence of 5-branes have two consecutive NS5-branes or $(1,k)$-5branes, the infrared theory will have extra $U(N)$ nodes without Chern-Simons term, nor Yang-Mills term, in which case the vector multiplet fields are auxiliary \cite{Gaiotto:2008sd, Imamura:2008dt}. Integrating out these auxiliary fields leads to a Chern-Simons gauge theory without these $U(N)$ nodes, coupled to sigma models with hyper-K\"ahler target space \cite{Gaiotto:2008sd}.\footnote{For all we know, there is no difference between the infrared limit of a Yang-Mills node and an auxiliary node. The matrix models computing their partition functions are identical, so that we are not able to distinguish between the two cases. }

\begin{figure}[th]
\centering
\includegraphics[scale=0.8]{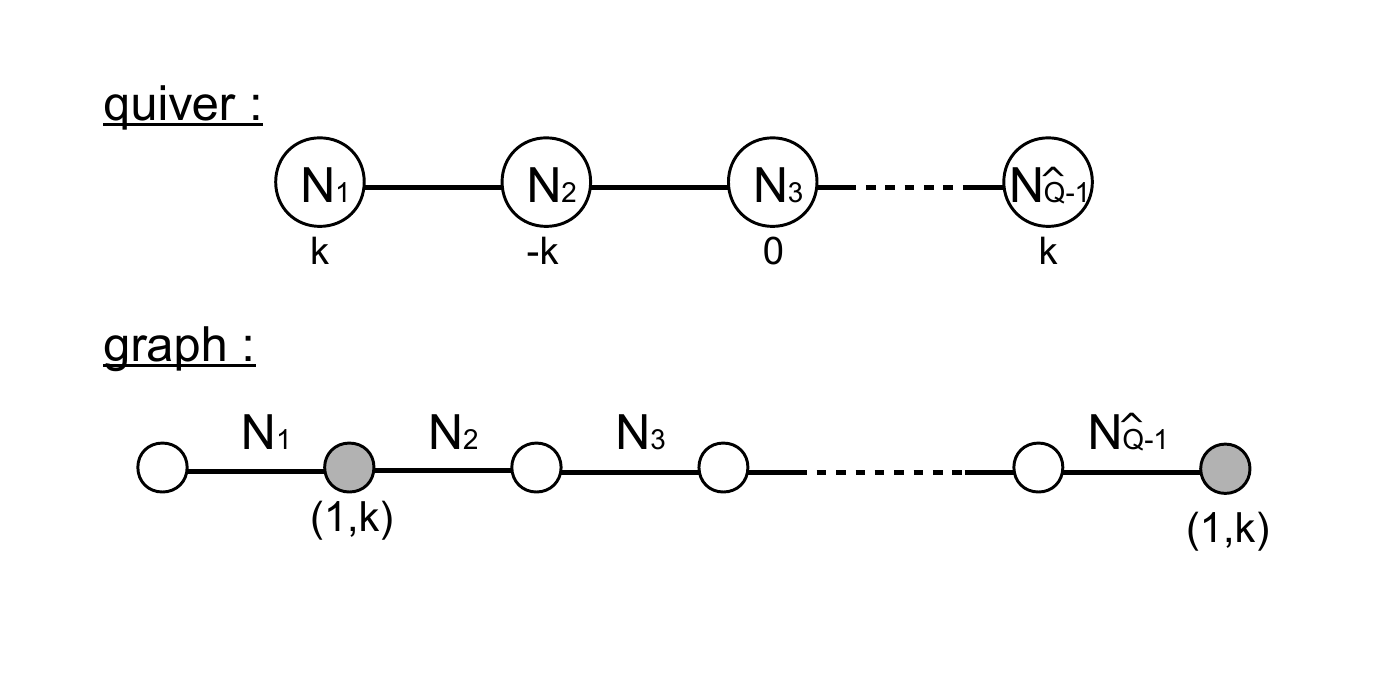}
\vspace{-1cm}
\caption{\footnotesize Chern-Simons quiver and its associated graph. The label below the nodes denotes the Chern-Simons level, zero meaning a node with auxiliary vector multiplet. The elements of the graph are white dots for NS5-branes and grey dots for $(1,k)$-5branes. $Q$ is the total number of 5-branes, or dots, in the brane picture, or graph. }
\label{CSquiv}
\end{figure}

For our purposes, it will be enough to describe the $\N=4$ infrared CFT living on a brane configuration associated to a sequence along $x^3$ of NS5-branes and $(1,k)$-5branes as the infrared fixed point of a quiver theory with:
\begin{itemize}
\item $U(N_i)_{\pm k}$ Chern-Simons gauge node for $N_i$ D3-segments extended between a NS5-brane and a $(1,k)$-5brane, with Chern-Simons level $+k$ if the NS5-brane is on the left of the $(1,k)$-5brane along the $x^3$ direction, and $-k$ otherwise.;
\item $U(N_i)_0$ auxiliary gauge node for $N_i$ D3-branes stretched between two NS5-branes or two $(1,k)$-5branes, with $\N=4$ auxiliary vector multiplet ;
\item A bifundamental hypermultiplet $(X_i,\ti X_i)$ in $( N_i, \overline N_{i+1}) \oplus (\overline N_i, N_{i+1})$ for each 5-brane with $N_i$ D3-branes on its left and $N_{i+1}$ D3-branes on its right.
\end{itemize}
This is summarized in figure \ref{CSquiv}, where we also introduce the corresponding graph with white dots for NS5-branes and grey dots for $(1,k)$-5branes.\\
The superpotential for a $U(N_i)_{k}$ Chern-Simons node is given by:
\begin{align}
W_i &= \frac{4\pi}{k} \tr \lp  \ti X_{i-1} X_{i-1} X_i \ti X_i \rp  \ .
\end{align}

The Chern-Simons SCFTs admit deformations by FI terms associated to the positions of the 5-branes in transverse space, namely if $t_j$ denotes the position of the $j$th 5-brane along $x^9$ (for NS5s) or $x^6$ (for $(1,k)$5s), then the FI parameter of the $j$th node is $\eta_j=t_j - t_{j+1}$.

\bigskip

\subsection{SCFTs realized with $(p,q)$ 5-branes (GW quivers):}
\label{ssec:GWquivers}
\bigskip

A natural extension of these brane constructions is to consider $\N=4$ SCFTs realized on brane configurations where D3-segments are stretched between $(p_1,q_1)$ and $(p_2,q_2)$-5branes oriented as the D5s and NS5s in table (\ref{tab:probeconfig}), with $p_1 \wedge q_1 =1$, $p_2 \wedge q_2 =1$ (coprime integers) and $D = p_1 q_2 - p_2 q_1 \neq 0$.
Acting with the appropriate $SL(2,\bZ)$ symmetry of type IIB string theory we can trade the $(p_1,q_1)$ and $(p_2,q_2)$-5-branes for NS5-branes and $(p,q)$-5brane, with $p \wedge q=1$ and $q \neq 0$.

Even this simpler configurations with D3-segments stretched between NS5-branes and $(p,q)$-5branes do not have a known Lagrangian description when $p>1$. It was argued in  \cite{Gaiotto:2008ak}(section 8) using the $SL(2,\bZ)$ symmetry of IIB string theory, that such theories have a dual description as the infrared fixed points of Chern-Simons quivers with $T(U(N))$ SCFTs ``interpolating" between gauge nodes. 

\begin{figure}[th]
\centering
\includegraphics[scale=0.8]{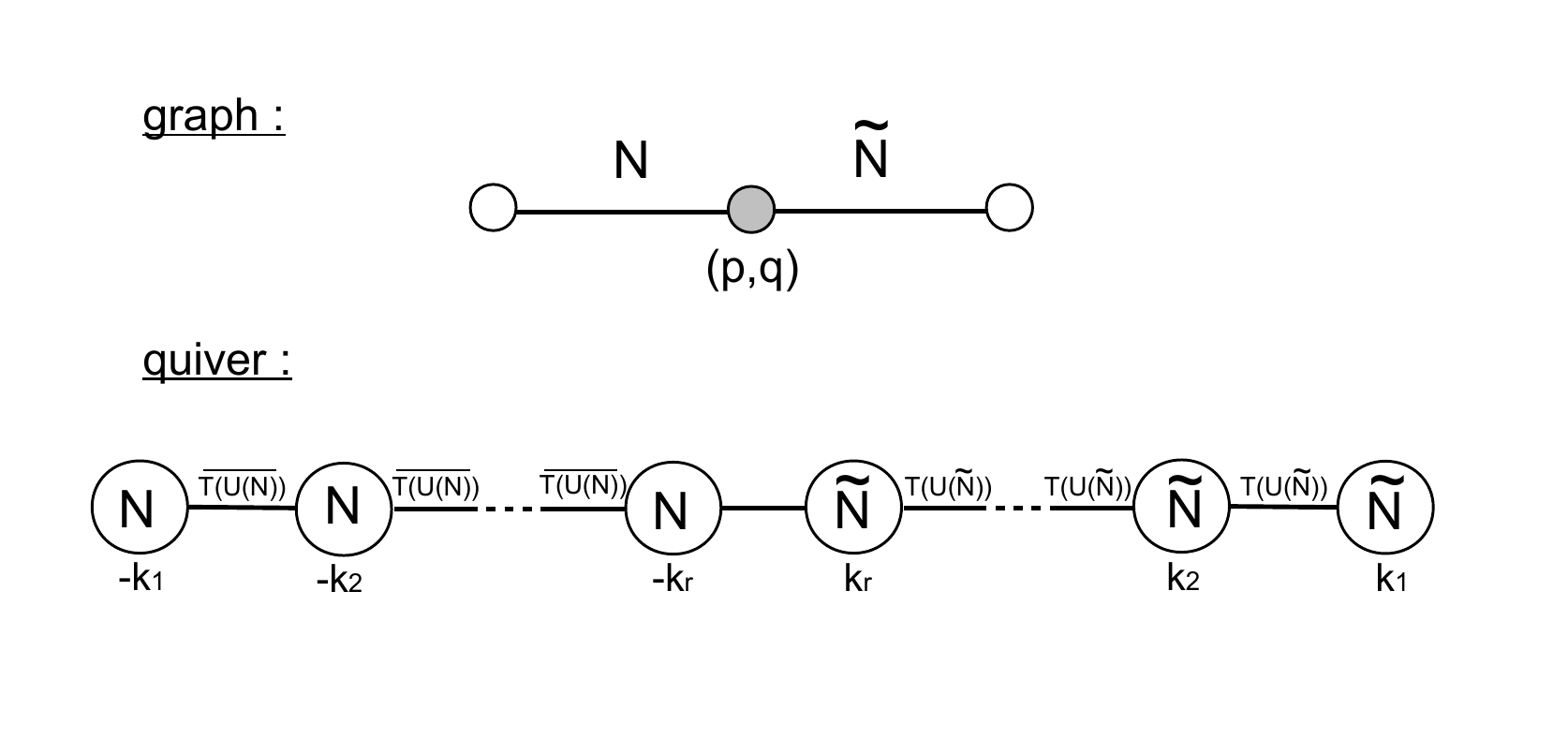}
\vspace{-1cm}
\caption{\footnotesize Graph corresponding to a sequence NS-$(p,q)$-NS of 5branes and the corresponding CS quiver theory with interpolating $T(U(\ti N))$ and $\overline{T(U(N))}$  couplings. The central link of the quiver is simply a bifundamental hypermultiplet.}
\label{GWquiv}
\end{figure}

The precise gauge theory description demands to find the sequence of $S$ and $T$ transformations that brings the $(p,q)$-5brane into a NS5-brane: $\binom{p}{q} = T^{k_1}ST^{k_2}S\cdots ST^{k_r}\binom{1}{0}$, where $S$ and $T$ are the generators of the $SL(2,\bZ)$ group (this will be reviewed in section \ref{subsec:SL2Z}). The relation between $p,q$ and the $k_i$ is
\begin{align}
\frac pq \ &= \ \frac{1}{k_1 - \frac{1}{k_2 - \frac{1}{\cdots - \frac{1}{k_r} }}} \quad .
\end{align} 
Then the SCFT corresponding to a sequence of 5-branes NS-$(p,q)$-NS, with $N$ D3-branes in the first segment and $\ti N$ D3-branes in the second, can be described as the infrared fixed point of a Chern-Simons quiver theory with $r-1$ ``interpolating" $T(U(\ti N))$ CFTs and $r-1$ ``interpolating" $\overline{T(U(N))}$ as shown in figure \ref{GWquiv}. In this picture a link labelled with $T(U(N))$ denotes a coupling between a $T(U(N))$ SCFT and the two adjacent $U(N)$ nodes by gauging the diagonal combination of the left $U(N)_L$ node with the $U(N)_F$ flavor symmetry of $T(U(N))$: $U(N)_L \times U(N)_F \rightarrow U(N)_{L'}$, and gauging the diagonal combination of the right $U(N)_R$ node with the $U(N)_J$ ``topological" symmetry of $T(U(N))$: $U(N)_R \times U(N)_J \rightarrow U(N)_{R'}$. The coupling between $T(U(N))$ and the left node $U(N)_L$ results in having a bifundamental hypermultiplet transforming in the $(N-1, \overline{N})\oplus (\overline{N-1},N)$ representation of $U(N-1)\times U(N)_{L'}$, with $U(N-1)$ the highest rank gauge node of $T(U(N))$, while the coupling between $T(U(N))$ and the right node $U(N)_{R}$ does not have a known microscopic description.

A link labelled with $\overline{T(U(N))}$ denotes a similar coupling between two adjacent $U(N)$ gauge nodes of the quiver theory and the $T(U(N))$ SCFT by the diagonal gaugings $U(N)_L \times [U(N)_F]^{\dagger} \rightarrow U(N)_{L'}$
\footnote{The diagonal gauging $U(N)_L \times U(N)_F \rightarrow U(N)_{L'}$ means that we gauge the transformation acting with the same matrix $M \in U(N)_L$ and $M \in U(N)_F$. The diagonal gauging $U(N)_L \times [U(N)_F]^{\dagger} \rightarrow U(N)_{L'}$ means that we gauge the transformation acting with the matrix $M \in U(N)_L$ and $M^{\dagger} \in U(N)_F$.}
and $U(N)_R \times U(N)_J \rightarrow U(N)_{R'}$.
This coupling between $T(U(N))$ and the left node $U(N)_L$ results in having a bifundamental hypermultiplet that transforms in the $(N-1, N)\oplus (\overline{N-1},\overline{N})$ representation of $U(N-1)\times U(N)_{L'}$.
\footnote{In \cite{Gaiotto:2008ak} the (subtle) difference between $T(U(N))$ and $\overline{T(U(N))}$ couplings was ignored.
}

\bigskip

A succession of $n$ $(p,q)$-5branes between two NS5-branes is associated to the same quiver theory as in figure \ref{GWquiv}, except that the central bifundamental hypermultiplet of $U(N)\times U(\ti N)$ is replaced by a sequence of $n$ Yang-Mills nodes connected by bifundamental hypermultiplets.

We will refer to these SCFTs realized with $(p,q)$-5branes as Gaiotto-Witten quiver theories (GW quivers). In general GW quiver theories do not admit a Lagrangian description, because of the interpolating $T(U(N))$ couplings, however this is not true for purely abelian theories. Abelian GW quivers involve interpolating $T(U(1))$ and $\overline{T(U(1))}$, that have a very simple Lagrangian description as BF couplings between adjacent $U(1)$ gauge nodes.


\section{$SL(2,\bZ)$ dualities}
\label{sec:SL2Z}

In this section we introduce the matrix model computing the partition function on $S^3$. We explain how to organize it into a sequence of 5-brane factors reproducing the sequence of the brane construction. Then we propose, following \cite{Gulotta:2011si}, an action of $SL(2,\bZ)$ duality on the 5-brane factors and show the equality of partition functions of $SL(2,\bZ)$-dual theories.

\subsection{Partition function on $S^3$}
\label{ssec:PartFunction}

In \cite{Kapustin:2009kz} it was shown that the partition function of three-dimensional SCFTs deformed by real mass and FI terms and defined on the round 3-sphere $S^3$ can be computed exactly and reduces to a simple matrix model. We summarize here the building blocks of the partition function matrix model on $S^3$ for $\N=4$ theories.

\bigskip

Through this paper we use the following short-hand notations
\begin{align*}
\sh(x) = 2\sinh(\pi x) \ , \quad \ch(x) = 2\cosh(\pi x) \ , \quad \thh(x) = \tanh(\pi x)
\end{align*}
\begin{align*}
x_{ij} = x_i - x_j \ , \quad \prod_j^N ... = \prod_{j=1}^N ... \quad , \quad  \prod_{j<k}^N ... = \prod_{1 \le j < k \le N } ... \quad , \quad
\prod_{j,k}^{N,M} ... = \prod_{j=1}^N \prod_{k=1}^M ...
\end{align*}
and similar notations for the sums $\Sigma_j^N =\Sigma_{j=1}^N, \cdots$.
\bigskip

The localisation on $S^3$ of the partition function $Z_{S^3}$ reduces the whole path integral to an integration over the Cartan subalgebra of the gauge group, divided by the order of the Weyl group $|\scW|$. We give here explicit formulas for a $U(N)$ gauge group.
\begin{align}
Z_{S^3} &=  \frac{1}{|\scW|} \int_{\textrm{Cartan}} d\sigma \ Z_{\rm classic} \,  Z_{\rm vector} \, Z_{\rm hyper} \ = \ \frac{1}{N!} \int \prod_{j}^N d\sigma_j  \ Z_{\rm classic} \,  Z_{\rm vector} \, Z_{\rm hyper} \ .
\label{3dZ}
\end{align}
The integrand receives contributions $Z_{\rm vector},Z_{\rm hyper}$ from the vector and hyper-multiplets of the theory, while $Z_{\rm classic}$ contains the contributions from Chern-Simons and FI terms. The $\sigma_j$ are called eigenvalues. For a $U(N)$ group, the $\sigma_j$ are the diagonal components of the hermitians matrices in the algebra $\mathfrak{u}(N)$.
The $\N=4$ vector multiplet contributes a factor
\begin{align}
 Z_{\rm vector} &= \textrm{det}_{\rm Adj} \Big( \sh (\sigma) \Big) \ = \  \prod_{i<j}^N \sh(\sigma_{ij})^2 \qquad , \ {\rm for} \ G=U(N) \ .
\end{align}
A hypermultiplet in a representation $R$ of the gauge group with real mass $m$  \footnote{It is meant that the two chiral multiplets transforming in $R$ and $\overline{R}$ have opposite real masses $-m$ and $m$ respectively.} contributes a factor
\begin{align}
Z_{\rm hyper} &= \textrm{det}_{R} \lp \frac{1}{\ch(\sigma - m)} \rp \no\\
&= \prod_j^N \frac{1}{\ch(\sigma_j - m)} \qquad , \ {\rm for} \ R \oplus \overline{R} = N \oplus \overline{N} \ {\rm of} \ U(N)  \\
 &=   \prod_{i,j}^{N,\ti N} \frac{1}{\ch(\sigma_i- \ti\sigma_j - m)}  \quad , \ {\rm for} \ R \oplus \overline{R} = (N,\overline{\ti N}) \oplus (\overline{N}, \ti N) \ {\rm of} \ U(N)\times U(\ti N)  \  .  \no
\end{align}
An $\N=3$ Chern-Simons term with level $k$ contributes a factor
\begin{align}
Z_{\rm CS} &= \textrm{det}_F \Big( e^{ i\pi k \sigma^2} \Big) \ = \ e^{i\pi k \sum_j^N \sigma_j^2  } \  .
\end{align}
A real Fayet-Iliopoulos term with parameter $\eta$ contributes a factor
\begin{align}
Z_{\rm FI} &= \textrm{det}_F \Big( e^{2 i \pi \eta \sigma} \Big) \ = \  e^{2i\pi \eta \sum_j^N \sigma_j } \ .
\end{align}
Here $\det_R$ is the
the determinant in the representation $R$. The indices $F$ and
${\rm Adj}$ refer to the fundamental and adjoint representations respectively.
\bigskip

\begin{figure}[th]
\centering
\includegraphics[scale=0.5]{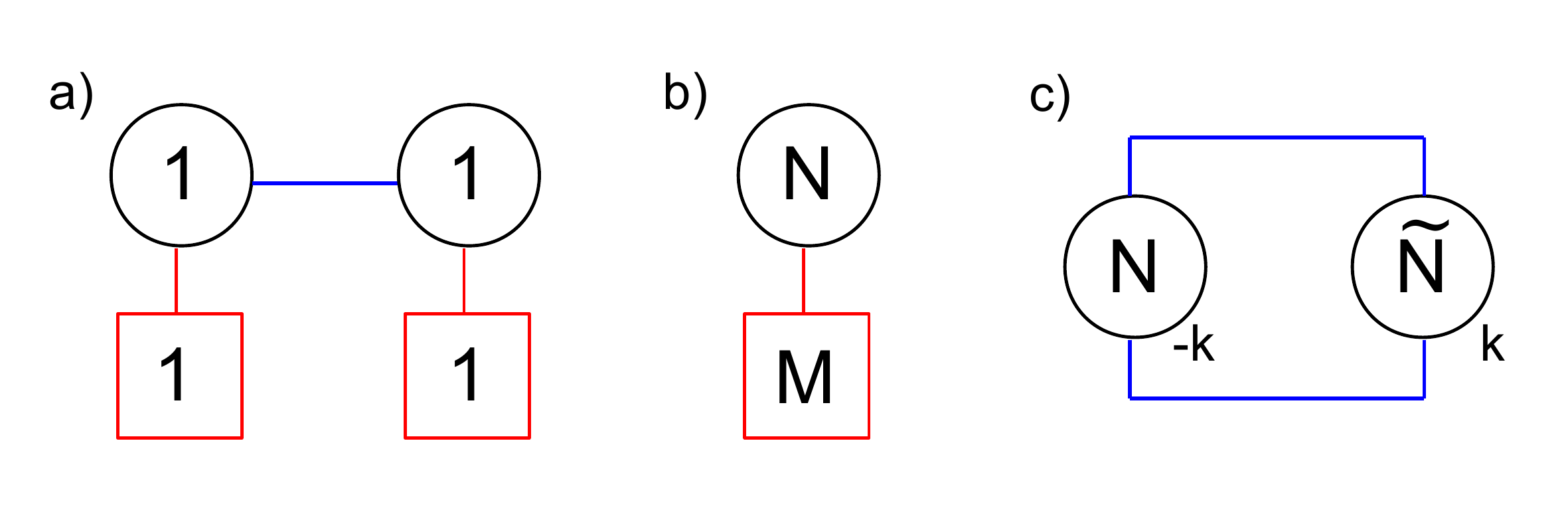}
\vspace{-0.5cm}
\caption{\footnotesize Some quiver theories.}
\label{quivexamples}
\end{figure}

Let us give a few examples. The partition functions of the quiver SCFTs shown in figure \ref{quivexamples} are given respectively by
\begin{align}
Z_{a} &= \int d\sigma d\ti\sigma \ \frac{e^{2\pi i \eta \sigma} \ e^{2\pi i \ti\eta \ti\sigma }}{\ch(\sigma -m) \, \ch(\sigma-\ti\sigma) \, \ch(\ti\sigma - \ti m)}  \\
Z_{b} &= \int \frac{d^N \sigma}{N!} \ e^{2\pi i \eta \sum_j^N \sigma_j} \frac{\prod_{i<j}^N \sh(\sigma_{ij})^2 }{ \prod_{a,j}^{M,N} \ch(\sigma_j-m_a)}  \\
Z_{c} &=  \int \frac{d^N \sigma}{N!} \frac{d^{\ti N} \sigma}{\ti N!} \ 
e^{-\pi i k \sum_j^N \sigma_j^2} \, e^{\pi i k \sum_j^{\ti N} \ti\sigma_j^2} \,
e^{2\pi i\eta \sum_j^N \sigma_j} \, e^{2\pi i \ti\eta \sum_j^{\ti N} \ti\sigma_j} 
\frac{\prod_{i<j}^N \sh(\sigma_{ij})^2 \prod_{i<j}^{\ti N} \sh(\ti\sigma_{ij})^2}{ [ \, \prod_{i,j}^{N, \ti N} \ch(\sigma_i-\ti\sigma_j)\, ]^2} \  ,
\end{align}
where $\eta,\ti\eta$ are real FI parameters, $m, \ti m, m_a$ are real mass parameters and $k, -k$ are Chern-Simons levels.

The partition function of the $T(SU(N))$ theory with fundamental hypermultiplet masses $m_a$ and FI parameters $\eta_p$ is given by
\begin{align}
& Z_{T(SU(N))} \\
&= \int \prod_{p=1}^{N-1} \lp \frac{d^p\sigma^{(p)}}{p!} \ \prod_{i<j}^p \sh \lp \sigma^{(p)}_{ij} \rp^2 \rp 
\frac{\prod_{p=1}^{N-1} e^{2\pi i \, \eta_p \sum_j^p \sigma^{(p)}_j} }{\prod_{p=1}^{N-2} \prod_{i,j}^{p,p+1}\ch[ \sigma^{(p)}_i - \sigma^{(p+1)}_j ] } \ 
\frac{1}{\prod_{a,j}^{N,N-1}\ch[ \sigma^{(N-1)}_j - m_a ] } \no
\end{align}
The partition function for the theory $T(U(N))$ is the same with the addition of a FI coupling for the flavor $U(N)_F$ global symmetry (background BF term):
\begin{align}
Z_{T(U(N))} & = e^{2\pi i \, t_N \sum_j^N m_j} \ Z_{T(SU(N))} \ ,
\end{align}
with $t_N$ the FI parameter.
The evaluation of this matrix model was carried out in \cite{Benvenuti:2011ga, Nishioka:2011dq} and gives
\begin{align}
 Z^{T(U(N))} &= (-i)^{\frac{N(N-1)}{2}} \frac{ \sum_{w \in \scS_N} (-1)^w \, e^{2i\pi \sum_j^N t_j m_{w(j)}}}{\prod_{j<k}^N \sh(t_j - t_k)\sh(m_j - m_k)}
 \label{ZTUN}
\end{align}
with $\scS_N$ the group of permutations of $N$ elements and the $t_j$ defined by the relations $\eta_j = t_j - t_{j+1}$, $1 \le j \le N-1$. $Z^{T(U(N))}$ is invariant under the exchange of the mass and FI parameters $m_j \leftrightarrow t_j$, as it is expected from mirror symmetry.

\subsection{Repackaging of matrix models}
\label{subsec:Conventions}

The partition function's matrix model of a given quiver theory can be recast into a sequence of elementary matrix factors of two types, that are naturally associated to the two types of 5branes entering into the brane realization of the quiver. This sequence of elementary matrix factors reproduces the sequence of 5-branes, or dots, of the brane realization, or graph.
We will treat the cases Yang-Mills and Chern-Simons quivers that involve NS5, D5 and $(1,k)$-5branes. The $(p,q)$-5branes matrix factors will appear later.

\bigskip

\begin{figure}[th]
\centering
\includegraphics[scale=0.8]{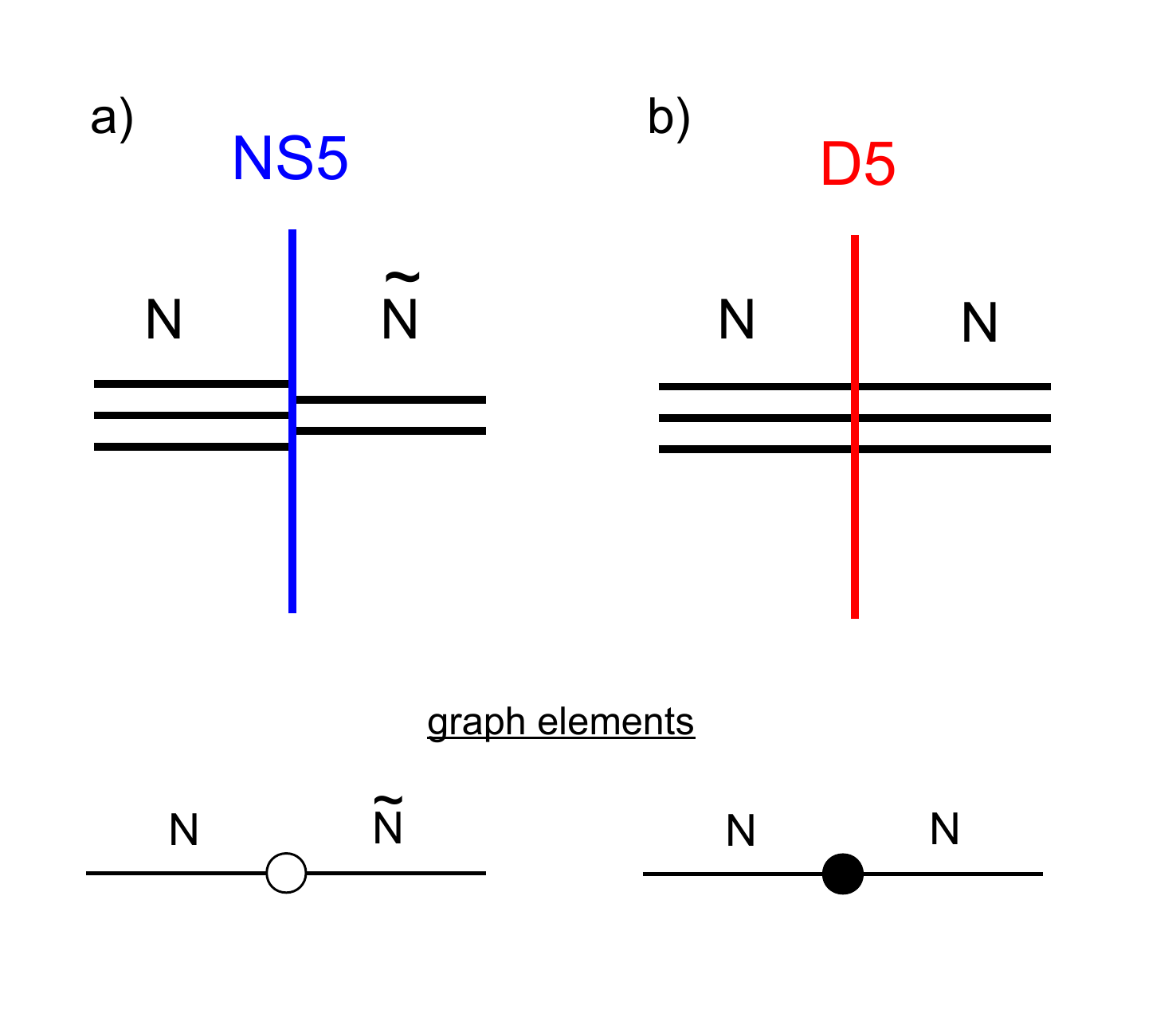}
\vspace{-1cm}
\caption{\footnotesize{ 5branes building blocks : a) NS5-brane. b) D5-brane. Below: corresponding graph elements in the graph description.}}
\label{braneblocks}
\end{figure}

Let us start with Yang-Mills quivers.
The brane realization and associated graph can be seen as a sequence made of two elementary 5-brane building blocks: one for the NS5-brane and one for the D5-brane as shown in figure \ref{braneblocks}.
Each 5-brane block can be associated with a matrix factor depending on two sets on eigenvalues $\{ \sigma_j , \ti \sigma_j \}$, which are associated to the D3-branes ending on the left and on the right of the 5-brane respectively. The matrix model computing the partition function of the quiver theory is obtained by assembling the matrix factors of 5-branes blocks in a sequence reproducing the sequence of blocks of the brane realization, then identifying the eigenvalues of adjacent blocks (those corresponding to the same D3-segment) and finally integrating over all eigenvalues.

The NS5 matrix factor is given by
\begin{align}
\label{NS5factor}
\pq{1}{0}_{\sigma \, \ti\sigma} &= \frac{1}{(N!\ti N!)^{1/2}} \, e^{-2i\pi \, t \lp \sum_j^N \sigma_j^2 - \sum_j^{\ti N} \ti\sigma_j^2 \rp}\frac{\prod_{i<j}^N \sh \, \sigma_{ij} \ \prod_{i<j}^{\ti N} \sh \, \ti\sigma_{ij}} {\prod_{i,j}^{N,\ti N} \ch(\sigma_i - \ti\sigma_j) } \ .
\end{align}
It contains a bifundamental hypermultiplet matrix factor,  ``half" the matrix factors of the $\sigma$ and $\ti\sigma$ vector multiplets and ``half" the matrix factors of the $\sigma$ and $\ti\sigma$ FI terms. In the brane picture $t$ is realted to the position of the NS5-brane in transverse space and it contributes to the FI terms of the two adjacent gauge nodes through the relations $\eta_j = t_j - t_{j+1}$, with $\eta_j$ the FI parameter of the $j$th node and $t_j$ the parameter of the $j$th NS5-brane. The factor $(N!\ti N!)^{-1/2}$ are ``half" the Weyl order of the left $U(N)$ and right $U(\ti N)$ groups.

The D5 matrix factor is given by
\begin{align}
\label{D5factor}
\pq{0}{1}_{\sigma \, \ti\sigma} &= \frac{1}{\prod_j^N \ch(\sigma_j -m)} \ \frac{1}{N!} \sum_{w \in \scS_N} (-1)^{w} \prod_j^N \delta(\sigma_j-\ti\sigma_{w(j)}) ,
\end{align}
where $\scS_N$ is the set of permutations of $N$ elements. The D5-factor contains the contribution of a fundamental hypermultiplet of mass $m$. The $\sigma$ and $\ti\sigma$ eigenvalues are associated to the $N$ D3-segments ending on the left and on the right of the D5-brane and are identified, as they should, with the $\delta$ functions. The averaging over permutations $w \in \scS_N$ ensures the anti-symmetrization of the factor with respect to permutations of the $\sigma$ or $\ti\sigma$ eigenvalues, without affecting the matrix model. This anti-symmetrization will be naturally present in all 5-brane factors. In the brane picture $m$ is here again related to the position of the D5-brane in transverse space.

\bigskip

The $(1,k)$-5brane matrix factor is given by
\begin{align}
\label{1k5factor}
\pq{1}{k}_{\sigma \, \ti\sigma} &= \frac{1}{(N!\ti N!)^{1/2}} \, e^{-2i\pi \, t \lp \sum_j^N \sigma_j - \sum_j^{\ti N} \ti\sigma_j \rp} \
e^{\pi i \, k \lp \sum_j^N \sigma_j^2 - \sum_j^{\ti N} \ti\sigma_j^2 \rp}  \
\frac{\prod_{i<j}^N \sh \, \sigma_{ij} \ \prod_{i<j}^{\ti N} \sh \, \ti\sigma_{ij} } {\prod_{i,j}^{N,\ti N} \ch(\sigma_i - \ti\sigma_j) } \ .
\end{align}
In the brane picture $t$ is the position of the 5-brane in transverse space and it contributes to the FI terms of the left and right gauge nodes as in the NS5-brane case, which corresponds to $k=0$.
As for the NS5-brane this factor contains the contribution of a bifundamental hypermultiplet and ``half" the contributions of the vector multiplet of the two nodes on the sides of the $(1,k)$-5brane. Moreover this factor includes the contribution from the Chern-Simons term induced by the 5-brane, with levels $\pm k$ in the two nodes.


\bigskip

To express conveniently matrix models, we define a ``matrix-like" product:
\begin{align}
\lp \pq{p_1}{q_1}{}_{(N)} \pq{p_2}{q_2}\rp_{\sigma \, \ti\sigma} &= \int d^N \sigma' \ \pq{p_1}{q_1}_{\sigma \, \sigma'} \pq{p_2}{q_2}_{\sigma' \, \ti\sigma} \ .
\label{blockproduct}
\end{align}
This product \footnote{We may sometimes use the even shorter notation $\pq{p_1}{q_1}{} \pq{p_2}{q_2}_{\sigma \, \ti\sigma}$for such a product.} is associative, but not commutative. Here $(p_i,q_i)$ can be $(0,1)$, $(1,0)$ or $(1,k)$, denoting a D5, NS5 or $(1,k)$-5brane factor. This ``matrix-like" product will extend naturally to other matrix factors that we define later in section \ref{subsec:SL2Z} (general $(p_i,q_i)$ 5-brane factors and duality-wall factors).
We also define a ``trace" for a factor depending on two sets of $N$ eigenvalues $F_{\sigma \, \ti\sigma}$ (``square matrix"):
\begin{align}
\tr \, F_{(N)} &= \int d^N \sigma \ F_{\sigma \, \sigma} \ .
\end{align}
\bigskip

\begin{figure}[th]
\centering
\includegraphics[scale=0.8]{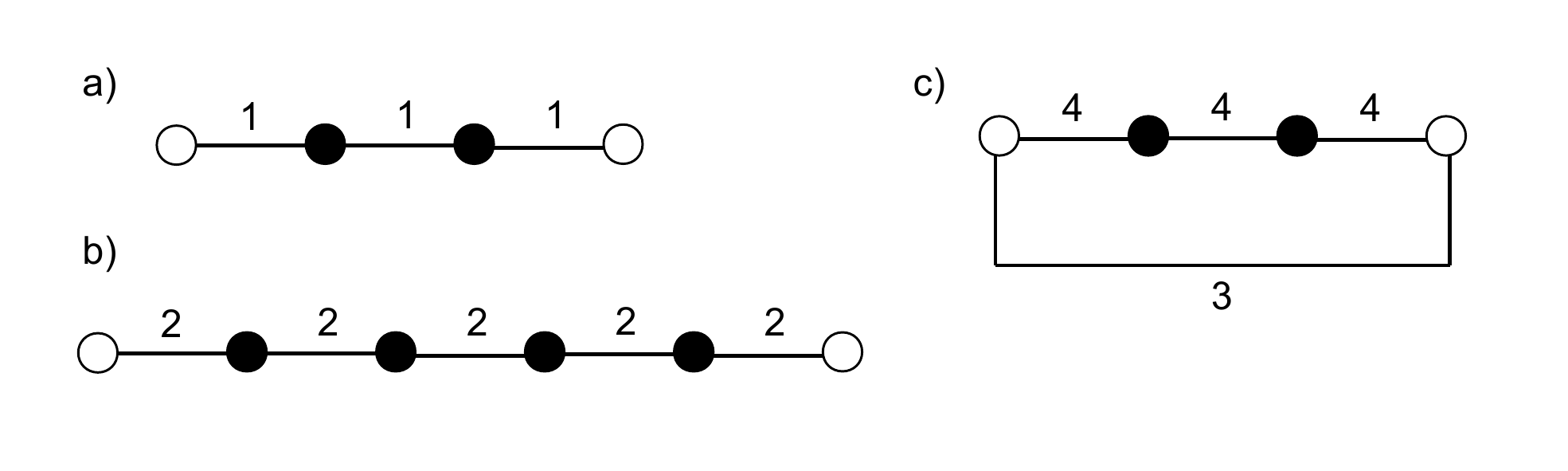}
\vspace{-0.5cm}
\caption{\footnotesize{a) Graph of $T(SU(2))$. b) Graph of $T^{(1111)}_{(22)}$. c) Graph of $U(3)\times U(4)$ circular quiver with two $U(4)$ fundamental hypermultiplets.}}
\label{graphexamples}
\end{figure}

This repackaging of the matrix model into elementary 5brane factors allows us to express the partition function's matrix model in a compact way, as a sequence of 5brane factors.
The partition function of a Yang-Mills linear quiver SCFT is expressed as\footnote{The sequence of factors here is just an example. In general it is given by its quiver data.}
\begin{align}
Z_{\rm YM} &=  \pq{1}{0}{}_{(N_1)} \pq{0}{1}{}_{(N_1)} \pq{0}{1}{}_{(N_1)} \pq{1}{0}{}_{(N_2)} \pq{0}{1}{}_{(N_2)} \pq{1}{0}{}_{(N_3)} \ \cdots \ {}_{(N_{\wat P -1})} \pq{1}{0} \ ,
\end{align}
where the first NS5-factor has no left-eigenvalues (no D3-segments on the left of the NS5-brane) and the last NS5-factor has no right-eigenvalues (no D3-segments on the right of the NS5-brane). The sequence of factors reproduces the sequence of NS5 and D5-branes of the quiver brane picture, or equivalently the sequence of white and black dots of the quiver's graph, shown in figure \ref{graphquiv}.\\
For Yang-Mills circular quivers the partition function is expressed as
\begin{align}
Z_{\rm YM} &= \tr  \  \pq{1}{0}{}_{(N_1)} \pq{0}{1}{}_{(N_1)} \pq{0}{1}{}_{(N_1)} \pq{1}{0}{}_{(N_2)} \pq{0}{1}{}_{(N_2)} \pq{1}{0}{}_{(N_3)} \ \cdots \ {}_{(N_{\wat P -1})} \pq{1}{0}{}_{(N_{\wat P})}  \ .
\end{align}
Again the sequence of factors reproduces the sequence of NS5 and D5-branes. \\
Let us give a few examples. The partition function of the $T(SU(2))$ theory is given by
\begin{align}
Z_{T(SU(2))} = \pq{1}{0}{}_{(1)} \pq{0}{1}{}_{(1)} \pq{0}{1}{}_{(1)} \pq{1}{0} \ ,
\end{align}
the partition function of the $U(2)$ theory with four fundamental hypermultiplets is given by:
\begin{align}
Z = \pq{1}{0}{}_{(2)} \pq{0}{1}{}_{(2)} \pq{0}{1}{}_{(2)} \pq{0}{1}{}_{(2)} \pq{0}{1}{}_{(2)} \pq{1}{0} \ ,
\end{align}
and the partition function of the $U(3)\times U(4)$ circular quiver theory with two fundamental hypermultiplets for the $U(4)$ node is given by:
\begin{align}
Z = \tr \  \pq{1}{0}{}_{(4)} \pq{0}{1}{}_{(4)} \pq{0}{1}{}_{(4)} \pq{1}{0}{}_{(3)} \ .
\end{align}
The corresponding graphs are shown in figure \ref{graphexamples}.

For Chern-Simons quivers, realized by a sequence of NS5-branes and $(1,k)$-5branes, the partition function is similarly expressed in the linear case by:
\begin{align}
Z_{\rm CS} &=  \pq{1}{0}{}_{(N_1)} \pq{1}{k}{}_{(N_1)} \pq{1}{k}{}_{(N_1)} \pq{1}{0}{}_{(N_2)} \pq{1}{k}{}_{(N_2)} \pq{1}{0}{}_{(N_3)} \ \cdots \ {}_{(N_{\wat P -1})} \pq{1}{0} \ ,
\end{align}
and for circular quivers:
\begin{align}
Z_{\rm CS} &= \tr  \  \pq{1}{0}{}_{(N_1)} \pq{1}{k}{}_{(N_1)} \pq{1}{k}{}_{(N_1)} \pq{1}{0}{}_{(N_2)} \pq{1}{k}{}_{(N_2)} \pq{1}{0}{}_{(N_3)} \ \cdots \ {}_{(N_{\wat P -1})} \pq{1}{0}{}_{(N_{\wat P})}  \ .
\end{align}
Again the sequence of 5brane blocks in $Z$ reproduces the sequence of 5branes, or dots, in the brane realization, or graph.

\begin{figure}[th]
\centering
\includegraphics[scale=0.8]{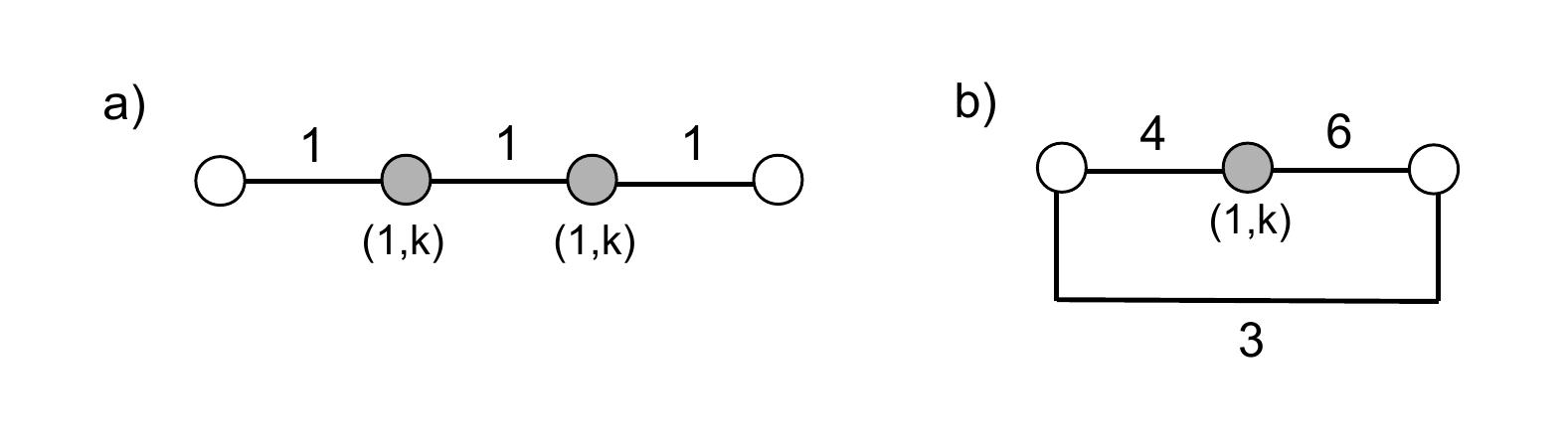}
\vspace{-0.5cm}
\caption{\footnotesize{a) Graph of $U(1)_k \times U(1)_{0} \times U(1)_{-k}$ linear CS quiver theory. b) Graph of $U(4)_k \times U(6)_{-k} \times U(3)_0$ circular CS quiver theory.}}
\label{CSgraphexamples}
\end{figure}

\noindent For instance the partition function of the linear quiver $U(1)_k \times U(1)_{0} \times U(1)_{-k}$ is given by
\begin{align}
Z =  \pq{1}{0}{}_{(1)} \pq{1}{k}{}_{(1)} \pq{1}{k}{}_{(1)} \pq{1}{0} \ ,
\end{align}
while the partition function of the circular quiver $U(4)_k \times U(6)_{-k} \times U(3)_0$ is given by
\begin{align}
Z = \tr \  \pq{1}{0}{}_{(4)} \pq{1}{k}{}_{(6)} \pq{1}{0}{}_{(3)} \ .
\end{align}
The corresponding graphs are shown in figure \ref{CSgraphexamples}.

\subsection{Local SL(2,\bZ) action on matrix models}
\label{subsec:SL2Z}

 The quiver SCFTs realized on brane configurations must inherit $SL(2,\bZ)$ dualities, from the $SL(2,\bZ)$ symmetry of type IIB string theory. For instance the standard mirror symmetry, discussed in \cite{Intriligator:1996ex}, is known to be related to the type IIB $S$-duality.

 In the case of standard mirror symmetry, dual pairs of quiver theories are found by identifying the brane realization of one theory with the $S$-dual of the brane realization of the other theory \cite{deBoer:1996ck, deBoer:1996mp}. One theory is realized by a sequence of NS5 and D5-branes, while the mirror-dual theory is realized with the same sequence of 5branes, with NS5 and D5-branes exchanged (or NS5 $\rightarrow$ D5, D5 $\rightarrow\, \overline{\rm NS5}$, see below). 
 
 More dualities can be found by identifying the brane realizations of quiver theories through general $SL(2,\bZ)$ transformations of IIB string theory. We will give evidence for these more general dualities by matching the partition function of dual SCFTs. The dualities will actually extend to ``local" $SL(2,\bZ)$ dualities, that were already discussed in \cite{Gulotta:2011si} and implicitly studied in \cite{Gaiotto:2008ak}.
 
\bigskip

Let us first remind and generalize the picture developed in \cite{Gulotta:2011si} of $SL(2,\bZ)$ action on the 5-brane factors of the matrix model.\\
The $SL(2,\bZ)$ action on the brane realization of the 3d theories is easily understood. $M \in SL(2,\bZ)$ leaves the D3 branes invariant and transforms a $(p,q)$ 5brane into a $(p',q')$ 5brane with
\begin{align}
M &= \abcd{a}{b}{c}{d} \ , \quad 
\binom{p'}{q'} = M \ \binom{p}{q} = \binom{ap + bq}{cp + dq} \ ,
\label{SL2Ztransfo}
\end{align}
where $ad-bc = 1$ and $p \wedge q =1$. 
The group $SL(2,\bZ)$ is generated by the transformations $S$ and $T$ given by
\begin{align*}
S &= \abcd{0}{-1}{1}{0} \ , \quad 
T = \abcd{1}{0}{1}{1} \ ,
\end{align*}
satifying $S^2=-1$ and $(ST)^3=1$.

\begin{figure}[th]
\centering
\includegraphics[scale=0.8]{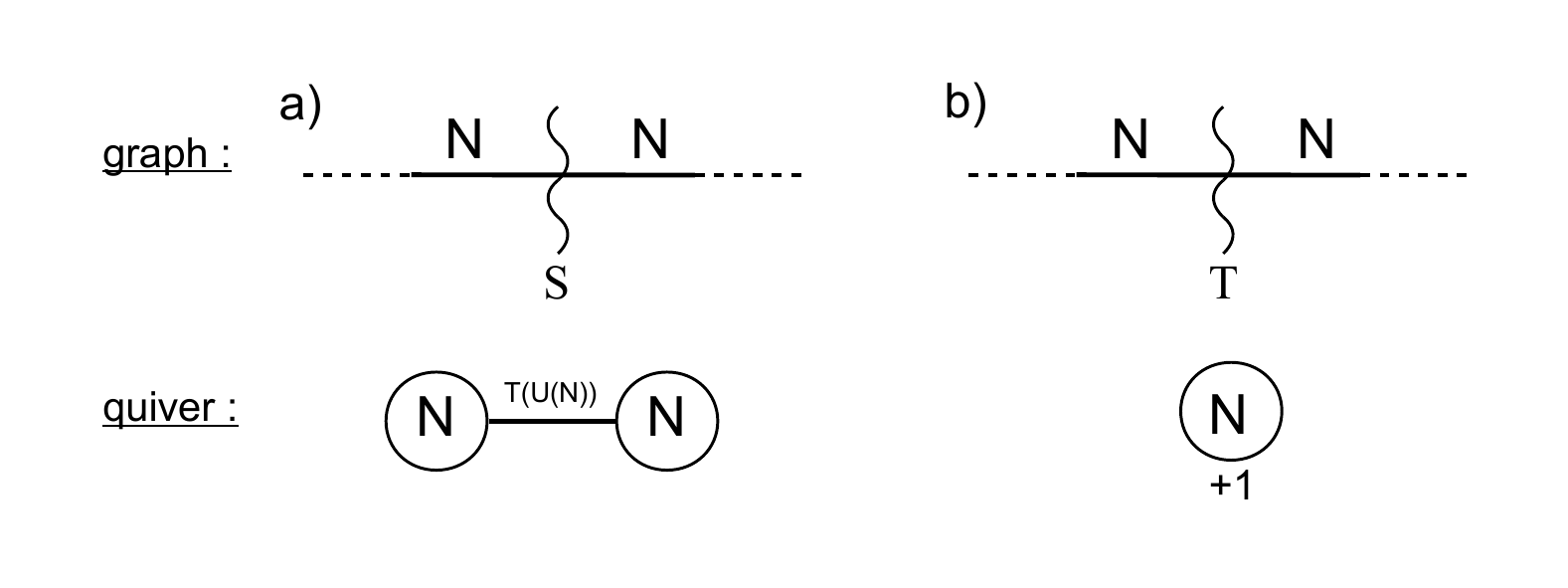}
\vspace{-1cm}
\caption{\footnotesize{Graph representations and quiver descriptions of the $S$-duality wall (a) and $T$-duality wall (b). The subscript $+1$ indices a Chern-Simons term at level $k=1$.}}
\label{STwall}
\end{figure}

In \cite{Gaiotto:2008ak} Gaiotto and Witten suggested that $SL(2,\bZ)$ transformations can be realized locally on the brane configuration, in a region  containing a single 5-brane (spanning an interval along $x^3$ around the 5-brane). 
The authors of \cite{Gulotta:2011si} reformulated their results in the following way. 
The local action of the transformation $S$ on a $(p,q)$-5brane trades the 
$(p,q)$-5brane for a $(-q,p)$-5brane and creates an $S$-duality wall on its right and a $S^{-1}$ duality wall on its left (hyperplanes at fixed $x^3$). The duality walls are the boundaries of the region of local $SL(2,\bZ)$ action and can be considered as new objects in the brane construction. 
From the analysis of \cite{Gaiotto:2008ak} it follows that a $S$-duality wall intersecting $N$ D3-branes induces in the quiver SCFT an interpolating $T(U(N))$ coupling between two $U(N)$ gauge nodes (see sec. \ref{ssec:GWquivers}), that are associated to the D3-branes on the left and on the right of the $S$-wall, as shown in figure \ref{STwall}. And the $S^{-1}$-wall intersecting $N$ D3-branes induces an interpolating $\overline{T(U(N))}$ coupling between the two $U(N)$ nodes.

Similarly the local action of $T$ on a $(p,q)$-5brane trades the 
$(p,q)$-5brane for a $(p,p+q)$-5brane with a $T$-duality wall on its right and a $T^{-1}$-duality wall on its left. From \cite{Gaiotto:2008ak} it follows that a $T$-duality wall intersecting $N$ D3-branes induces in the quiver SCFT a Chern-Simons term with level $k=+1$ for a $U(N)$ node, associated with the $N$ D3-branes crossing the $T$-wall, as shown in figure \ref{STwall}. The $T^{-1}$-wall induces a Chern-Simons term with coupling $k=-1$.

\noindent The iterations of $S$ and $T$ local actions bringing a $(p,q)$-5brane to a NS5-brane lead to the quiver theory description of SCFTs realized with $(p,q)$-5branes, presented in section \ref{ssec:GWquivers}.

It is rather easy now to associate matrix factors to the $S$- and $T$-duality walls, as we did for the 5branes. The $S$-wall with $N$ D3-branes on both sides must be associated to the contribution of an interpolating $T(U(N))$, plus ``half" the contribution of the left and right $U(N)$ vector multiplets and some phase $e^{i \theta}$ to be fixed:
\begin{align}
S_{\sigma \ti\sigma} = \frac{e^{i \theta}}{N!} \, \prod_{i<j}^N \sh \sigma_{ij} \, Z_{T(U(N))}[\sigma,\ti\sigma] \, \prod_{i<j}^N \sh \ti\sigma_{ij} \ &= \ \frac{e^{i \theta} (-i)^{\frac{N(N-1)}{2}}}{N!} \, \sum_{w \in \scS_N} e^{2\pi i \sum_{j=1}^N \sigma_j \ti\sigma_{w(j)}} \no\\
&  \longrightarrow \quad  e^{i \theta_{S}} \ e^{2\pi i \sum_{j=1}^N \sigma_j \ti\sigma_j} \ ,
\end{align} 
where $Z_{T(U(N))}[\sigma,\ti\sigma]$ denotes the partition function of $T(U(N))$ with mass parameters $\sigma_j$ and FI parameters $\ti\sigma_j$. These are identified with the eigenvalues of the $U(N)$ left and right gauge nodes, as required by the diagonal gaugings \footnote{The masses $m_j$ of fundamental hypermultiplets are background values of the real scalar in a $U(N)_F$ vector multiplet, as $\sigma_j$ are background values for the scalar in the $U(N)_L$ vector multiplet. The diagonal gauging $U(N)_L \times U(N)_F \rightarrow U(N)_{L'}$ amounts to identifying the eigenvalues $\sigma_j$ with the masses $m_j$. Similarly the other diagonal gauging leads to the indentification of the eigenvalues $\ti\sigma_j$ with the FI parameters $t_j$.}. The second equality is obtained after plugging in~\eqref{ZTUN}. Since the 5-branes factors, that necessarily stand on the left and on the right of the $S$-wall in the graph, are anti-symmetric under permutations of the $\sigma_j$ or $\ti\sigma_j$ eigenvalues, it is does not affect the matrix models to replace the average over permutations $w \in \scS_N$ above by a single term $e^{2\pi i \sum_{j=1}^N \sigma_j \ti\sigma_j}$, that we can associate to the $S$-duality wall. We named $\theta_S$ the overall phase of $S_{\sigma \ti\sigma}$.

\noindent For the $T$-duality wall with $N$ D3-branes on both sides we can associate the factor
 \begin{align}
T_{\sigma \, \ti\sigma} = e^{i \theta_T} \ e^{\pi i \sum_{j=1}^N \sigma_j^2} \ \prod_{j=1}^N \delta(\sigma_j - \ti\sigma_j) \ ,
\end{align}
corresponding to the addition of a Chern-Simons term contribution at level $k=1$ and the identifications of the eigenvalues of the left and right nodes $\sigma=\ti\sigma$, so that the quiver theory has a single node for the $N$ D3-branes crossing the $T$-wall. The phases $e^{i \theta_T}$ and $e^{i \theta_S}$ must be fixed for consistency with the $SL(2,\bZ)$ group relations as explained below.

To summarize, the $S$- and $T$-wall elements of the graph (or brane realization) are associated with the elementary matrix factors:
\begin{align}
S_{\sigma \, \ti\sigma} &= e^{i \theta_S} \ e^{2\pi i \sum_{j=1}^N \sigma_j \ti\sigma_j} \ , \qquad 
T_{\sigma \, \ti\sigma} = e^{i \theta_T} \ e^{\pi i \sum_{j=1}^N \sigma_j^2} \ \prod_{j=1}^N \delta(\sigma_j - \ti\sigma_j) \ .
\label{STfactors}
\end{align}
These duality-wall factors were already proposed in \cite{Gulotta:2011si}. Our derivation emphasizes that these factors follow from the description of GW quivers of \cite{Gaiotto:2008ak}.

\noindent Upon fixing $e^{4i\theta_S}=1$ and $e^{3i(\theta_T+\theta_S)} = e^{\frac{\pi i N}{4}}$, the matrix factors obey the $SL(2,\bZ)$ relations $(S^2)_{\sigma \, \ti\sigma}=(-1)_{\sigma \, \ti\sigma}$ and $(ST)^3_{\sigma \, \ti\sigma}=1_{\sigma \, \ti\sigma}$ with $(-1)_{\sigma \, \ti\sigma} =e^{2i \theta_S} \prod_{j=1}^N \delta(\sigma_j + \ti\sigma_j)$ and $1_{\sigma \, \ti\sigma} = \prod_{j=1}^N \delta(\sigma_j - \ti\sigma_j)$. $\pm 1_{\sigma \, \ti\sigma}$ are the matrix factors for $\pm \mathbf{1} \in SL(2,\bZ)$. 


\bigskip

The local action of an arbitrary transformation $M \in SL(2,\bZ)$ can be decomposed into a sequence of $S$ and $T$ actions. The matrix factor $M_{\sigma \, \ti\sigma}$ for the $M$-wall is given by the product of the $S_{\sigma \, \ti\sigma}$ and $T_{\sigma \, \ti\sigma}$ factors, using~\eqref{blockproduct}, reproducing the sequence of $S$ and $T$ transformations. In particular this gives
\begin{align}
S^{-1}_{\sigma \, \ti\sigma} &= e^{-i \theta_S} \ e^{-2\pi i \sum_{j=1}^N \sigma_j \ti\sigma_j} \ , \quad 
T^{-1}_{\sigma \, \ti\sigma} = e^{-i \theta_T} \ e^{-\pi i \sum_{j=1}^N \sigma_j^2} \ \prod_{j=1}^N \delta(\sigma_j - \ti\sigma_j)  \ ,
\end{align}
which are in agreement with the quiver description as interpolating $\overline{T(U(N))}$ coupling for the $S^{-1}$-wall and additional Chern-Simons coupling with level $k=-1$ for the $T^{-1}$-wall.\\

To be concrete let us show explicitly the computation of the factors for the $S^{-1}$-wall and $T^{-1}$-wall coupled to $U(N)$ nodes. We have $S^{-1} = TSTST$ and $T^{-1} = STSTS$, so the matrix factors are given by:\footnote{In the computations there is a sign ambiguity in the evaluation of the integrals $\int d\sigma e^{i\pi k \sigma^2} = \pm e^{-i \, {\rm sgn}(k)\pi/4} \sqrt{\frac{1}{|k|}}$, for $k \neq 0$ real. We choose $\int d\sigma e^{i\pi k \sigma^2} = e^{-i \, {\rm sgn}(k)\pi/4} \sqrt{\frac{1}{|k|}}$, keeping in mind that the overall sign may not be uniquely defined.}
\begin{align*}
 S^{-1}_{\sigma \, \ti\sigma} &= (TSTST)_{\sigma \, \ti\sigma} \ , \qquad T^{-1}_{\sigma \, \ti\sigma} = (STSTS)_{\sigma \, \ti\sigma} \\
 (ST)_{\sigma \, \ti\sigma}&= e^{i (\theta_T + \theta_S)} \int d^N \sigma' \ e^{2\pi i \sum_{j=1}^N \sigma_j \sigma'_j} \ e^{\pi i \sum_{j=1}^N \sigma'_j{}^2} \ \prod_{j=1}^N \delta(\sigma'_j - \ti\sigma_j) \\
& = \ e^{i(\theta_T+\theta_S)} \ e^{2\pi i \sum_{j=1}^N \sigma_j \ti\sigma_j} \ e^{\pi i \sum_{j=1}^N \ti\sigma_j{}^2} \\
 ((ST)^2)_{\sigma \, \ti\sigma} &= e^{2i (\theta_T+ \theta_S)} \int d^N \sigma' \ e^{2\pi i \sum_{j=1}^N \sigma_j \sigma'_j} \ e^{\pi i \sum_{j=1}^N \sigma'_j{}^2} e^{2\pi i \sum_{j=1}^N \sigma'_j \ti\sigma_j} \ e^{\pi i \sum_{j=1}^N \ti\sigma_j{}^2}
\\ & =  e^{2i (\theta_T+\theta_S)-\frac{\pi i N}{4}} \ e^{-\pi i \sum_{j=1}^N \sigma_j{}^2} e^{-2\pi i \sum_{j=1}^N \sigma_j \ti\sigma_j} \\
 S^{-1}_{\sigma \, \ti\sigma} &=  e^{i (3\theta_T+2\theta_S)-\frac{\pi i N}{4}} \int d^N \sigma' \, e^{\pi i \sum_{j=1}^N \sigma_j{}^2} \, \prod_{j=1}^N \delta(\sigma_j - \sigma'_j) \ e^{-\pi i \sum_{j=1}^N \sigma'_j{}^2} e^{-2\pi i \sum_{j=1}^N \sigma'_j \ti\sigma_j} \\
& = \ e^{-i\theta_S} \, e^{-2\pi i \sum_{j=1}^N \sigma_j \ti\sigma_j} \\
 T^{-1}_{\sigma \, \ti\sigma} &= \ e^{i (2\theta_T +3\theta_S)-\frac{\pi i N}{4}} \int d^N \sigma' \ e^{-\pi i \sum_{j=1}^N \sigma_j{}^2} e^{-2\pi i \sum_{j=1}^N \sigma_j \sigma'_j} e^{2\pi i \sum_{j=1}^N \sigma'_j \ti\sigma_j} \\
& = \ e^{-i\theta_T} e^{-\pi i \sum_{j=1}^N \sigma_j{}^2} \prod_{j=1}^N \delta(\sigma_j - \ti\sigma_j) \ .
\end{align*}
From now on we will assume the particular values:
\begin{align}
\theta_S = 0 \quad , \quad \theta_T = \frac{\pi N}{12} \ ,
\end{align}
which are compatible with the conditions $e^{4i\theta_S}=1$ and $e^{3i(\theta_T+\theta_S)} = e^{\frac{\pi i N}{4}}$.
\bigskip

\begin{figure}[th]
\centering
\includegraphics[scale=0.8]{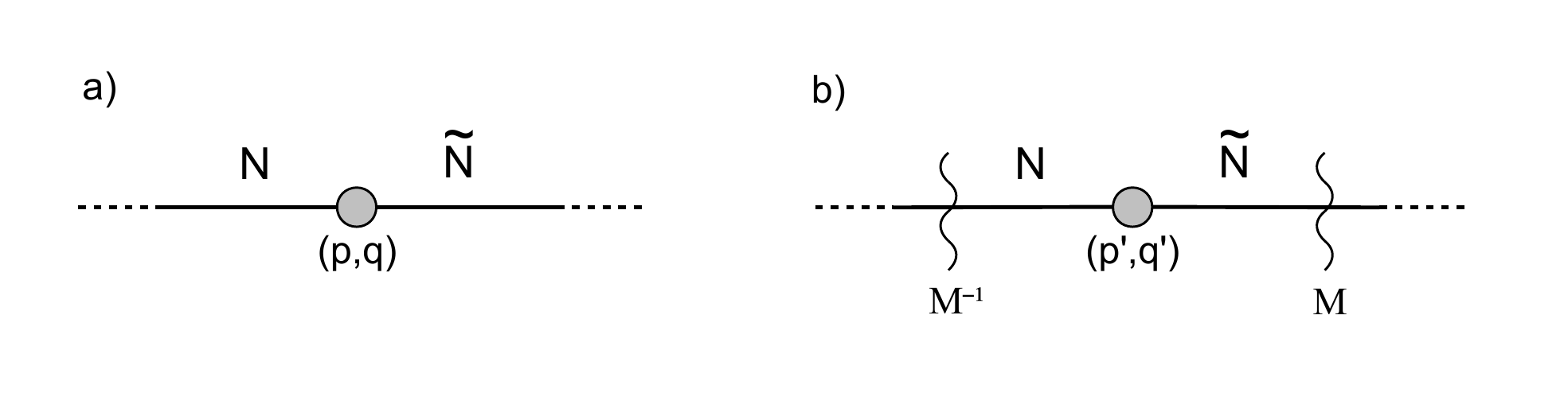}
\vspace{-1cm}
\caption{\footnotesize{a) Part of a graph with a $(p',q')$-5brane. b) Graph after the local $M$-duality. The curly lines denote the presence of duality walls.}}
\label{localSL2Z}
\end{figure}

The local $SL(2,\bZ)$ actions on the brane configurations imply dualities between the 3d gauge theories that they realize.
At the level of the matrix models of the 3d gauge theories, these local $SL(2,\bZ)$ dualities should be expressed by the relation:
\begin{align}
\pq{p}{q}_{\sigma \, \ti\sigma} \stackrel{?}{=} \  \lp M^{-1} \pq{p'}{q'} M \rp_{\sigma \, \ti\sigma} \ , \ {\rm with} \ t=t' \ ,
\label{SL2Zaction}
\end{align}
with $M \in SL(2,\bZ)$ and $(p',q')$ given by~\eqref{SL2Ztransfo} and arbitrary numbers $N, \ti N$ of eigenvalues $\{\sigma_j\},\{ \ti\sigma_j\}$. $t$ and $t'$ denote the deformation parameters of the $(p,q)$- and $(p',q')$-5branes respectively. Figure \ref{localSL2Z} displays the graphs of the two theories related by a local $M$-duality. We have used the symbol $\stackrel{?}{=}$ to emphasize that we have not proven this relation. Indeed we will show that the relation holds only up to a phase. 

Checking the relation~\eqref{SL2Zaction} for $(p,q),(p',q') \in \{ (1,0), (0,1), (1,k) \}$ constitutes a non-trivial test of the local $SL(2,\bZ)$ dualities. For other $(p,q)$,~\eqref{SL2Zaction} can be used to derive a matrix factor for an arbitrary $(p,q)$-5brane with $p\wedge q =1$.


\subsubsection{$(p,q)$-5brane matrix factor}
\label{subsubsec:pq5factor}

\begin{figure}[th]
\centering
\includegraphics[scale=0.8]{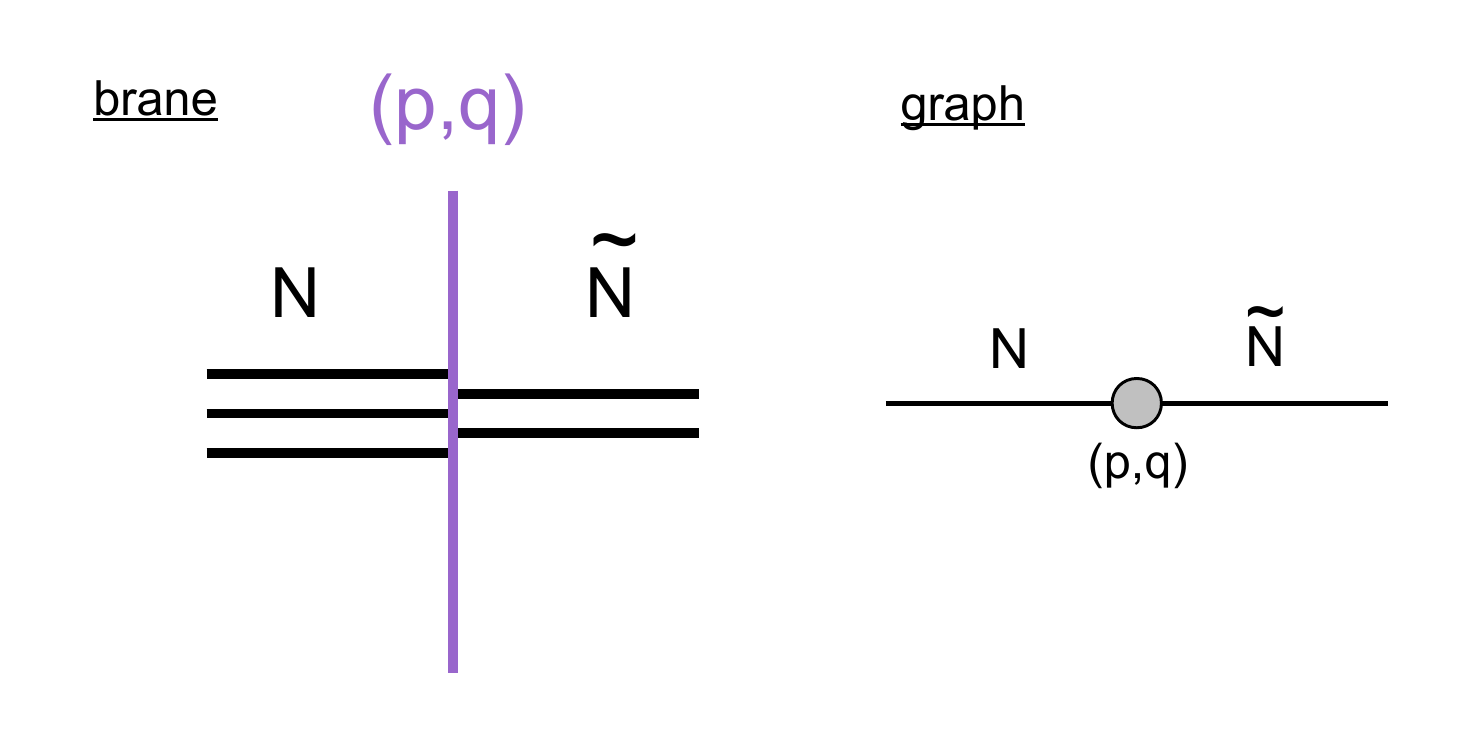}
\vspace{-0.5cm}
\caption{\footnotesize{ $(p,q)$-5brane building block and corresponding graph element in the graph description.}}
\label{pqblock}
\end{figure}

With $p,q$ being two integers such that $p \neq 0$ and $p \wedge q = 1$, we propose the matrix model factor for a $(p,q)$ 5-brane (figure \ref{pqblock}):
\begin{align}
\label{pq5factor}
\pq{p}{q}_{\sigma \, \ti\sigma} &= \frac{|p|^{-\mu} }{(N!\ti N!)^{1/2}} e^{-2\pi i \frac{t}{p} \lp \sum_j^N \sigma_j - \sum_j^{\ti N} \ti\sigma_j \rp} \
e^{\pi i \, \frac{q}{p} \lp \sum_j^N \sigma_j^2 - \sum_j^{\ti N} \ti\sigma_j^2 \rp}  \
\frac{\prod_{i<j}^N \sh [ \, p^{-1} \sigma_{ij}] \ \prod_{i<j}^{\ti N} \sh [\,p^{-1} \ti\sigma_{ij}] } {\prod_{i,j}^{N,\ti N} \ch[ \, p^{-1}(\sigma_i - \ti\sigma_j)] } \ ,
\end{align}
where $\mu =\frac{N+\ti N}{2}$. In analogy with the previous 5-brane factors, $t$ should be associated to the position of the 5-brane in transverse space. This factor is compatible with the cases of NS5-brane and $(1,k)$-5brane factors~\eqref{NS5factor}~\eqref{1k5factor}. 
It generalizes the proposal of \cite{Gulotta:2011si} to the cases of unequal numbers of D3-segments ending on the left and right of the $(p,q)$-5brane, $N \neq \ti N$ and by the addition of the parameter $t$ dependence.

We have to show that this factor obeys the relation~\eqref{SL2Zaction} and it is enough to consider the cases $M=S$ and $M=T$ and prove the inverted relations:
\begin{align}
\pq{-q}{p}_{\sigma \, \ti\sigma} \stackrel{?}{=} \lp S \pq{p}{q} S^{-1} \rp_{\sigma \, \ti\sigma} \ , \qquad 
\pq{p}{p+q}_{\sigma \, \ti\sigma} \stackrel{?}{=} \lp T \pq{p}{q} T^{-1} \rp_{\sigma \, \ti\sigma} \ ,
\label{STaction}
\end{align}

\noindent Consider the sets of $N$ eigenvalues $\{\sigma_j\}$ and $\ti N$ eigenvalues $\{\ti\sigma_j\}$ and a couple of coprime integers $(p,q)$ with $p \neq 0$. For $M=T$, we have
\begin{align}
& \lp T \pq{p}{q} T^{-1}\rp_{\sigma \, \ti\sigma} = \int d^N\sigma' d^{\ti N}\ti\sigma' \ e^{\frac{\pi i N}{12}} \ e^{\pi i \sum_{j=1}^N \sigma_j^2} \ \prod_{j=1}^N \delta(\sigma_j - \sigma'_j) \no\\
& \qquad \ .  \frac{|p|^{-\mu}}{N!} e^{-2i\pi \frac{t}{p} \lp \sum_j^N \sigma'_j - \sum_j^{\ti N} \ti\sigma'_j \rp} \  e^{\pi i \, \frac{q}{p} \lp \sum_j^N \sigma'_j{}^2 - \sum_j^{\ti N} \ti\sigma'_j{}^2 \rp}  \
\frac{\prod_{i<j}^N \sh [ \, p^{-1} \sigma'_{ij}] \ \prod_{i<j}^{\ti N} \sh [\,p^{-1} \ti\sigma'_{ij}] } {\prod_{i,j}^{N,\ti N} \ch[ \, p^{-1}(\sigma'_i - \ti\sigma'_j)] } \no\\
& \qquad \ . e^{-\frac{\pi i \ti N}{12}} \, e^{-\pi i \sum_{j=1}^{\ti N} \ti\sigma'_j{}^2} \ \prod_{j=1}^{\ti N} \delta(\ti\sigma'_j - \ti\sigma_j) \no\\
&=  e^{\frac{\pi i \Delta}{12}} \ \frac{|p|^{-\mu}}{N!} e^{-2i\pi \frac{t}{p} \lp \sum_j^N \sigma_j - \sum_j^{\ti N} \ti\sigma_j \rp} \  e^{\pi i \, \frac{p+q}{p} \lp \sum_j^N \sigma_j{}^2 - \sum_j^{\ti N} \ti\sigma_j{}^2 \rp}  \
\frac{\prod_{i<j}^N \sh [ \, p^{-1} \sigma_{ij}] \ \prod_{i<j}^{\ti N} \sh [\,p^{-1} \ti\sigma_{ij}] } {\prod_{i,j}^{N,\ti N} \ch[ \, p^{-1}(\sigma_i - \ti\sigma_j)] } \no\\
&= e^{\frac{\pi i \Delta}{12}} \ \ \pq{p}{p+q}_{\sigma \, \ti\sigma} \ ,
\end{align}  
where $\mu = \frac{N + \ti N}{2}$ and $\Delta =N-\ti N$.

\noindent For $M=S$ the computation is given in appendix \ref{app:Stransfo}. In total we obtain:
\begin{align}
\lp S \pq{p}{q} S^{-1} \rp_{\sigma \, \ti\sigma} &= \ e^{\frac{\pi i}{12\, pq}\Delta(\Delta^2-1)} \  e^{-\frac{i \pi \Delta t^2}{pq}} \pq{-q}{p}_{\sigma \, \ti\sigma} \quad , \quad  pq \neq 0  \no\\
\lp T \pq{p}{q} T^{-1}\rp_{\sigma \, \ti\sigma} &= \ e^{\frac{\pi i \Delta}{12}} \ \ \pq{p}{p+q}_{\sigma \, \ti\sigma}  \hspace{2.5cm} , \quad  p \neq 0  \ .
\label{STrel}
\end{align}
These relations have extra phases compared to the expected relations~\eqref{STaction}. The phases $ e^{\frac{\pi i}{12\, pq}\Delta(\Delta^2-1)}$ and $e^{\frac{\pi i \Delta}{12}}$ are a priori irrelevant to the dynamics of the theories.\footnote{It is possible that these phases are related to a change of frame of the Chern-Simons actions on $S^3$ (see \cite{Marino:2011nm}).} 
On the other hand the phase $e^{-\frac{i \pi \Delta t^2}{pq}}$ can be attributed to the presence of a background Chern-Simons term for a weakly gauged $U(1)$ symmetry associated to the deformation parameter $t$, with Chern-Simons level $k= \frac{\Delta}{pq}$. It was shown in \cite{Closset:2012vg,Closset:2012vp} that the background Chern-Simons terms with integer coefficients are local counter-terms and characterize the ambiguity of the partition function. In the case at hand the Chern-Simons level $\frac{\Delta}{pq}$ is generically fractional, but it is not clear what conclusions should be drawn from this observation, since the theories realized with $(p,q)$-5branes are rather peculiar (non-Langrangian).
However there are simpler cases when $\frac{\Delta}{pq}$ is an integer. For instance when $N=\ti N$ or when $|p|=|q|=1$. In those cases the $S$-action on the matrix factors is directly compatible with local S-duality of the brane configuration.

\bigskip

We are left with the untreated case when the relations involve D5-factors or $\overline{\rm D5}$-factors $(p,q)=(0,\pm 1)$. The D5-factor was given in~\eqref{D5factor} when $N=\ti N$, but we have not specified it when $N \neq \ti N$, because it was not needed to describe the matrix models of quiver theories. Let us review first this simpler case.
When $N=\ti N$, we have directly:
\begin{align}
\lp T \pq{0}{1} T^{-1} \rp_{\sigma \, \ti\sigma} &= \ \pq{0}{1}_{\sigma \, \ti\sigma} \ ,
\end{align}
and
\begin{align}
\lp S \pq{0}{1} S^{-1} \rp_{\sigma \, \ti\sigma} &= \int d^N\sigma' d^{N}\ti\sigma' \,
\frac{e^{2\pi i \sum_{j=1}^N \sigma_j \sigma'_j}}{\prod_j^N \ch(\sigma'_j -m)} \, 
 \lp \sum_{w \in \scS_N} \frac{(-1)^{w}}{N!} \prod_j^N \delta(\sigma'_j-\ti\sigma'_{w(j)}) \rp
 e^{-2\pi i \sum_{j=1}^N \ti\sigma'_j \ti\sigma_j}  \no\\
&= \frac{1}{N!} \sum_{w \in \scS_N} (-1)^{w} \int d^{N}\sigma' \
\frac{e^{2\pi i \sum_{j=1}^N \sigma_j \sigma'_j}}{\prod_j^N \ch(\sigma'_j -m)} \ e^{-2\pi i \sum_{j=1}^N \sigma'_j \ti\sigma_{w(j)} }  \no\\
&= \frac{e^{2\pi i m \sum_j^N (\sigma_j - \ti\sigma_j)}}{N!} \sum_{w \in \scS_N} (-1)^{w} 
\frac{1}{\prod_j^N \ch(\sigma_j - \ti\sigma_{w(j)})}  \no\\
&= \pq{-1}{0}_{\sigma \, \ti\sigma} \ ,
\end{align}
where the last equality follows from the Cauchy formula~\eqref{Cauchyformula}. 
The $\overline{\rm D5}$-factor is defined as the D5-factor~\eqref{D5factor} with $m$ replaced by $-m$, and the above relations hold with the roles of D5 and NS5 exchanged for $\overline{\rm D5}$ and $\overline{\rm NS5}$ .
It will be important to test mirror symmetry on arbitrary quivers in section \ref{sec:MirrorSym} to provide a D5-factor when $N \neq \ti N$, so we address this question now.


\subsubsection{D5-factor with $N\neq \ti N$}
\label{subsubsec:D5factor}

To complete the picture it is possible to derive a D5-factor when $N \neq \ti N$ (different numbers of D3-branes ending on the left and on the right of the D5-brane) by assuming the local $S$-duality relation for arbitrary $N, \ti N$:
\begin{align}
\pq{0}{1}_{\sigma \, \ti\sigma} = \lp S \pq{1}{0} S^{-1} \rp_{\sigma \, \ti\sigma} \ .
\end{align}
Then we only need to check~\eqref{SL2Zaction} for $M=T$ on this D5-factor to complete the proof that~\eqref{SL2Zaction} holds for all $M \in SL(2,\bZ)$ and $(p,q)$-factors.
With $\Delta = N - \ti N$, $\mu = \frac{N+\ti N}{2}$ and $ N > \ti N$, the explicit D5-factor is 
\begin{align*}
 \pq{0}{1}_{\sigma \, \ti\sigma} & =  \ \lp S \pq{1}{0} S^{-1} \rp_{\sigma \, \ti\sigma} \\
&= \int d^N\tau d^{\ti N} \ti\tau \ e^{2\pi i \sum_j^N \sigma_j \tau_j} \ \pq{1}{0}_{\tau \ti\tau}  \ e^{-2\pi i \sum_j^{\ti N} \ti\tau_j \ti\sigma_j} \\
&= \frac{(-1)^{\Delta \ti N}}{(N!\ti N!)^{1/2}} \sum_{w \in \scS_N} (-1)^{w} \int d^N\tau d^{\ti N} \ti\tau \   e^{2\pi i \sum_j^N \tau_j (\sigma_j-m)}  \ e^{-2\pi i \sum_j^{\ti N} \ti\tau_j (\ti\sigma_j-m)}  \\
 & \hspace{4cm} . \prod_{j=1}^{\ti N} \frac{e^{-\pi\Delta(\tau_{w(j)} - \ti\tau_j)}}{\ch(\tau_{w(j)} - \ti\tau_j)}  
 \prod_{j=\ti N +1}^{N} e^{2\pi\tau_{w(j)} \lp \mu + \half - j\rp }    
\end{align*}
where we used the generalized Cauchy formula (\ref{Cauchyformula3}) to replace the NS5 factor. $m$ is the 5-brane deformation parameter. The result factorizes into a product of $\ti N$ single integrals over the $\ti\tau_j$. These integrals are not convergent, however it is possible to evaluate them as the analytical continuation to complex $y$ of the standard integral $\int dx \, \frac{e^{2\pi i \, x \, y}}{\ch \, x} = \frac{1}{\ch \, y}$.\footnote{A more rigorous approach is to consider expanding the integrand as $\frac{e^{-\pi\Delta x_j}}{\ch(x_j)}= e^{\pi(\Delta-1)x_j} - e^{\pi(\Delta-3)x_j} + \cdots + \frac{(-1)^{\Delta/2}}{\ch(x_j)} $ for $\Delta \in 2\bZ$, or $\frac{e^{-\pi\Delta x_j}}{\ch(x_j)}= e^{\pi(\Delta-1)x_j} - e^{\pi(\Delta-3)x_j} + \cdots + \frac{(-1)^{(\Delta-1)/2}}{2}(1+\thh(x_j))$ for $\Delta \in 2\bZ +1$, where $x_j =\tau_{w(j)} - \ti\tau_j$, and then notice that all contributions vanish because of the $\tau_j$ anti-symmetrization, except the contribution from $\frac{(-1)^{\Delta/2}}{\ch(x_j)} $ or $\frac{(-1)^{(\Delta-1)/2}}{2}\thh(x_j)$. The integrals over these factors are convergent and give the same result as the analytical continuation proposed in the text. }
\begin{align*}
 &= \frac{(-1)^{\Delta \ti N}}{(N!\ti N!)^{1/2}} \sum_{w \in \scS^N} (-1)^{w} \int d^N\tau \ e^{2\pi i \sum_j^N \tau_j (\sigma_j-m)}  
 \ \prod_{j=1}^{\ti N} \frac{ e^{2\pi i (m-\ti\sigma_j)\tau_{w(j)} } }{\ch(m-\ti\sigma_j -i \frac{\Delta}{2}) }  \  
 \prod_{j=\ti N +1}^{N} e^{2\pi\tau_{w(j)} \lp \mu + \half - j\rp }   \\
  &=   
  \frac{(-1)^{\Delta \ti N} \, (N!\ti N!)^{-1/2}}{\prod_{j=1}^{\ti N} \ch(\ti\sigma_j -m +i \frac{\Delta}{2})}
  \sum_{w \in \scS^N} (-1)^{w} 
    \int d^N\tau \ \prod_{j=1}^{\ti N} e^{2\pi i \tau_j (\sigma_{w(j)}-\ti\sigma_j)}  
    \prod_{j=\ti N+1}^N e^{2\pi i \tau_j \lp \sigma_{w(j)} - m - i \lp \mu + \half - j \rp \rp }  
\end{align*}
where we have relabelled $\tau_{w(j)} \rightarrow \tau_j$ in each integral.
We are left with the integrations over the $\tau_j$ which correspond to usual $\delta$ functions and special $\delta$ functions of a complex variable that we denote $\hat\delta(.)$.
\begin{align*}
  &=  \frac{(N!\ti N!)^{-1/2}}{\prod_{j=1}^{\ti N} \ch \lp \ti\sigma_j -m -i \frac{\Delta}{2} \rp } \  \sum_{w \in \scS^N}(-1)^{w}
  \prod_{j=1}^{\ti N} \delta(\sigma_{w(j)}-\ti\sigma_j) \ \prod_{j=\ti N +1}^N \hat\delta[\sigma_{w(j)} -m^{(\Delta)}_{j-\ti N}]
\end{align*}
with $m^{(\Delta)}_j = m + i \lp \frac{\Delta+1}{2} - j \rp$ for $j =1, \cdots, \Delta$ and we have used $\ch(x + i n) = (-1)^n \ch \, x$, for $n \in \bZ$.
The $\hat\delta$ must be defined as a special distribution that generalizes the usual $\delta$ distribution for the class of integrals that we treat.
\footnote{When dealing with divergent matrix integrals, a standard recipe is to rely on analytic continuation of the integrals \cite{vandeBult2007} with respect to some parameters of the integrand. In our case such an analytical continuation is not possible because the the function to be analytically continued is the Dirac delta function. This is why we do not speak about analytic continuation, but about "generalization" of the delta function. It might be surprising that we have to define such an exotic object, however one should remember that this defines a matrix factor for a D5-brane with $N \neq \ti N$, which is not a natural factor arising in the matrix model of a gauge theory. When gluing together all the factors of a "good" matrix model, one should be able to integrate out the D5-factors and remain with a standard converging matrix model.} 
Formally it should correspond to:
\begin{align}
\hat\delta(y) = \int_{\bR} dx \ e^{2\pi i y x} \ ,
\end{align}
 for $y \in \bC$. 
A heuristic computation presented in appendix \ref{app:deltahat} lead us to the following definition.
For $z_0 \in \bZ$, $\hat\delta_{z_0} \equiv \hat\delta(\, . \, - z_0)$ is defined by its action on
 a meromorphic function $f$ with simple poles (away from $z_0$):
\begin{align}
 \int_{\bR} dx \ \hat\delta(x-z_0)f(x) &= f(z_0) \  + \ 2\pi i \epsilon(z_0) \sum_{j}\hat\delta(u_j-z_0) \hat f(u_j) 
\label{defdelta0}
\end{align}
where $\epsilon(z_0)= \ $sgn$({\rm Im}(z_0))$,  $u_j$ are the poles of $f$ in the region $0\le {\rm Im}(u)\le {\rm Im}(z_0)$ (or ${\rm Im}(z_0) \le {\rm Im}(u) \le 0$). When $0 < |{\rm Im}(u_j)| < |{\rm Im}(z_0)|$, $\hat f(u_j)$ is the residue at the pole $u_j$, when ${\rm Im}(u_j) = 0$ or ${\rm Im}(u_j) = {\rm Im}(z_0)$, $\hat f(u_j)$ is half the residue at the pole $u_j$. 
We give details about these exotic $\hat\delta$ in appendix \ref{app:deltahat}.
\bigskip

\noindent To conclude we have obtained a generalized D5-factor:
\begin{center}
\fbox{
\begin{minipage}{14cm}
\begin{align}
& \pq{0}{1}_{\sigma \, \ti\sigma} = 
  \frac{(N!\ti N!)^{-1/2}}{\prod_{j=1}^{\ti N} \ch \lp \ti\sigma_j -m - i \frac{\Delta}{2} \rp} \  \sum_{w \in \scS^N}(-1)^{w}
  \prod_{j=1}^{\ti N} \delta(\sigma_{w(j)}-\ti\sigma_j) \ \prod_{j=\ti N +1}^N \hat\delta \big[ \sigma_{w(j)} -m^{(\Delta)}_{j-\ti N} \big]  \no\\
& \quad  \textrm{with} \quad  m^{(\Delta)}_j =  m + i \, \frac{\Delta-2j+1}{2}  \ , \quad  {\rm for} \quad N \ge \ti N
\label{D5factor2}
\end{align}
\end{minipage}
}
\end{center}
For $N = \ti N$ it reduces to~\eqref{D5factor} with mass parameter $m$.
When $N < \ti N$ the D5-factor is given by the above formula with $N \leftrightarrow \ti N$, $\sigma \leftrightarrow -\ti\sigma$ and $m \rightarrow - m$. This yields
\begin{align}
& \pq{0}{1}_{\sigma \, \ti\sigma} = 
 \frac{(N!\ti N!)^{-1/2}}{\prod_{j=1}^{N} \ch \lp \ti\sigma_j -m + i \frac{\ti\Delta}{2} \rp} \  \sum_{w \in \scS^{\ti N}}(-1)^{w}
  \prod_{j=1}^{N} \delta(\ti\sigma_{w(j)}-\sigma_j) \ \prod_{j= N +1}^{\ti N} \hat\delta(\ti\sigma_{w(j)} - m^{(\ti\Delta) \ast}_{j-N})  \ ,
\label{D5factor3}
\end{align}
with $\ti\Delta= \ti N - N$ and $\ast$ denotes complex conjugation.

\bigskip

\noindent Let us see what becomes of the relations~\eqref{STaction} for the D5-factor.
\begin{align}
\lp S \pq{0}{1} S^{-1} \rp_{\sigma \, \ti\sigma} &=  \lp  S^2 \pq{1}{0} S^{-2}  \rp_{\sigma \, \ti\sigma} = \lp (-1) \pq{1}{0} (-1) \rp_{\sigma \, \ti\sigma}  = \pq{-1}{0}_{\sigma \, \ti\sigma}  \ .
\end{align}
This equality can also be checked directly using the D5-factor we have derived.
Then for $N \ge \ti N$,
\begin{align}
& \lp T \pq{0}{1} T^{-1} \rp_{\sigma \, \ti\sigma} = \ e^{\frac{\pi i \Delta}{12}} \int d^N\tau d^{\ti N}\ti\tau \  e^{\pi i \sum_{j=1}^N \sigma_j^2} \ \prod_{j=1}^N \delta(\sigma_j - \tau_j) \ \pq{0}{1}_{\tau \, \ti\tau}
 \ e^{-\pi i \sum_{j=1}^{\ti N} \ti\tau_j^2} \ \prod_{j=1}^{\ti N} \delta(\ti\tau_j - \ti\sigma_j) \no \\
  &= \ e^{\frac{\pi i \Delta}{12}} \ e^{\pi i \sum_{j=1}^N \sigma_j^2} \ \pq{0}{1}_{\sigma \, \ti\sigma}
 \ e^{-\pi i \sum_{j=1}^{\ti N} \ti\sigma_j^2} \no\\
 &=  \frac{e^{\frac{\pi i \Delta}{12}}  \, (N!\ti N!)^{-1/2}}{\prod_{j=1}^{\ti N} \ch \lp \ti\sigma_j -m -i \frac{\Delta}{2} \rp} \  \sum_{w \in \scS^N}(-1)^{w}
  \prod_{j=1}^{\ti N} \delta(\sigma_{w(j)}-\ti\sigma_j) \ \prod_{j=\ti N +1}^N \hat\delta \big[ \sigma_{w(j)} -m^{(\Delta)}_{j-\ti N} \big] \
 e^{\pi i \sum_{j}^{\Delta} (m^{(\Delta)}_j{})^2} \no\\
 &= e^{\frac{\pi i \Delta}{12}} \, e^{\frac{\pi i}{12}\Delta(\Delta^2-1)} \, e^{\pi i \Delta m^2} \ \pq{0}{1}_{\sigma \, \ti\sigma} \ ,
\end{align}
at the third line we have used the property~\eqref{NoPoleFunction}.
The relations~\eqref{STrel} are then completed with:
\begin{align}
\lp S \pq{1}{0} S^{-1} \rp_{\sigma \, \ti\sigma} &= \ \pq{0}{1}_{\sigma \, \ti\sigma}   \quad , \quad
\lp S \pq{0}{1} S^{-1} \rp_{\sigma \, \ti\sigma} = \ \pq{-1}{0}_{\sigma \, \ti\sigma}   \no\\
\lp T \pq{0}{1} T^{-1}\rp_{\sigma \, \ti\sigma} &= \  e^{\frac{\pi i}{12}\Delta^3} \ e^{\pi i \Delta m^2} \ \pq{0}{1}_{\sigma \, \ti\sigma}  \ .
\label{STrel2}
\end{align}
The phase $e^{\pi i \Delta m^2}$ accounts for a background Chern-Simons term with integer level $\Delta$, which does not affect the physics of the theory \cite{Closset:2012vg, Closset:2012vp}. \\
The dualities we will test involve theories realized with D5-, NS5- and $(1,\pm 1)$-5branes and thus the additional phases will contain only unphysical integer level background CS terms.

\subsection{Tests of SL(2,\bZ) dualities}
\label{ssec:TestSL2Z}

In this section we show  how the machinery of local $SL(2,\bZ)$ transformations produces a powerful test of $SL(2,\bZ)$ dualities, by matching the exact partition functions on $S^3$ of dual SCFTs.

\begin{figure}[th]
\centering
\includegraphics[scale=0.8]{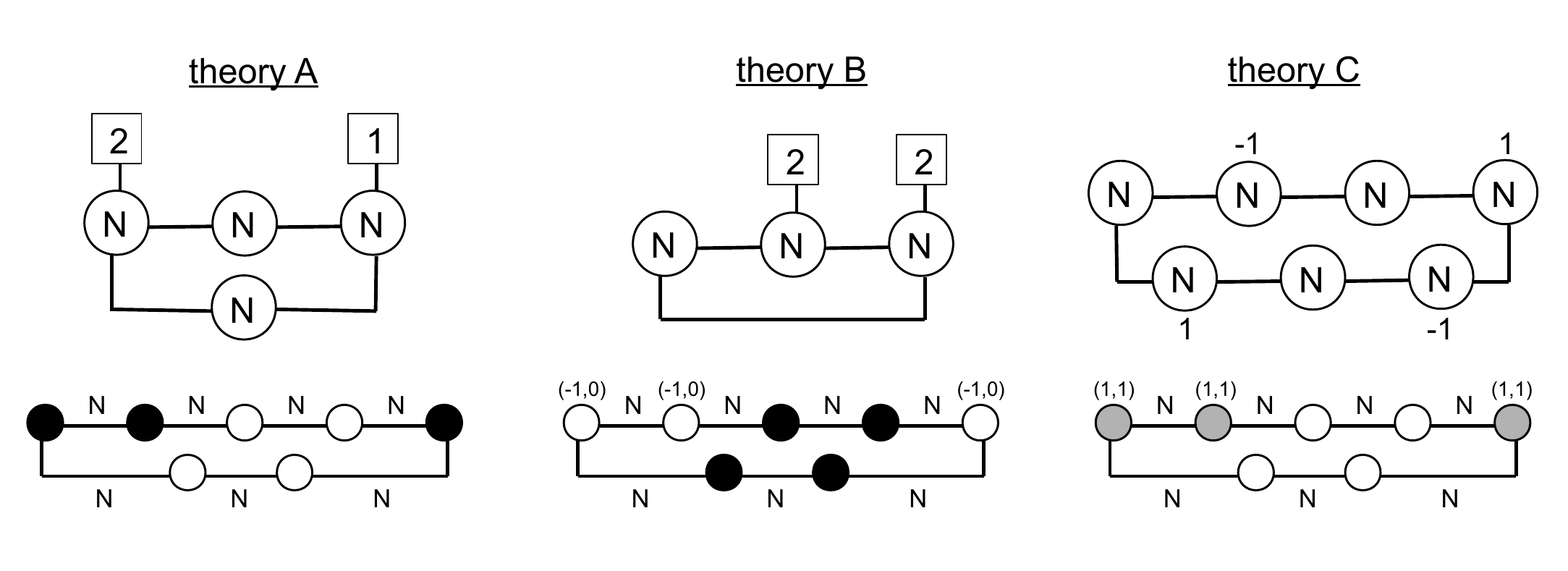}
\vspace{-0.5cm}
\caption{\footnotesize{Triplet of $SL(2,\bZ)$ dual theories. Theories A and B are Yang-Mills quiver SCFTs, while theory C is a Chern-Simons quiver SCFT. Theory B is the $S$-dual of theory A, it coincides with the mirror dual of theory A. Theory C is the $T^T$-dual of theory A, where $T^T$ is the transposed of the matrix $T$. The subscript $(-1,0)$ above the white dots indicates a $\overline{\rm NS5}$ brane.}}
\label{equalrankduals}
\end{figure}

Let us first re-derive the test of mirror symmetry for Yang-Mills circular quivers with nodes of equal rank presented in \cite{Kapustin:2010xq}. Consider a circular quiver theory $A$ with nodes of equal ranks. It has a gauge group $G_A=\prod_{j=1}^{\wat P}U(N)$ with bifundamental hypermultiplets and $M_j$ fundamental hypermultiplets in the $j$th node. Its brane realization involves $N$ D3-branes on a circle crossing a sequence of NS5 and D5-branes.
The mirror-dual theory B is realized by the S-dual brane configuration changing the NS5 into D5-branes and the D5 into $\overline{\rm NS5}$-branes. Theory B is also a Yang-Mills circular quiver, with gauge group $G_B=\prod_{j=1}^{P} U(N)$ and $\ti M_j$ fundamental hypermultiplets in the $j$th node. The relations between the A and B quivers data can be read from the brane configurations. 
The partition function of theory A is given by its sequence of NS5 and D5 factors:
\begin{align}
Z_A &= \tr  \  \pq{1}{0}{}_{(N)} \pq{0}{1}{}_{(N)} \pq{0}{1}{}_{(N)} \pq{1}{0}{}_{(N)} \cdots {}_{(N)} \pq{0}{1}{}_{(N)} \no\\
&= \tr  \  \pq{1}{0} \pq{0}{1} \pq{0}{1} \pq{1}{0} \cdots  \pq{0}{1}  \ .
\end{align}
Each 5-brane factor can be replaced by its local $S$-dual using~\eqref{SL2Zaction}:
\begin{align}
Z_A &= \tr  \  S^{-1}\pq{0}{1}S \,  S^{-1}\pq{-1}{0}S \,  S^{-1}\pq{-1}{0}S \,  S^{-1}\pq{0}{1}S \,  \cdots \, S^{-1}\pq{-1}{0}S  \no\\
&= \tr  \  \pq{0}{1} \, \pq{-1}{0} \, \pq{-1}{0} \, \pq{0}{1} \,  \cdots \, \pq{-1}{0} \no \\
&= Z_B \ ,
\end{align}
where the second equality follows from $(S \,  S^{-1})_{\sigma \ti\sigma} = 1_{\sigma \ti\sigma}$. 
The matrix model obtained after these simple manipulations is the matrix model for the partition function $Z_B$ of the $S$-dual theory $B$. The parameter $t_j$ of the $j$-th 5brane in the sequence of $Z_A$ is mapped to the parameter $\hat t_j$ of the $j$-th 5brane in the sequence of $Z_B$. This means that the mass parameters $m_i$ and $t_j$, associated to D5- and NS5-branes, are exchanged with the parameters $\hat t_i$ and $\hat m_j$ of the mirror theory, associated with $\overline{\rm NS5}$ and D5-branes. For these quiver theories, $S$-duality coincides with mirror symmetry.
Theories A and B in figure \ref{equalrankduals} are an example of YM mirror theories.

In \cite{Hanany:1996ie}, mirror symmetry is associated to a $S$-duality transformation, together with a rotation that brings $x^{j}$ into $-x^{j+3}$ and $x^{j+3}$ into $x^{j}$ for $j=4,5,6$, changing the $\overline{\rm NS5}$ into NS5-branes.
At the level of the matrix model, this extra rotation can be implemented by reversing the sequence of 5-brane factors and reversing the NS5-charge of the factors $(p,q) \rightarrow (-p,q)$. This operation does not change the matrix model, since it is equivalent to a simple relabelling of the nodes from right to left. In the ``matrix language" it would be the equivalent of the property $\tr (M_1 ... M_k) = \tr(M_k^{\rm T} ... M_1^{\rm T})$ with the ``transposition" implementing the NS5-charge conjugation. We obtain:
\begin{align}
Z_B &= \tr  \  \pq{0}{1} \, \pq{-1}{0} \, \pq{-1}{0} \, \pq{0}{1} \,  \cdots \, \pq{-1}{0} \no \\
&= \tr  \  \pq{1}{0} \, \cdots \, \pq{0}{1} \, \pq{1}{0} \, \pq{1}{0} \,  \pq{0}{1} \ .
\end{align}
This completes the usual mirror symmetry transformation.


This simple $S$-transformation of blocks allowed us to match the partition functions of $S$-dual pairs of Yang-Mills $\N=4$ SCFTs for all the circular quivers with nodes of equal rank.
The same manipulation would work for any pair of theories A and B, whose brane realizations are related by a transformation $M \in SL(2,\bZ)$. A theory A realized by a sequence of $(p_1,q_1)$- and $(p_2,q_2)$-5branes has a partition function given generically by
\begin{align}
Z_A &= \tr  \  \pq{p_1}{q_1}{}_{(N_1)} \pq{p_2}{q_2}{}_{(N_2)} \pq{p_2}{q_2}{}_{(N_3)} \pq{p_1}{q_1}{}_{(N_4)} \cdots {}_{(N_{Q-1})} \pq{p_2}{q_2}{}_{(N_Q)} \no\\
&= e^{\pi i \, \varphi_{AB}} \ \tr  \  M^{-1}\pq{p'_1}{q'_1}M \,  M^{-1}\pq{p'_2}{q'_2}M \,  M^{-1}\pq{p'_2}{q'_2}M \,  M^{-1}\pq{p'_1}{q'_1}M \,  \cdots \, M^{-1}\pq{p'_2}{q'_2}M  \no\\
&= e^{\pi i \, \varphi_{AB}} \ \tr  \  \pq{p'_1}{q'_1}{}_{(N_1)} \pq{p'_2}{q'_2}{}_{(N_2)} \pq{p'_2}{q'_2} {}_{(N_3)} \pq{p'_1}{q'_1}{}_{(N_4)}  \cdots {}_{(N_{Q-1})}\pq{p'_2}{q'_2}{}_{(N_Q)} \no \\
&= e^{\pi i \, \varphi_{AB}} \ Z_B 
\end{align}
where $M \in SL(2,\bZ)$ and the dual theory B is realized by the same sequence
 of 5branes as theory A, but with $(p'_i,q'_i)= M (p_i,q_i)$. The two partition functions are not equal but only equal up to a pure phase $e^{\pi i \, \varphi_{AB}}$ that follows from the relations~\eqref{STrel}, \eqref{STrel2}.
This extra phase is irrelevant as long as it can be understood as a background Chern-Simons term with integer level (see discussion above). This happens for instance when the gauge nodes have equal ranks $N_1=N_2=\cdots=N_Q$, in which case the phase vanishes.

This simple computation constitutes a non-trivial test of the global $M$-duality symmetry between quiver theories A and B.
Although we have tested successfully general $SL(2,\bZ)$ dualities, all the dualities that involve Gaiotto-Witten quivers (\ref{ssec:GWquivers}), realized with $(p,q)$-5branes with $|p|>1$ are empty, because the GW quiver SCFTs are already described by using (local) $SL(2,\bZ)$ duality. The theories for which we have an independent description are those involving D5-, NS5- and $(1,k)$-5branes, corresponding to YM or CS quiver SCFTs. Non-trivial $SL(2,\bZ)$ dualities acting in this subclass of SCFTs can be found for dual theories realized with  D5-, NS5- and $(1,\pm 1)$-5branes. An example of a triplet $A,B,C$ of dual SCFTs is described in figure \ref{equalrankduals}. Theory A and B are mirror dual YM quiver SCFTs and theory C is a dual CS quiver SCFT, obtained by acting with the $SL(2,\bZ)$ transformation $T^{\rm T}$ that changes D5-branes into $(1,1)$-5branes.
 
\bigskip
 
\begin{figure}[th]
\centering
\includegraphics[scale=0.8]{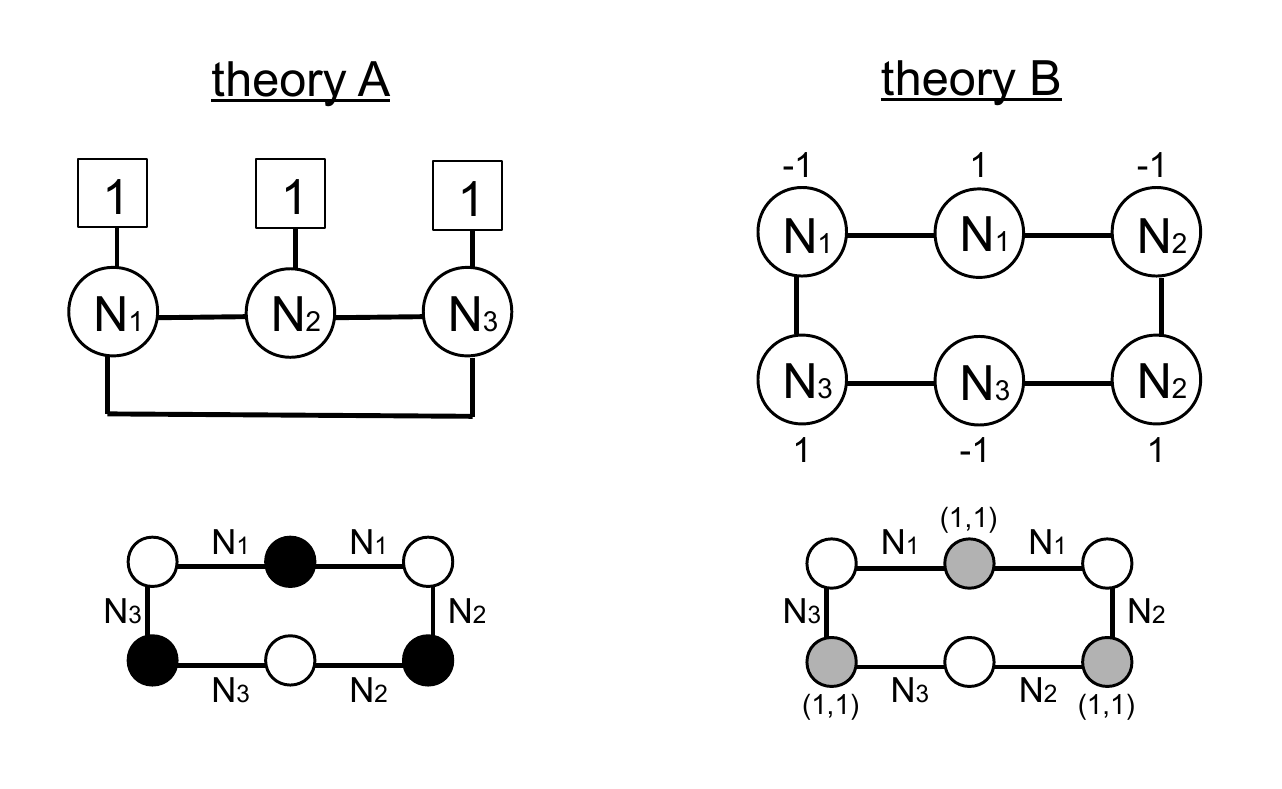}
\vspace{-0.5cm}
\caption{\footnotesize{ Pure Chern-Simons theory B is the $T^{\rm T}$-dual of Yang-Mills theory A.}}
\label{duals2}
\end{figure}

 Let us detail another example, where a Yang-Mills SCFT is mapped to a pure Chern-Simons SCFT (non-vanishing CS levels at all nodes):
\begin{itemize}
\item \underline{theory A :} \ Yang-Mills quiver with gauge group  $G_A=\prod_{j=1}^{\wat P} U(N_j)$ with bifundamental hypermultiplets and one fundamental hypermultiplet in each node. The brane sequence has $\wat P$ NS5-branes and $\wat P$ D5-branes alternating. The deformation parameters of the theory are the fundamental hypermultiplet masses $m_j$ and the FI parameters $t_j-t_{j+1}$ for the $j$th node.
\item \underline{theory B :} Chern-Simons quiver with gauge group  $G_B=\prod_{j=1}^{\wat P} \big[ U(N_j)_{-1} \times U(N_j)_{1} \big]$ with bifundamental hypermultiplets and Chern-Simons level $+1$ and $-1$ alternating from one node to its neighbours. The brane sequence has $\wat P$ NS5-branes and $\wat P$ $(1,1)$-5branes alternating. The deformation parameters of the theory are the FI parameters $\ti t_j- \ti t_{j+1}$ for the $j$th node.
\end{itemize}
Theory B is the $T^{\rm T}$-dual of theory A, with 
$T^{\rm T} = \abcd{1}{1}{0}{1} = - TST$, transforming D5-branes into $(1,1)$-5branes, while leaving the NS5-branes invariant. 
The map between deformation parameters is given by $\ti t_{2j} = m_j$, $\ti t_{2j-1}=t_j$ for $j=1, \cdots, N$. An illustration in given in figure \ref{duals2}

\bigskip

It should be noticed however that $SL(2,\bZ)$ dualities are not sufficient to test mirror symmetry for quivers with nodes of varying ranks. Acting with $S$-duality on a generic YM quiver SCFT, one obtains a brane realization involving D5-branes with different numbers of D3-branes ending on their left and on their right. These brane configurations do not have a simple gauge theory description. To arrive at the mirror dual brane configuration, one has to move the D5-branes along the $x^3$ direction, crossing NS5-branes and making the number of D3-branes vary, until a configuration corresponding to a YM quiver is reached. To be able to check mirror symmetry, it is necessary to understand how the 5-brane moves and the D3-brane creation effect is reproduced in matrix models. We turn now to this question.

\section{Mirror symmetry and other dualities involving HW moves}
\label{sec:MirrorSym}

\begin{figure}[th]
\centering
\includegraphics[scale=0.8]{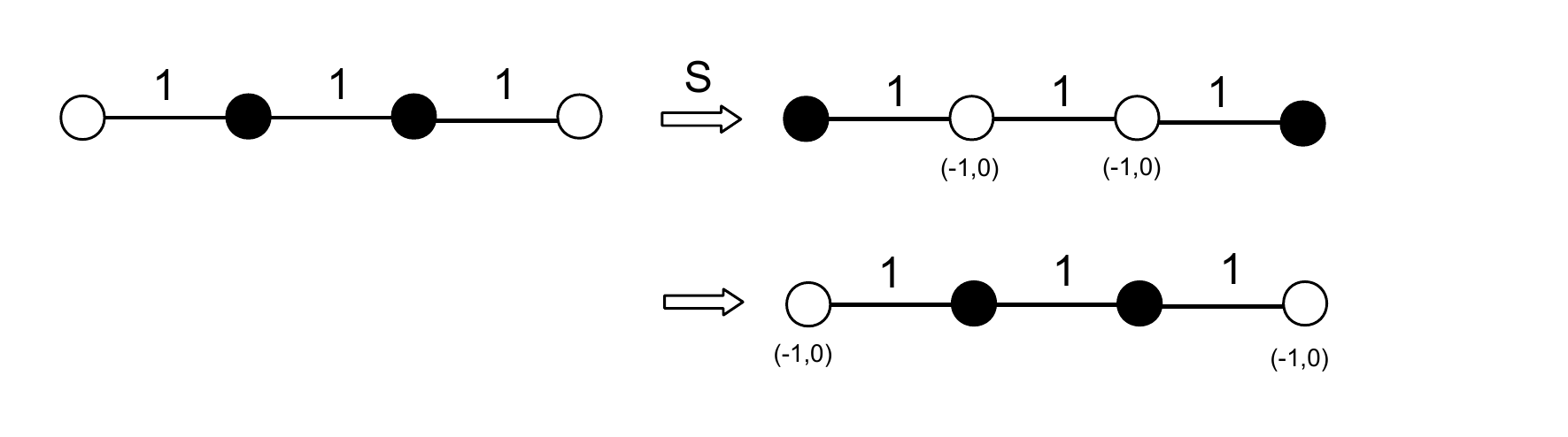}
\vspace{-0.5cm}
\caption{\footnotesize{ On the left is the graph of the $T(SU(2))$ theory, then the the graph resulting from the action of $S$-duality and below is the graph obtained after moving the external D5-branes (black dots) in-between the two NS5-branes (white dots), recovering the graph of $T(SU(2))$, which is a self mirror theory}}
\label{TSU2graph}
\end{figure}

To be able to test mirror symmetry we need to combine the $S$-duality transformation of the matrix models with another type of transformation that corresponds to interchanging the positions of 5-branes in the brane realization. For instance consider the case of the self-mirror $T(SU(2))$ theory, which is $\N=4$ SQED with two fundamental hypermultiplets. Its brane realization involves a single D3-brane stretched between two NS5-branes and intersecting two D5-branes (figure \ref{TSU2graph}). After $S$-duality the NS5-branes have become D5-branes and the D5-branes have been turned into $\overline{\rm NS5}$-branes. However we do not know which SCFT is realized by such a brane configuration. To reach a brane configuration that we understand, one has to move the D5-branes in between the two $\overline{\rm NS5}$-branes (figure \ref{TSU2graph}) to recover a brane configuration realizing again $T(SU(2))$. When two 5-branes of different types pass through each other, the number of D3-branes stretched between vary. This is known has the Hanany-Witten effect and it is not supposed to affect the IR theory living on the D3-branes. This is why one can use these 5-brane moves to reach the desired brane configuration, as in the case of the self-mirror $T(SU(2))$.

 In this section we show that the 5-brane moves, with brane creation effect, are reproduced by identities in the matrix model, allowing us to prove the equality of partition functions for mirror dual theories with nodes of arbitrary ranks. Moreover these HW-identities can also be used to test various dualities between CS quiver SCFTs involving 5-brane moves. Combined with $SL(2,\bZ)$ dualities, they generate a rich web of dualities for $\N=4$ quiver SCFTs.

\subsection{Hanany-Witten moves}
\label{ssec:HWmoves}

\begin{figure}[h]
\centering
\includegraphics[scale=0.8]{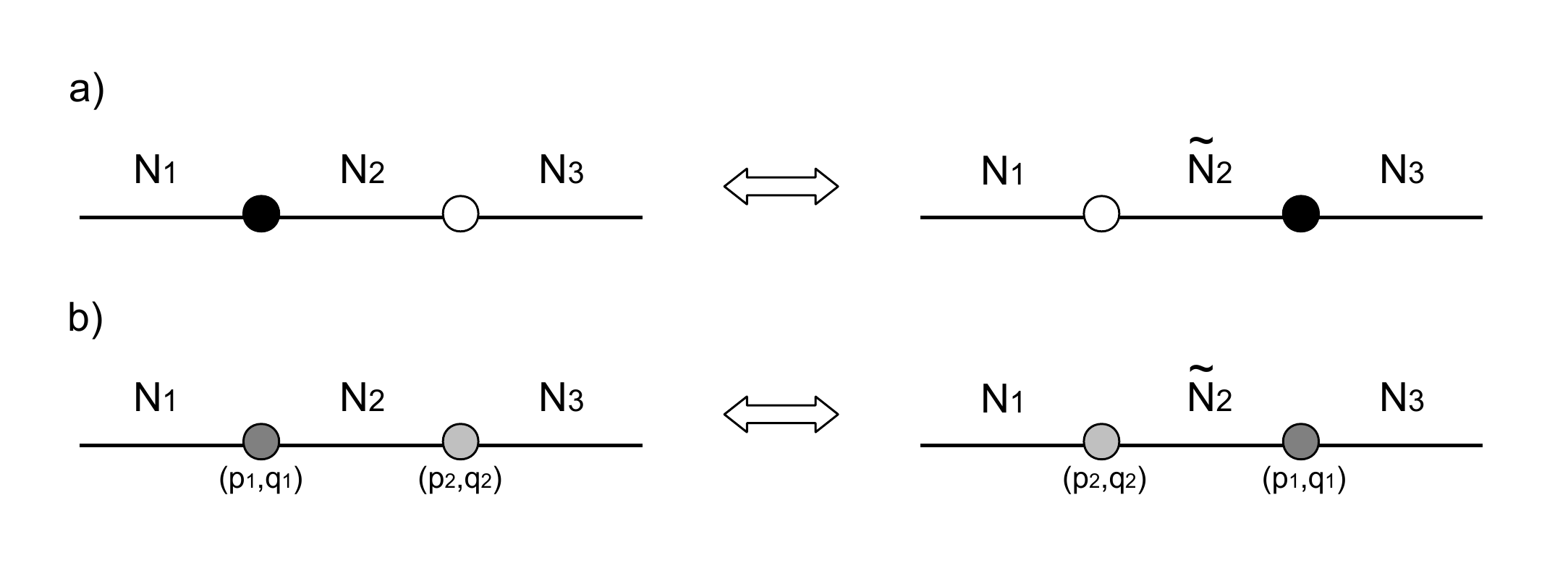}
\vskip -5mm
\caption{\footnotesize  Parts of graphs of two theories related by a Hanany-Witten move. a) A D5-brane (black dot) and a NS5-brane (white dot)  are exchanged and the rank of the middle link is changed from $N_2$ to $\ti N_2 = N_1+N_3-N_2+1$. b) Generic $(p_1,q_1)$- and $(p_2,q_2)$-5branes are exchanged, with $D =|p_1q_2-p_2 q_1| > 0$. In this case $\ti N_2 = N_1 + N_3 - N_2 +D$.}
\label{HWmove}
\end{figure}

Hanany-Witten 5-brane moves refer to the situation when a D5-brane and a NS5-brane pass through each other by moving along the $x^3$ direction. The conservation of 3-form fluxes on the 5-brane worldvolumes implies the creation of a D3-brane stretched between them \cite{Hanany:1996ie}. In the meantime the D3-branes that were initially stretched between the D5- and NS5-branes have their orientation reversed an become anti-D3-branes, annihilating with other D3-branes. In total we arrive at the following rule: if the initial configuration has a D5-brane with $N_1$ D3-branes ending on its left, a NS5-brane with $N_3$ D3-branes ending on its right and $N_2$ D3-branes are stretched between the two 5-branes, then after the HW move the positions of the 5-branes are exchanged and the number of D3-branes stretched between the two has changed to $\ti N_2 = N_1+N_3-N_2+1$. This is summarized in the graphs of figure \ref{HWmove}-a. The HW 5-brane move has a generalization when one considers the exchange of a $(p_1,q_1)$-5brane and a $(p_2,q_2)$-5brane, with $D =|p_1q_2-p_2q_1| > 0$, in which case $D$ new D3-branes are created and the final number of D3-branes stretched between the two 5-branes is $\ti N_2 = N_1 + N_3 -N_2 +D$ \cite{Kitao:1998mf} (see figure \ref{HWmove}-b). Note the we do not consider the exchange of 5-branes of the same type, which would have $D=0$, and for which the argument developed in \cite{Hanany:1996ie} does not apply. These ``forbidden" 5-brane moves are related to $\N=4$ Seiberg-like dualities and involve subtleties that we hope to address in the future.

It is important to notice that the number $N_2$ of D3-branes stretched between a D5-brane and a NS5-brane obeys the constraint $N_2 \le N_1 + N_3+1$ to ensure unbroken supersymmetry. This follows from the ``s-rule" (\cite{Hanany:1996ie,Bergman:1999na}) that says that supersymmetry is broken if more than one D3-branes are stretched between a D5-brane and a NS5-brane. The counterpart for $(p_1,q_1)$- and $(p_2,q_2)$-5branes is that there must not  be more than $D$ D3-branes stretched bewteen the two 5branes, leading to $N_2 \le N_1 + N_3+D$. This ensures $\ti N_2 \ge 0$. 

The claim of \cite{Hanany:2003hp} is that the two brane configurations represented by the graphs in figure \ref{HWmove} flow to equivalent IR theories. For the matrix models this imply the following identity for $(p_1,q_1)$- and $(p_2,q_2)$-5brane factors:
\begin{align}
\lp \pq{p_1}{q_1} {}_{(N_2)} \pq{p_2}{q_2} \rp_{\sigma \ti\sigma} & \stackrel{?}{=} \lp \pq{p_2}{q_2} {}_{(\ti N_2)} \pq{p_1}{q_1} \rp_{\sigma \ti\sigma} \ .
\label{HWidentity0}
\end{align}
This identity would ensure that the SCFT realized with the $(p_1,q_1)$- and $(p_2,q_2)$-5branes has the same partition function has the theory realized with the same brane configuration, but with these two 5-branes exchanged. Again we used the symbol $\stackrel{?}{=}$ to point out that we have not proven this relation yet and that it will be modified ultimately by the addition of some phase.

The identity~\eqref{HWidentity0} is difficult to prove, however we can use the local $SL(2,\bZ)$ action on matrix models to trade the $(p_1,q_1)$-5brane factor and $(p_2,q_2)$-5brane factor for a D5-brane factor and a generic $(p,q)$-5brane factor with $p \neq 0$. The identity to prove is reduced to
\begin{align}
\lp \pq{0}{1} {}_{(N_2)} \pq{p}{q} \rp_{\sigma \ti\sigma} & \stackrel{?}{=} \lp \pq{p}{q} {}_{(\ti N_2)} \pq{0}{1} \rp_{\sigma \ti\sigma} \ ,
\label{HWidentity1}
\end{align}
with $\ti N_2 = N_1+N_3-N_2+|p|$.
Deriving this identity involves various tricks and manipulations that we present in appendix \ref{app:HWmove}. The computation relies on properties of the special $\hat\delta$ distributions that enter into the D5-brane matrix factor\eqref{D5factor3} and requires several mathematical identities given in appendix \ref{app:formulas}.  Moreover it is necessary to deal with various cases corresponding to different ordering of $N_1, N_2, N_3, N_1+p, N_3+p$. In appendix \ref{app:HWmove}, we discuss only the case $N_1+p \le N_2 \le N_3$, which contains the tricks necessary to deal with the other cases. There are however cases for which we were not able to finish the computations due to some additional complications appearing, corresponding to 
$ 0 < N_2 < |p|$ and $N_1+N_3 < N_2 < N_1 + N_3 + |p|$ \footnote{We believe however that the identities hold also for these cases.}. In the dualities that we discuss in this work, we only consider $p=\pm 1$ for which these cases do not exist.

Our final result is~\eqref{finalrelation} \footnote{In our computation we have not taken care of overall factors of $i$, so our result is only valid up to factors of $i$. This can be justified by the fact that the partition function is generally complex and that it is not clear wether the initial matrix model summarized in \ref{ssec:PartFunction} contains the correct factors of $i$ or not (see \cite{Marino:2011nm} for a related discussion).}
\begin{align}
\lp \pq{0}{1}{}_{(N_2)} \pq{p}{q} \rp_{\sigma \ti\sigma} & =  \ e^{\mp 2\pi i t m} \  e^{\pm \pi i q m^2 } \  e^{\frac{\pi i}{12} \frac q p \Phi} \ 
\lp \pq{p}{q}{}_{(\ti N_2)} \pq{0}{1} \rp_{\sigma \ti\sigma} \ ,
\label{HWidentity2}
\end{align}
where $\pm$ is the sign of $p$ (and $\mp$ its opposite) and $\Phi = |\Delta_{21}|(\Delta_{21}^2-1) - |\Delta_{3\ti 2}|(\Delta_{3\ti 2}^2-1)$, with $\Delta_{21} = N_2-N_1$ and $\Delta_{3 \ti 2} = N_3 - \ti N_2$.
The extra phase in the relation~\eqref{HWidentity2} contains matrix factors for a background CS term with level $\pm q$ for the $U(1)$ global symmetry associated to $m$ and a background BF coupling between the two $U(1)$ global symmetries associated to $m$ and $t$, which can be viewed as a mixed CS coupling with level $\mp 1$. The CS levels are integers, implying that such background terms do not affect the physics of the theory \cite{Closset:2012vp}. 

We are now in a position to argue for the equality of the partition functions, up to irrelevant phases, of mirror dual theories for YM quiver SCFTs with nodes of arbitrary ranks.

\subsection{Mirror symmetry}
\label{ssec:MirrorSym}

Mirror symmetry is a duality that relates pairs of $\N=4$ Yang-Mills quiver SCFTs, satisfying the ``good" conditions of \cite{Gaiotto:2008ak}, namely at each node $M_j + N_{j+1} + N_{j-1} \ge 2 N_j$, where $M_j$ is the number of fundamental hypermultiplets of the node $U(N_j)$. 
The mirror dual of a YM quiver theory A is found by first considering the brane configuration realizing A, with D5 and NS5-branes, then taking the $S$-dual of this brane configuration and implementing HW moves until the D5-branes have the same number of D3-branes on each side. One obtains the brane configuration realizing a YM quiver theory B, which is the mirror dual of the theory A.
For this algorithm to work, it is necessary that, after some HW moves, each D5-brane arrives at a position where it has zero {\it net number} of D3-branes ending on it, where by {\it net number} we mean the number of D3-branes ending on its right, minus the number of D3-branes ending on its left. Although it is not completely straightforward, this is ensured by the conditions $M_j + N_{j+1} + N_{j-1} \ge 2 N_j$ that define a ``good" quiver. Note also that the mirror theory B is also a ``good" quiver theory.

The partition functions of mirror dual theories A and B can be mapped using the tools we have developed. The partition function $Z_A$ of the theory A is given by the sequence of 5-brane factors mimicking its brane realization, for instance:
\begin{align}
Z_A &= \tr  \  \pq{1}{0}{}_{(N_1)} \pq{0}{1}{}_{(N_1)} \pq{0}{1}{}_{(N_1)} \pq{1}{0}{}_{(N_2)} \pq{0}{1}{}_{(N_2)} \pq{1}{0}{}_{(N_3)}\cdots {}_{(N_{\wat P})} \pq{0}{1}{}_{(N_{\wat P})} \ .
\end{align}
Each 5-brane factor can be replaced by its local $S$-dual using~\eqref{STrel2}:
\begin{align}
Z_A &= \tr  \  S^{-1}\pq{0}{1}S \,  S^{-1}\pq{-1}{0}S \,  S^{-1}\pq{-1}{0}S \,  S^{-1}\pq{0}{1}S \, S^{-1}\pq{-1}{0}S \, \cdots \, S^{-1}\pq{-1}{0}S  \no\\
&=  \tr  \  \pq{0}{1}{}_{(N_1)} \pq{-1}{0}{}_{(N_1)} \pq{-1}{0}{}_{(N_1)} \pq{0}{1}{}_{(N_2)} \pq{-1}{0}{}_{(N_2)} \pq{0}{1}{}_{(N_3)}\cdots {}_{(N_{\wat P})} \pq{-1}{0}{}_{(N_{\wat P})} \ , \no\\
\end{align}
At this stage the matrix model is given by the sequence of D5 and $\overline{\rm NS5}$ factors, but the associated brane configuration has D5-branes with different numbers of D3-branes on each side (except when all $N_i$ are equal). Then we can use the identity~\eqref{HWidentity2} to exchange  D5- and $\overline{\rm NS5}$ factors. For instance if $N_1 > N_{\hat P} \, (\equiv N_0)$, we have to exchange the two first factors:
\begin{align}
Z_A &= \ e^{i \varphi} \ \tr  \  \pq{-1}{0}{}_{(\ti N_1)} \pq{0}{1}{}_{(N_1)} \pq{-1}{0}{}_{(N_1)} \pq{0}{1}{}_{(N_2)} \cdots {}_{(N_{\wat P})} \pq{-1}{0}{}_{(N_{\wat P})} \ , \no\\
\end{align}
where $\ti N_1 = N_1 + N_{\hat P} - N_1 +1 = N_{\hat P} + 1 $ and $e^{i \varphi}$ is a phase depending on the background parameters. Note that the net number of D3-branes ending on the D5-brane has decreased by one unit. If $\ti N_1 < N_1$, we have to continue moving the same D5-factor to the right of the chain of factors by permuting with $\overline{\rm NS5}$ factors, until the net number of D3-branes ending on it is zero (it decreases by one unit at each permutation). The fact that there are enough $\overline{\rm NS5}$ factors is ensured by the conditions defining a ``good" quiver. The same algorithm must be applied to all D5-factors, so that in the end they all have zero net number of D3-branes ending on them.
\begin{align}
Z_A &= e^{i \varphi_{AB}} \ \tr  \   \pq{-1}{0}{}_{(\ti N_1)} \ \cdots  \ {}_{(N_{\wat P})} \ = \ e^{i \phi_{AB}} \ Z_B  \ . \no\\
\end{align}
 The final sequence of D5 and $\overline{\rm NS5}$ factors is obviously very different from the initial sequence associated to the theory A, instead it corresponds precisely to the sequence associated to the brane configuration of theory B. This is because the permutations of 5-brane factors reproduce exactly the HW 5-brane moves that are needed to go to the brane configuration realizing the mirror theory B. 
 
 As discussed in section \ref{ssec:TestSL2Z}, it is common to complete the transformation of the brane configuration by a ``rotation" that changes the $\overline{\rm NS5}$-branes into NS5-branes. At the level of the matrix model, this corresponds to reversing the sequence of 5-brane factors and trading the $\overline{\rm NS5}$ factors for NS5-factors. This is actually a simple relabelling of the eigenvalues (from right to left) that does not affect the matrix model. 
 
In the end we find that the partition functions $Z_A$ and $Z_B$ differ only by a phase $e^{i \varphi_{AB}}$, corresponding to unphysical background CS terms. If $\{m_i, t_j\}$, $1 \le i \le P$, $1 \le j \le \hat P$,  are the mass and FI deformation parameters of theory A, then deformation parameters of theory B are the FI parameters $\hat t_i = m_i$ and the mass parameters $\hat m_j = t_j$.

The map between the exact partition functions of the SCFTs A and B, with FI and mass deformation terms, provides a important new test of mirror symmetry for all non-abelian circular and linear\footnote{The test for linear quivers corresponds to having $N_{\wat P}=0$ in the above argument.} YM quiver SCFTs.

\subsection{Level-rank and YM-CS dualities}
\label{ssec:YMCSDuals}

The identity~\eqref{HWidentity2} can be used to test other dualities that involve 5-brane moves. Such dualities have already been described for a few abelian SCFTs in \cite{Jafferis:2008em}. Instead of trying to describe all possible dualities, we will concentrate on specific interesting cases corresponding to $\N=4$ level-rank dualities for CS quiver theories and dualities mapping YM quiver SCFTs and pure Chern-Simons SCFTs.

\begin{figure}[h]
\centering
\includegraphics[scale=0.8]{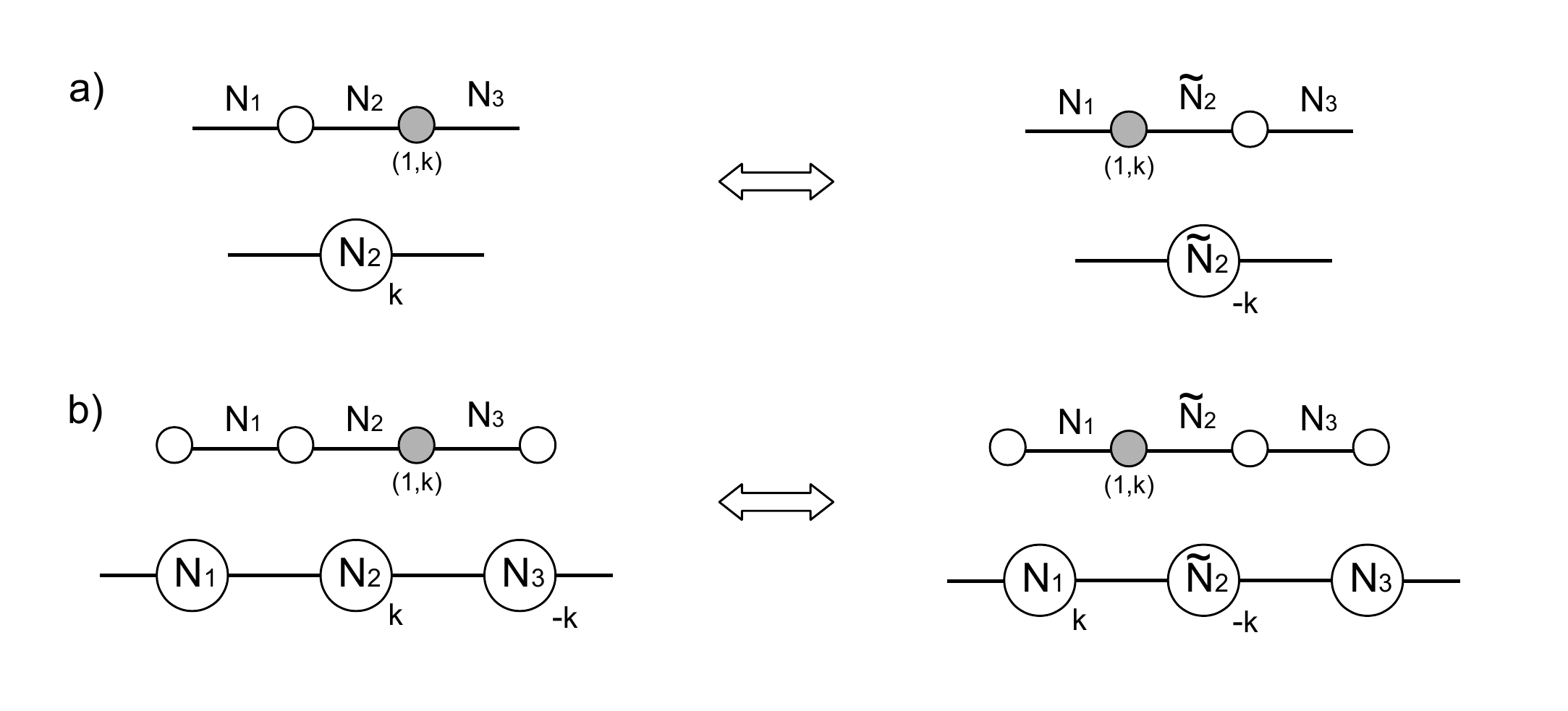}
\vskip -5mm
\caption{\footnotesize  a) A HW-move in CS quiver theories. The rank and CS level of the central node are changed for $\ti N_2 = N_1 + N_3 - N_2 + |k|$ and $-k$. b) Example of dual theories obtained from a single HW move. Only the element of the quivers changing under the duality are shown, the rest of the quiver is identical for the two theories. }
\label{HWmoveCS}
\end{figure}

CS quiver theories are realized with brane configurations corresponding sequences of NS5-branes and $(1,k)$-5branes (see section \ref{ssec:CSquivers}). Permuting a NS5-brane and a $(1,k)$-5brane, adjacent in a brane sequence, leads to the brane configuration of a dual CS quiver theory, as illustrated in figure \ref{HWmoveCS}. The 5-brane exchange affects three consecutive gauge nodes of the quiver, in particular the central node is changed from $U(N_2)_{k}$ to $U(\ti N_2)_{-k}$ with $\ti N_2 = N_1 + N_3 - N_2 + |k|$. This can be seen as a generalization to $\N=4$ quiver SCFTs of the Giveon-Kutasov duality for $\N=2$ SCFTs \cite {Giveon:2008zn} (tests of the Giveon-Kutasov duality using exact partition functions were presented in \cite{Kapustin:2010mh}), which is already a generalization of the level-rank duality of simple Chern-Simons theory without matter.

The equality of matrix models of the two dual theories, up to a phase, is ensured by the HW relation:
\begin{align}
{}_{(N_1)}  \pq{1}{0} {}_{(N_2)} \pq{1}{k} {}_{(N_3)}  & = \ e^{- \pi i \,  (t-s)^2 } \ {}_{(N_1)}  \pq{1}{k} {}_{(\ti N_2)} \pq{1}{0} {}_{(N_3)}  \ ,
\label{HWidentityCS}
\end{align}
which is obtained using~\eqref{STrel},~\eqref{STrel2} and ~\eqref{HWidentity2} in the following way:
\footnote{we omit here the phases of the form $e^{\frac{i \pi}{12} (...) }$ that are independent of the parameters of the theory and play no role in the derivation of the dualities.}
\begin{align*}
{}_{(N_1)} \pq{1}{0} {}_{(N_2)} \pq{1}{k}  {}_{(N_3)} & =  \ e^{\frac{\pi i}{k} \Delta_{23} \, s^2 } \  S^{-1} \pq{0}{1} {}_{(N_2)}  \pq{-k}{1}  S \\
& =  \ e^{\frac{\pi i}{k} \Delta_{23} \, s^2 } \ e^{2\pi i \, t s} \ e^{-\pi i \, t^2} \  S^{-1} \pq{-k}{1} {}_{(\ti N_2)}  \pq{0}{1}  S \\
& =  \ e^{- \pi i \,  s^2 } \ e^{2\pi i \, t s} \ e^{-\pi i \, t^2} \  {}_{(N_1)} \pq{1}{k} {}_{(\ti N_2)} \pq{1}{0}  {}_{(N_3)} \ ,
\end{align*}
where $t$ and $s$ are the deformation parameters associated to the NS5 and $(1,k)$-5brane respectively and we have used $\Delta_{23} - \Delta_{1 \ti 2} = k$. Note that again the extra phase in~\eqref{HWidentityCS} appears as a background CS term with integer level, that does not affect our conclusions regarding dualities.
The map between parameters of the dual theories indicates that the FI parameter of the middle node is reversed $\eta_2 = t- s \ \rightarrow \ti \eta_2 = s - t = - \eta_2$. The FI parameters of the two exterior nodes $U(N_1)$ and $U(N_3)$ are also affected.

This simple 5-brane permutation can be repeated for all couple of adjacent NS5 and $(1,k)$-5brane, leading to a web of dualities between CS quiver SCFTs. The identity~\eqref{HWidentityCS} proves that the exact partition functions of the theories related by these HW-moves are equal, up to an irrelevant phases. 

\bigskip

The dualities following from HW 5brane moves can be combined with the $SL(2,\bZ)$ dualities to generate interesting dualities. Mirror symmetry between YM SCFTs is one example of such combined dualities. Another interesting case would be a duality mapping Yang-Mills SCFTs to pure Chern-Simons SCFTs, where by pure Chern-Simons we mean that all nodes of the CS quiver have non-vanishing CS levels. The Chern-Simons theories have superconformal Lagrangians and would be understood as an explicit description of the infrared fixed points of the ``dual" Yang-Mills quiver theories. 

We have already seen in section \ref{ssec:TestSL2Z} that YM theories have CS duals related through $T^{\rm T}$-duality, which changes D5-branes into $(1,1)$-5-branes. However, the CS duals have generically $U(N)_0$ auxiliary nodes. To obtain a pure CS dual theory we can try to use HW-moves. The necessary and sufficient condition to obtain a pure CS dual theory is that the number of D5-branes and NS5-branes in the brane realization of the initial YM theory are equal for circular quivers, or differ by at most one for linear quivers. In this case, the $(1,1)$-5branes of the $T^{\rm T}$-dual CS theory can be moved along the $x^3$ direction to reach a sequence of alternating NS5 and $(1,1)$-5branes. The corresponding SCFT is then a pure CS theory with alternating Chern-Simons levels $\pm 1$.

\begin{figure}[h]
\centering
\includegraphics[scale=0.8]{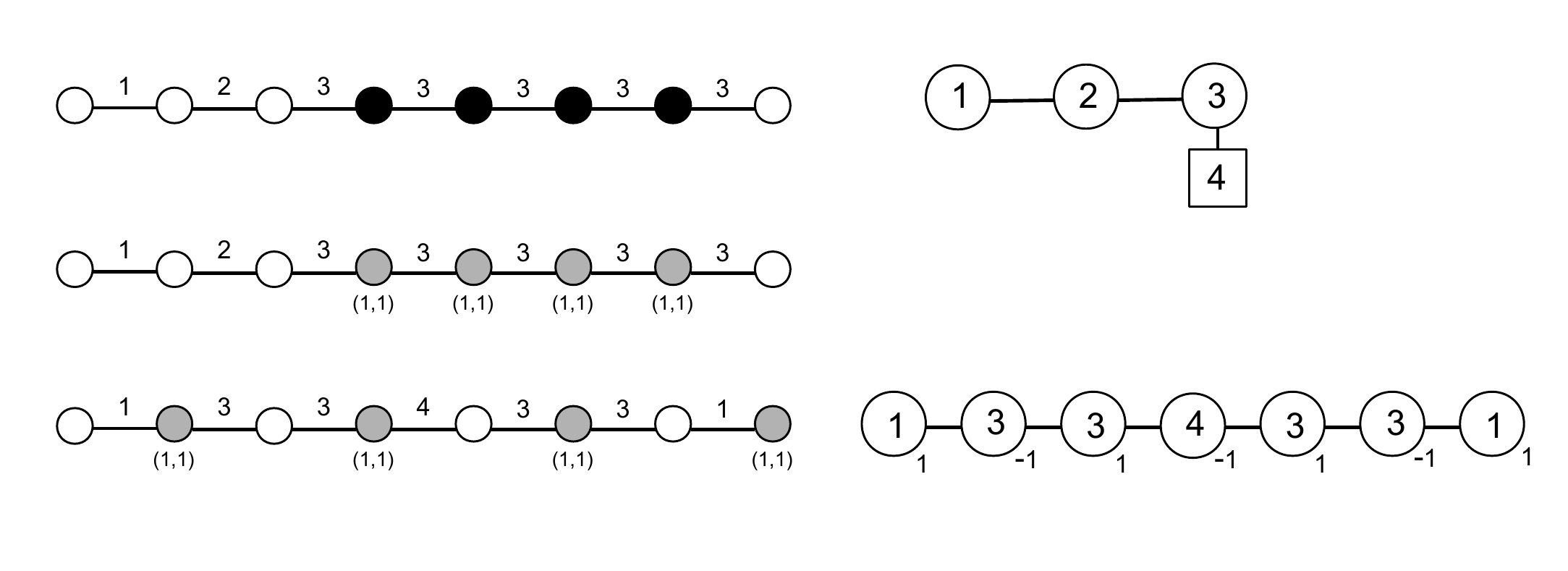}
\vskip -5mm
\caption{\footnotesize  From top to bottom: quiver and graph of $T(SU(4))$ theory; graph after $T^{\rm T}$ action; graph and quiver of the pure CS dual theory, obtained after HW moves.}
\label{TSU4duals}
\end{figure}

The $T(SU(N))$ theories belong to this class of YM theories with pure CS duals. Let us see how the duality works by transforming the partition function of $T(SU(4))$ to its Chern-Simons dual. The partition function of $T(SU(4))$ with deformation parameters $t_{1,2,3,4}, m_{1,2,3,4}$ is given by the sequence of 5brane factors:
\begin{align}
Z^{T(SU(4))} &= \pq{1}{0} {}_{(1)} \pq{1}{0} {}_{(2)} \pq{1}{0} {}_{(3)}  \pq{0}{1} {}_{(3)} \pq{0}{1} {}_{(3)} \pq{0}{1} {}_{(3)} \pq{0}{1} {}_{(3)}  \pq{1}{0} \ .
\end{align}
After acting (locally on each 5-brane factor) with $T^{\rm T}$-duality, the matrix model becomes
\begin{align}
Z^{T(SU(4))} &= e^{\pi i ( t_1^2 + t_2^2 + t_3^2 - 3 \, t_4^2)}  \ \pq{1}{0} {}_{(1)} \pq{1}{0} {}_{(2)} \pq{1}{0} {}_{(3)}  \pq{1}{1} {}_{(3)}  \pq{1}{1} {}_{(3)} \pq{1}{1} {}_{(3)} \pq{1}{1} {}_{(3)}  \pq{1}{0} \ ,
\end{align}
where the extra phase is derived from the identities~\eqref{STrel2} and the decomposition $T^{\rm T} = - TST$. 
Next we use the identity~\eqref{HWidentityCS} to rearrange the sequence of factors, so that NS5 and $(1,1)$-5brane factors alternate:
\begin{align}
Z^{T(SU(4))} &= e^{\pi i \, \varphi}  \ \pq{1}{0} {}_{(1)} \pq{1}{1} {}_{(3)} \pq{1}{0} {}_{(3)}  \pq{1}{1} {}_{(4)}  \pq{1}{0} {}_{(3)} \pq{1}{1}  {}_{(3)}  \pq{1}{0} {}_{(1)} \pq{1}{1} \ = \ Z_{\rm CS},
\end{align}
with $\varphi =  t_1^2 + t_2^2 + t_3^2 - 3 \, t_4^2 - (t_2-m_1)^2 - (t_3-m_1)^2 - (t_3-m_2)^2 + (t_4-m_4)^2$. The final matrix model correspond to a pure Chern-Simons theory with gauge group $U(1)_{1} \times U(3)_{-1} \times U(3)_{1} \times U(4)_{-1} \times U(3)_{1} \times U(3)_{-1} \times U(1)_{1}$. The transformation is summarized in figure \ref{TSU4duals}. 

The necessary and sufficient condition for a YM fixed point to admit a pure CS description boils down to having a number of nodes equal to the number of fundamental hypermultiplets for circular quivers (or different by at most one for linear quivers). It must also be noticed that the pure Chern-Simons duals of Yang-Mills theories have only nodes with Chern levels $\pm 1$ and thus are never weakly coupled, since the effective gauge coupling in Chern-Simons theory is $\lambda = \frac N k$ and it can become small only in the large $k$ limit.

\subsubsection{A check by direct computations}
\label{sssec:DirectCheck}

\begin{figure}[h]
\centering
\includegraphics[scale=0.8]{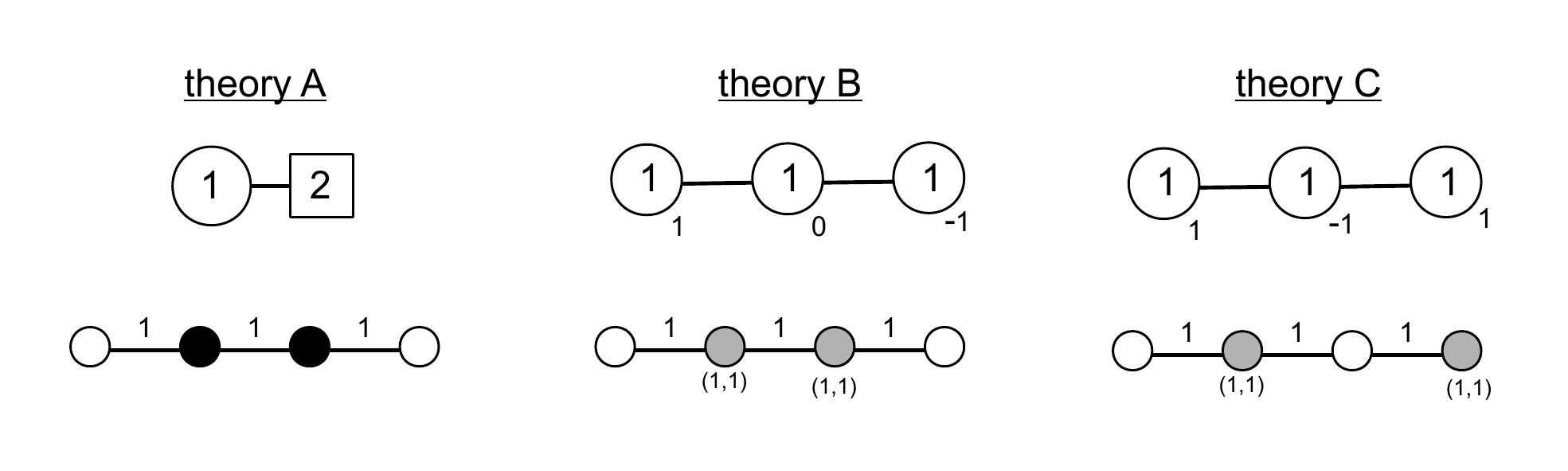}
\vskip -5mm
\caption{\footnotesize  Quivers and graphs of three abelian dual theories.}
\label{AbelianDuals}
\end{figure}

In the argumentation that we have developed, we were able to map partition functions of dual theories without explicitly computing the matrix models. We would like to provide some consistency check of our results by computing matrix models of simple dual theories. 
Let us consider the three dual abelian theories:
\begin{itemize}
\item $T(SU(2))$, which is a Yang-Mills theory with gauge group $U(1)$ and two fundamental hypermultiplets. The deformation parameters are the FI parameter $t_1-t_2$ and the  real masses $m_1,m_2$. We denote its partition function $Z_A$;
\item the CS theory with gauge group $U(1)_{1} \times U(1)_0 \times U(1)_{-1}$, with FI parameters $\hat t_1-\hat t_2, \hat t_2-\hat t_3, \hat t_3-\hat t_4$. We denote its partition function $Z_B$;
\item the pure CS theory with gauge group $U(1)_{1} \times U(1)_{-1} \times U(1)_{1}$, with FI parameters $\check t_1-\check t_2, \check t_2-\check t_3, \check t_3-\check t_4$. We denote its partition function $Z_C$.
\end{itemize}
These abelian duals were already proposed in \cite{Jafferis:2008em}, where it was shown that their moduli spaces coincide. In addition our analysis provides the following map between parameters
\begin{align}
t_1 &= \ \hat t_1 \ = \ \check t_1 \quad , \quad  m_1 = \ \hat t_2 \ = \ \check t_2   \quad , \quad 
m_2 \ = \ \hat t_3 \ = \ \check t_4 \quad , \quad  t_2 = \ \hat t_4 \ = \ \check t_3  \ .
\label{abelianparammap}
\end{align}
The gauge quivers and corresponding graphs are presented in figure \ref{AbelianDuals}. The theories A and B be are related by $T^{\rm T}$-duality, while the theories B and C are related by a HW move or level-rank duality.

The map between partition functions obtained by acting with $T^{\rm T}$ duality on the matrix model, using (\ref{STrel},\ref{STrel2}),  is given by:
\begin{align}
Z_A = e^{ i \pi (t_1^2 - t_2^2)} \, Z_B = e^{i \pi  [ t_1^2 - t_2^2 + (t_2-m_2)^2 ]} \, Z_C \ ,
\label{abeliandualZ}
\end{align}
up to overall phases independent of the deformation parameters.

The matrix model for the $T(SU(2))$ theory is given by:
\begin{align*}
Z_A & = \ Z^{T(SU(2))} \ = \  \int  d\sigma \ \frac{e^{2 \pi i (t_1 - t_2) \sigma}}{\ch(\sigma-m_1)\ch(\sigma-m_2)} 
\    = \  (-i) \, \frac{e^{2\pi i (t_1 - t_2) m_1} - e^{2\pi i (t_1 - t_2) m_2} }{\sh(t_1-t_2) \sh(m_1-m_2)} \  , 
\end{align*}
where the integral was computed by deforming the contour to $i \, \infty$ and summing over residues.
The partition function for the CS theory B is given by:
\begin{align*}
Z_B & = \  \int  d\sigma_1 d\sigma_2 d\sigma_3 \  e^{\pi i (\sigma_1^2 -\sigma_3^2)}  \ \frac{e^{2 \pi i [(\hat t_1 - \hat t_2) \sigma_1+ (\hat t_2 - \hat t_3) \sigma_2 + (\hat t_3 - \hat t_4) \sigma_3]}}{\ch(\sigma_1- \sigma_2)\ch(\sigma_2- \sigma_3)}  \  .
\end{align*}
To evaluate these integrals, we change variables $\sigma_1 \rightarrow \sigma_1 + \sigma_2$, $\sigma_3 \rightarrow \sigma_3 + \sigma_2$ . Then the integration over $\sigma_2$ yields a delta function that can be used to integrate over $\sigma_1$, leading to
\begin{align*}
Z_B & = \  e^{\pi i (\hat t_4^2 -\hat t_1^2)} \ e^{2\pi i (\hat t_1 -\hat t_4)\hat t_2}  
\int  d\sigma_3 \ \frac{e^{2 \pi i  (\hat t_3 - \hat  t_2) \sigma_3}}{\ch(\sigma_3- \hat t_1 + \hat t_4)\ch(\sigma_3)} 
\\
& = \ e^{\pi i (\hat t_4^2- \hat t_1^2)} \  (-i) \, \frac{e^{2\pi i (\hat t_1 - \hat t_4) \hat t_2} - e^{2\pi i (\hat t_1 - \hat t_4) \hat t_3} }{\sh(\hat t_1 - \hat t_4) \sh(\hat t_2 - \hat t_3)} \  .
\end{align*}
The partition function of the pure Chern-Simons theory C can be evaluated using similar ideas, giving
\begin{align*}
Z_C & = \  \int  d\sigma_1 d\sigma_2 d\sigma_3 \  e^{\pi i (\sigma_1^2 - \sigma_2^2 + \sigma_3^2)}  \, \frac{e^{2 \pi i [(\check t_1 - \check t_2) \sigma_1+ (\check t_2 - \check t_3) \sigma_2 + (\check t_3 - \check t_4) \sigma_3]}}{\ch(\sigma_1- \sigma_2)\ch(\sigma_2- \sigma_3)}  \\
& \ = \ e^{-\frac{\pi i}{4}}  \ e^{-\pi i (\check t_1^2 + \check t_4^2 - 2 \check t_3 \check t_4)} \  (-i) \, \frac{e^{2\pi i (\check t_1 - \check t_3) \check t_2} - e^{2\pi i (\check t_1 - \check t_3) \check t_4} }{\sh(\check t_1 - \check t_3) \sh(\check t_2 - \check t_4)}  \ . 
\end{align*}
These explicit results match the relations ~\eqref{abeliandualZ} with the parameter mapping~\eqref{abelianparammap}, providing a direct check of the dualities.

\section{Explicit partition functions} 
\label{sec:ExplicitResults}

Finally we can make use of our matrix model machinery to derive some explicit evaluation of partition functions.

In \cite{Nishioka:2011dq} the authors conjectured an explicit formula for the partition function of an arbitrary YM linear quiver SCFT deformed by mass and FI terms. The formula is expressed in terms of two partitions $(\rho,\hat\rho)$ of a positive integer $N$, that encode the linear quiver data.

\noindent An invariant way of encoding a brane configuration -- and the corresponding quiver gauge theory -- is by specifying the {\it linking numbers} of the five-branes.
They can be defined as follows
\begin{align}\
l_a &= - n_a + R_a^{\rm NS5} \qquad (a=1,...,P) \cr
\hat l_b &= \hat n_b + L_b^{\rm D5} \qquad (b=1,...,\hat P) \ ,
\label{defnlinking}
\end{align}
where $n_a$ is the number of D3 branes ending on the $a$th D5 brane from the right minus the number ending from the left, $\hat n_b$ is the same quantity for the $b$th NS5 brane, $R_a^{\rm NS5}$ is the number of NS5 branes lying to the right of the $a$th D5 brane and $L_b^{\rm D5}$ is the number of D5 branes lying to the left of the $b$th NS5 brane.  These numbers  are conserved under Hanany-Witten moves \cite{Hanany:1996ie}, which correspond to moving a D5-brane across a NS5-brane with a D3-brane creation or annihilation. Since the extreme infrared limit is expected to be  insensitive to these moves, it is  convenient to label the infrared dynamics in terms of the linking numbers of the 5-branes.

\begin{figure}[th]
\centering
\includegraphics[scale=0.8]{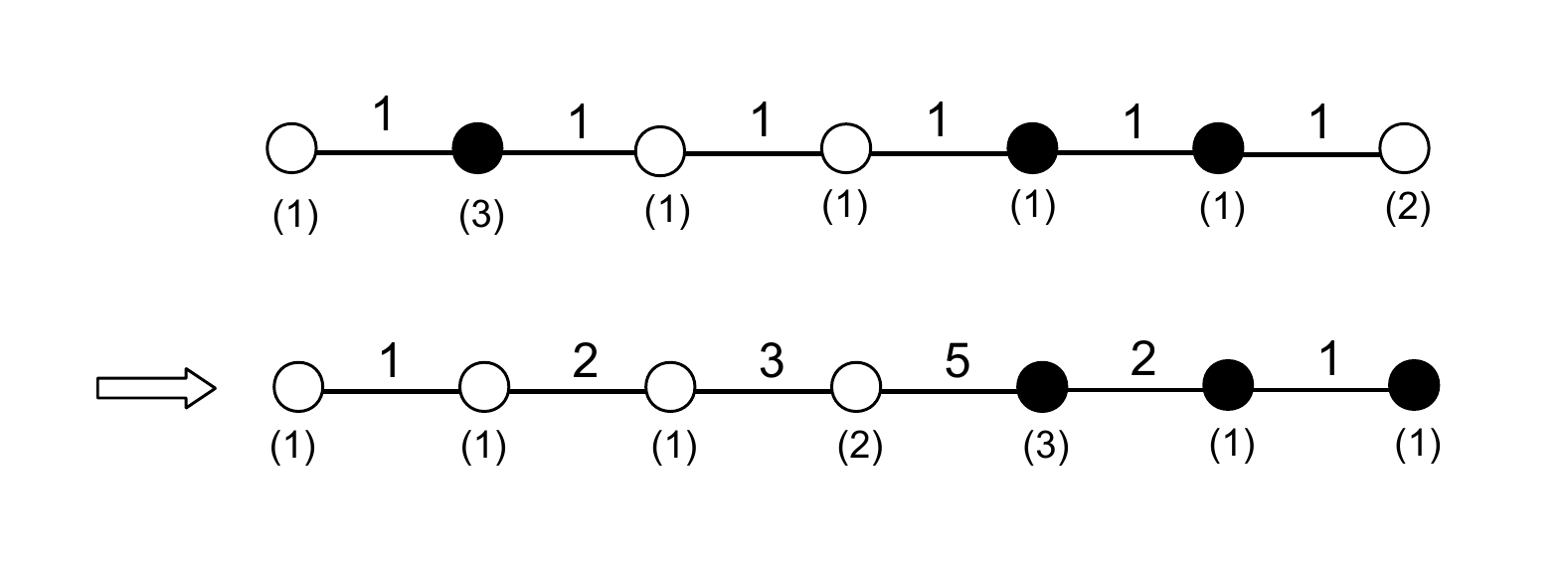}
\vspace{-0.5cm}
\caption{\footnotesize Pushing all D5-branes (black dots) to the right of all NS5-branes (white dots) makes it easy to read the linking numbers (indicated below each 5-brane), as
the net number of D3-branes ending on each five-brane.
In this example  $\rho = (1,1,3)$ and $\hat\rho = (1,1,1,2)$. The IR fixed point is $T^{(113)}_{(1112)}(SU(5))$. }
\label{separategraph0}
\end{figure}

We may move all the NS5-branes to the left and all the D5-branes to the right, noting that a new D3-brane is created every time that a
D5 crosses a NS5. In the end, all the D3 branes will be suspended between a NS5 brane on the left and
 a D5 brane on the right (see figure \ref{separategraph0} for an example), so  that the linking numbers satisfy the sum rule $\sum_{a=1}^{P} l_a = \sum_{b=1}^{\hat P} \hat l_b    \equiv N $,
where $N$ is the total number of  suspended D3 branes. This implies that the
  two sets of five-brane linking numbers define two partitions of $N$. 
 This is the repackaging of the quiver data in terms of partitions $\rho, \hat \rho$ of $N$ mentioned above with
\begin{align}
\rho &= (l_1,l_2,...,l_P) \quad , \quad l_1 \le l_2 \le ... \le l_P  \ , \no\\
\hat \rho &= (\hat l_1, \hat l_2, ..., \hat l_{\hat P})   \quad , \quad  \hat l_1 \le \hat l_2 \le ... \le \hat l_{\hat P} \ .
\end{align}
We adopt the convention that the linking numbers in the partitions are ordered non-decreasingly. 
\footnote{We point out, to try to avoid confusions, that this is the opposite choice compared to the convention of \cite{Nishioka:2011dq} where the linking numbers are ordered non-increasingly.}
 The IR fixed point SCFT of the theory labelled by  $(\rho,\hat \rho)$ is called $T^{\rho}_{\hat \rho}(SU(N))$.

\smallskip

In the original configuration of figure \ref{linquiv} the D5-brane linking numbers are, by construction, positive and  non-decreasing
but this is not automatic for the linking numbers of the NS5 branes.  
Requiring that the NS5-brane linking numbers be non-decreasing
 is equivalent to the conditions for the quiver to be a ``good" theory.
\smallskip

In the presence of mass and FI deformation terms, each D5-brane is associated to a real mass parameter $m_a$ and each NS5-brane with a FI parameter $t_b$. It is convenient to define {\it deformed partitions}, that we call again $\rho$ and $\hat \rho$, as
\begin{align}
 \rho & := \Big( (l_1,m_1),(l_2,m_2), \ ... \ ,(l_P,m_P) \Big) \no\\
 \hat \rho & := \Big( (\hat l_1,t_1),(\hat l_2,t_2), \ ... \ ,(\hat l_{\hat P},t_{\hat P}) \Big) \ .
\end{align}
With these definitions, the mirror dual of $T^{\rho}_{\hat \rho}(SU(N))$ is $T^{\hat \rho}_{\rho}(SU(N))$. It is implied here that the masses $m_a$ label the D5-branes from right to left (which is the opposite of the convention we had adopted up to now), whereas the FI parameters $t_b$ label the NS5-branes from left to right as before.
\smallskip

\noindent To express our results it is convenient to define ``$N$-vectors" as
\begin{align}
\label{Nvec}
 M &= \big( \ \textrm{coord}( \vec m_1) \  , \ \textrm{coord}( \vec m_2) \ , \ ... \ , \ \textrm{coord}( \vec m_P) \ \big) \no \\
   & \textrm{with} \quad \vec m_a = \left\{ \ m_a + i \Big(\frac{l_a+1}{2}-1 \Big), \ m_a + i \Big(\frac{l_a+1}{2}-2 \Big), \ ... \ , \  m_a + i \Big(\frac{l_a+1}{2}-l_a \Big) \ \right\} \no \\
 T &= \big( \ \textrm{coord}( \vec t_1) \  , \ \textrm{coord}( \vec t_2) \ , \ ... \ , \ \textrm{coord}( \vec t_{\hat P}) \  \big) \\
   & \textrm{with} \quad \vec t_b = \bigg\{ \ t_b + i \Big(\frac{\hat l_b+1}{2}-1\Big), \ t_b + i \Big(\frac{\hat l_b+1}{2}-2\Big), \ ... \ , \ t_b + i\Big(\frac{\hat l_b+1}{2}- \hat l_b\Big) \ \bigg\}  \ , \no
\end{align}
where coord$(\vec v) = v_1 , v_2 , v_3, ... , v_p$  for a vector $\vec v$ with $p$ coordinates.
Note that $ M$ and $T$ are vectors with $N$ coordinates, while $\vec m_a$ and $\vec t_b$ are vectors with $l_a$ and $\hat l_b$ coordinates respectively.
For instance for the linear quiver described by the graph of figure \ref{separategraph0}, we have
\begin{align}
 \rho & = \Big( (1,m_1),(1,m_2), (3,m_3) \Big) \ , \quad 
 \hat \rho  = \Big( (1,t_1),(1,t_2),(1,t_3),(2,t_4) \Big) \no\\
 M & = \Big( m_1 \ , \ m_2 \ , \ m_3 + i \ , \ m_3 \ , \ m_3 - i \Big) \ , \quad
 T  = \left( t_1 \ , \ t_2 \ , \ t_3 \ , \ t_4 + \frac{i}{2} \ , \ t_4 - \frac{i}{2} \right) \ .
 \label{Nvec_ex}
\end{align}

\bigskip

The exact formula for the partition function of linear quivers can be proven in a simple way using the tools we have developed.

\begin{figure}[th]
\centering
\includegraphics[scale=0.8]{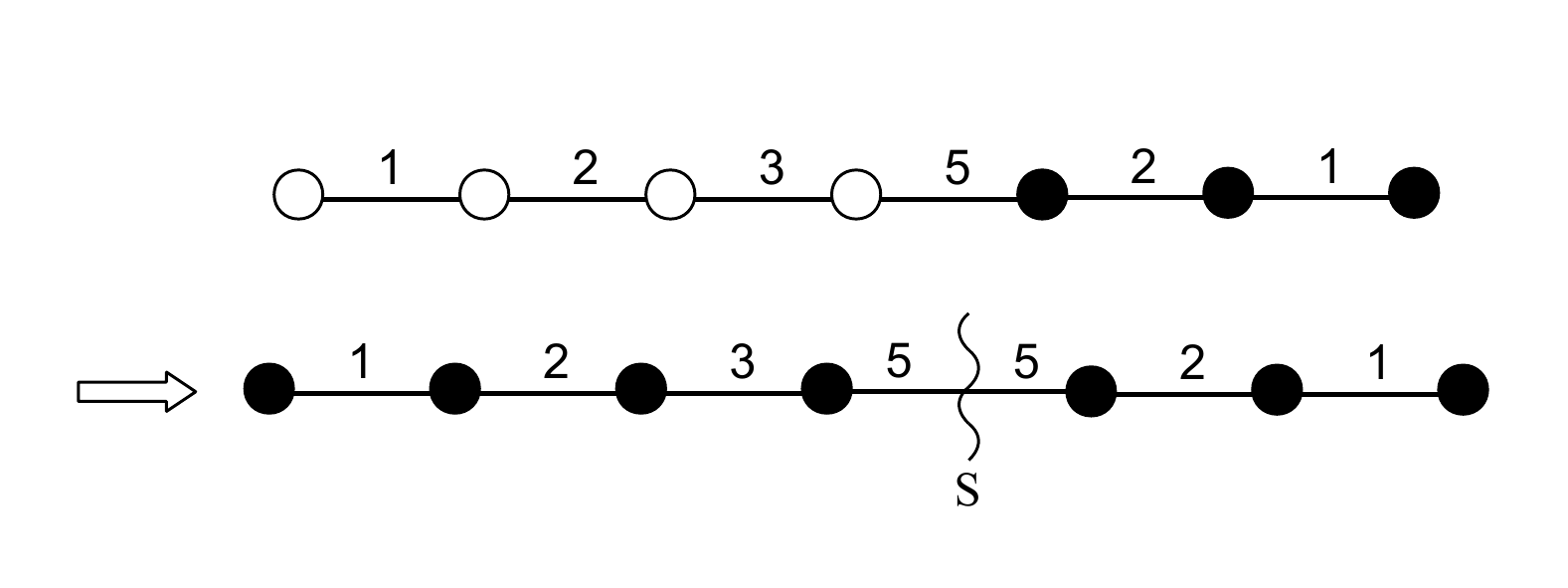}
\vspace{-0.5cm}
\caption{\footnotesize Acting with $S$-duality on the left part of the graph of figure \ref{separategraph0}, one obtains a graph with only black dots and a $S$-duality wall. }
\label{separategraph1}
\end{figure}

Starting form an arbitrary linear quiver $T^{\rho}_{\hat\rho}(SU(N))$, one can move all the white dots to the left and all the black dots to the right and then act locally with $S$-duality on the left part of the graph, transforming all the white dots into black dots, as shown in the example of figure \ref{separategraph1}. The resulting graph has only black dots and a $S$-wall separating the graph into two branches. The left branch of the graph is characterized by the partition $\hat \rho$, the right branch by the partition $\rho$ and the two branches are glued together through the $S$-wall.
The partition function can be read from this graph and has the following structure:
\begin{align}
Z^{(\rho,\hat\rho)} &= \int d^N\ti\sigma d^N\sigma \ Z_{\rm branch}^{\hat\rho}[\ti\sigma] \   S_{\ti\sigma \sigma} \ Z_{\rm branch}^{\rho}[\sigma]
\label{Z2branches}
\end{align}
where $Z_{\rm branch}^{\rho}[\sigma]$ is the partition function of a single black dots-branch, associated to the partition $\rho$ and the set of eigenvalues $\{\sigma_j\}$.
As in \ref{ssec:MirrorSym}, the partition function of the original theory $Z_{\rm quiver}$ and the partition function associated with the graph with two branches $Z^{(\rho,\hat\rho)}$ are equal up to an unphysical phase:
\begin{align}
Z_{\rm quiver} & = e^{i \, \varphi} \ Z^{(\rho,\hat\rho)} \ ,
\label{PhaseGap}
\end{align}
with $\varphi = - 2\pi \sum_{a \prec b} m_a t_b$ and $a \prec b$ indicates that the D5-brane with parameter $m_a$ is placed to the left of the NS5-brane with parameter $t_b$ in the initial brane sequence realizing the theory.

The single branch partition function for a partition $\rho=[ \, (l_a, m_a) \ , \ 1 \le a \le P \ ]$ of $N$ evaluates to:
\begin{align}
Z_{\rm branch}^{\rho}[\sigma] &= \frac{1}{\sqrt{N!}} \sum_{w \in \scS^N} (-1)^{w} \prod_{a=1}^P \lp \frac{1 }{ \prod_{j=1}^{L_{a-1}} \ch(\sigma_{w(j)} - m_a - i \, \frac{l_a}{2}) } \ \prod_{j=1}^{l_a} \hat\delta \lp \sigma_{w(j+L_{a-1})} - m_{a,j} \rp  \ \rp
\label{Zbranch}
\end{align}
with $L_a \equiv \sum_{c=1}^{a} l_c$ ($L_P = N$ , $L_0=0$) and $m_{a,j} = m_a + i \, \frac{l_a +1 -2j}{2}$.

This result can be proved by induction. For $P=1$ the branch has a single black-dot and (\ref{Zbranch}) reproduced the associated D5-factor.
Then suppose (\ref{Zbranch}) is true for a value $P$ and take a partition of $\rho' = [ \, (m_a, l_a) \ , \ 1 \le a \le P+1 \ ]$. $\rho'$ is a partition of $L_{P+1} = L_{P} + l_{P+1}$, corresponding to adding one black dot to the graph with $P$ dots. Using \eqref{Zbranch}, the corresponding matrix model is:
\begin{align*}
Z_{\rm branch}^{\rho'}[\sigma'] =  &  \sum_{w' \in \scS^{L_{P+1}}} (-1)^{w'} \int d^{L_{P}}\sigma \  \prod_{a=1}^P  \lp \frac{1 }{ \prod_{j=1}^{L_{a-1}} \ch(\sigma_{j} - m_a - i \, \frac{l_a}{2}) } \ \prod_{j=1}^{l_a} \hat\delta \lp \sigma_{j+L_{a-1}} - m_{a,j} \rp  \rp  \\
& \quad . \ \frac{1}{ \sqrt{L_{P+1}!}}  \  \frac{\prod_{j=1}^{L_P} \delta \lp \sigma'_{w'(j)} - \sigma_j \rp}{\prod_{j=1}^{L_P} \ch(\sigma_j - m_{P+1} - i \, \frac{l_{P+1}}{2} ) }  \  
\prod_{j=1}^{l_{P+1}} \hat\delta \lp \sigma'_{w'(j+L_{P})} - m_{P+1,j} \rp 
\end{align*}
where we have de-antisymmetrized the matrix factor (\ref{Zbranch}) by permuting the eigenvalues $\sigma_j$ in the integrand, cancelling an overall $1/L_P!$ factor.
Integrating over the $\sigma_j$ using the $\delta(.)$ we recover the result (\ref{Zbranch}) at level $P+1$, which completes the proof.
\smallskip

We can turn now to the partition function (\ref{Z2branches}) of a linear quiver with partitions 
$\rho=[ \, (m_a, l_a) \ , \ 1 \le a \le P \ ]$, $\hat\rho=[ \, (t_b, \hat l_b) \ , \ 1 \le b \le \hat P \ ]$ of $N$. De-antisymmetrizing the two branch factors by permuting the $\sigma_j$ and $\ti\sigma_j$ in the integrand leads to:
\begin{align*}
Z^{(\rho,\hat\rho)} &=  \sum_{w \in \scS_N} (-1)^{w}   \int d^N\hat\sigma d^N\sigma \ 
\prod_{b=1}^{\hat P} \lp \frac{1 }{ \prod_{j=1}^{\hat L_{b-1}} \ch(\hat \sigma_{j} - t_b - i \, \frac{\hat l_b}{2}) } \ \prod_{j=1}^{\hat l_b} \hat\delta \lp \hat\sigma_{j+\hat L_{b-1}} - t_{b,j} \rp \rp \\
& \quad . \  e^{2\pi i \sum_j^N \hat\sigma_j \sigma_{w(j)}} \
\prod_{a=1}^P \lp \frac{1 }{ \prod_{j=1}^{L_{a-1}} \ch(\sigma_{j} - m_a - i \, \frac{l_a}{2}) } \ \prod_{j=1}^{l_a} \hat\delta \lp \sigma_{j+L_{a-1}} - m_{a,j} \rp  \rp
\end{align*}
The integration must now be carried out using the $\hat\delta(.)$. The result of such integration contains generally a regular piece obtained by treating the $\hat\delta$ as simple $\delta$ functions plus a complicated sum of terms involving products of $\hat\delta(.)$. However in our case we know that the matrix model evaluates to a regular finite result as it is equal to the matrix model of a good linear quiver. This means that the singular terms with $\hat\delta(.)$ all cancel in the anti-symetrization by $\sum_{w \in \scS_N}$, so we can evaluate the result by treating the $\hat\delta(.)$ as usual $\delta(.)$ functions.\footnote{We have verified this property explicitly in simple examples.}
 This leads directly to the NTY conjectured formula:\footnote{Our formula correct some overall factor mistake in the NTY formula.}

\begin{center}
\fbox{
\begin{minipage}{16cm}
\begin{align}
\label{NTY2}
& Z^{(\rho,\hat\rho)} =  \frac{\sum_{w \in \scS^N} (-1)^{w} e^{2\pi i \, \sum_j^N T_j M_{w(j)}} } {\Delta(T) \Delta(M)} \\
 & \textrm{with} \quad \Delta(M) = \prod_{a=1}^{P} \prod_{j=1}^{L_{a-1}} \ch \lp M_j - m_a - i \, \frac{l_a}{2} \rp \ , \quad  \Delta(T) = \prod_{b=1}^{\hat P} \prod_{j=1}^{\hat L_{b-1}} \ch \lp T_j - t_b - i \, \frac{\hat l_b}{2} \rp    \no 
\end{align}
\end{minipage}
}
\end{center}
and $M$ and $T$ are the $N$-vectors defined in \eqref{Nvec}.
As pointed in \cite{Nishioka:2011dq}, the result is explicitly symmetric under the exchange of the deformed partitions $\rho$ and $\hat\rho$.

As a simple consistency check of the validity of our manipulations of matrix models, we provide in appendix \ref{app:Check} an explicit computation mapping the matrix model of a separated graph (black dots on the right, white dots on the left) to the matrix model of the initial quiver theory.

\section{Perspectives}
\label{sec:Discussion}

We have shown how the matrix model giving the exact partition function of an  $\N=4$ theory on $S^3$ can be expressed as a sequence of elementary factors, mimicking the sequence of 5-branes of its brane realization in type IIB string theory.
We have described the action of $SL(2,\bZ)$ dualities on these 5-brane factors and shown that the partition functions of $SL(2,\bZ)$ dual theories are equal, up to a phase, which was unphysical in all cases we studied. We found that the Hanany-Witten 5-brane move is expressed by a non-trivial identity for 5-brane blocks that we proved. This allowed us to map the partition functions of Yang-Mills mirror-dual theories 
for linear and circular quivers with unitary nodes of arbitrary ranks, providing a significant extension of the results of \cite{Kapustin:2010xq}.
In addition our results go beyond simple mirror symmetry and provide maps between the partition functions for a large web of dualities, generated by $SL(2,\bZ)$ actions and 5-brane moves. These involve $\N=4$ level-rank dualities and dualities relating Yang-Mills to Chern-Simons quiver theories.


There are several extensions one can think of. It would be nice to consider the cases of quivers with orthogonal and symplectic gauge nodes realized by brane configurations involving orientifold planes \cite{Dey:2014dwa} and see if a similar story exists.  At the technical level one can try to consider the partition functions on a squashed 3-sphere $S^3_b$ instead of the round $S^3$, providing stronger tests of the dualities.
An interesting question, which we did not address at all, is the question of how the moduli spaces of the $SL(2,\bZ)$ dual theories are mapped and how they are affected or not by quantum corrections. Usual mirror symmetry is known to exchange the Coulomb and Higgs branches of dual Yang-Mills theories, the Higgs branch being classically exact (no quantum correction). It would be natural to investigate the properties of the moduli space of dual Chern-Simons theories, as was done in \cite{Jafferis:2008em}.


\subsection*{Acknowledgments}
\noindent  
We thank Jaume Gomis for sharing ideas, in particular on the possible Chern-Simons description of Yang-Mills fixed points, and for many interesting discussions. We thank Nadav Drukker for comments on the manuscript.
We thank the Perimeter Institute for  hospitality during a visit.
This work was supported by the ERC Starting Grant N. 304806, ``The Gauge/Gravity Duality and Geometry in String Theory''.
Research at the Perimeter Institute is supported in part by the Government of Canada through NSERC and by the Province of Ontario through MRI.


\appendix

\section{$\hat\delta$ distributions}
\label{app:deltahat}

In this paper we consider integrations of meromorphic functions with simple poles, whose integral on the real line is convergent. When a pole lies on the integration contour we choose the principal value prescription to integrate.

Formally, the function $x \rightarrow \hat\delta(x-z_0)$, with $x \in \bR, z_0 \in \bC$, is meant to be the Fourier transform of the function $x \rightarrow e^{-2\pi i z_0 x}$, generalizing the usual Dirac distribution to complex $z_0$:
\begin{align}
\hat\delta (x - z_0) = \int_{\bR} dy \ e^{2\pi i y(x-z_0)} \quad , \quad x \in \bR \, , \, z_0 \in \bC \ .
\end{align}
With $f$ a meromorphic function on $\bC$ with simple poles away from $z_0$.
Concretely we define the $\hat\delta_{z_0} \equiv \hat\delta(\, . - z_0)$ distribution  by the recursion relation:
\begin{align}
\int_{\bR} dx \ \hat\delta(x-z_0)f(x) &= f(z_0) \  + \ 2\pi i \epsilon(z_0) \sum_{j}\hat\delta(u_j-z_0) \hat f(u_j) 
\label{defdelta}
\end{align}
where $\epsilon(z_0)= \ $sgn(Im$(z_0)$) and $u_j$ are the poles of $f$ in the region $0\le $Im$(u)\le $Im$(z_0)$ for $\epsilon(z_0)=+1$ or the region Im$(z_0) \le $Im$(u) \le 0$ for $\epsilon(z_0)=-1$. When $0 < |$Im$(u_j)| < |$Im$(z_0)|$, $\hat f(u_j)$ is the residue at the pole $u_j$. When Im$(u_j) = 0$ or Im$(u_j) = $Im$(z_0)$, $\hat f(u_j)$ is half the residue at the pole $u_j$. 
This definition comes down to shifting the contour of the $x$ integration from $\bR$ to $\bR+i $Im$(z_0)$, picking pole residues, as can be understood from the heuristic derivation:
\begin{align}
 \int_{\bR} dx \ \hat\delta(x-z_0) & f(x) \ =  \ \int_{\bR} dx \int_{\bR} d\sigma \ e^{2\pi i (x-z_0) \sigma} f(x)  \ 
   = \int_{\bR} d\sigma \int_{\bR} dx \ e^{2\pi i (x-z_0) \sigma} f(x) \no\\
  &= \int_{\bR} d\sigma \lp \int_{\bR} dx \ e^{2\pi i (x-Re(z_0)) \sigma} f(x + i {\rm Im}(z_0))  + 2\pi i \epsilon(z_0) \sum_{j} e^{2\pi i (u_j-z_0) \sigma} \hat f(u_j) \rp \no\\
  &=  f(z_0)  + 2\pi i \epsilon(z_0) \sum_{j}\hat\delta(u_j-z_0) \hat f(u_j) \no
\end{align}
where at the second line we have moved the contour of integration of $x$ to $\bR + i $Im$(z_0)$, picking pole residues, and made a change of variable $x \rightarrow x + i $Im$(z_0)$.

 The definition~\eqref{defdelta} implies
\begin{align}
\hat\delta(x-z_0) g(x) = \hat\delta(x-z_0) g(z_0) \ ,
\label{NoPoleFunction}
\end{align}
for any function $g$ without poles on the complex plane (or simply without poles in the region $0\le $Im$(u)\le $Im$(z_0)$).
\footnote{This formula can be applied in an integral as $\int dx \, f(x) g(x) \hat\delta(x-z_0) = \int dx \,  f(x) g(z_0) \hat\delta(x-z_0)$, as long as $f$ does not have a pole at $z_0$. It may happen that $g(z_0)=0$ and $f$ has a pole at $z_0$, so that the integral yields a finite result, in which case \eqref{NoPoleFunction} is not valid.}

When dealing with multiple integrals, the $\hat\delta$ distribution can lead to shifts of contours of integration. For instance:
\begin{align}
\int_{\bR} dx \int_{\bR} d\ti x \ \hat\delta(\ti x - x - i y_0) f(x) g(\ti x) =  \int_{\scC(y_0)} dx \ f(x) g_{y_0}(x) \ ,
\end{align}
where  $y_0 \in \bR$, $g_{y_0}(x) = g(x + i y_0)$, 
 and the final contour of integration $\scC(y_0)$ of $x$ is obtained by deforming the real line $\bR$ to a curve in $\bC$ in such a way that, for $y_0 \ge 0$, the contour passes below the poles $u$ of $g_{y_0}$ in the region $-y_0 \le $Im$(u)\le 0$, or for $y_0 \le 0$, the contour passes above the poles $u$ of $g_{y_0}$ in the region $0\le $Im$(u_g)\le -y_0$, while the poles of $f$ stay on the same side of the contour.

\vspace{4mm}


%

\section{Formulas}
\label{app:formulas}
This appendix contains some formulas we used in our compuations.
\begin{itemize}
\item Weyl denominator formula
\begin{align}
\label{Weylformula}
\prod_{i<j}^N \sh(\sigma_i-\sigma_j) = \sum_{w \in \scS^N} (-1)^w \ e^{2\pi \sum_{j=1}^N \sigma_j W^{(N)}_{w(j)}}
\end{align}
where $W^{(N)}$ is the Weyl vector of $U(N)$: $W^{(N)}_j = \frac{N+1-2j}{2}$, $j=1,\cdots , N$.

\item Cauchy determinant formula
\begin{align}
\label{Cauchyformula}
 \frac{\prod_{i<j}^N \sh(\sigma_i - \sigma_j)\prod_{i<j}^N \sh(\ti \sigma_i - \ti\sigma_j)} {\prod_{i,j}^N \ch(\sigma_i - \ti\sigma_j) } =
 \sum_{ w \in \scS^N} (-1)^{w} \frac{1}{\prod_j^N \ch(\sigma_{w(j)} - \ti\sigma_j)} \ .
\end{align}
A generalized version of this formula is, for $N \geq \ti N$, $\Delta \equiv N-\ti N$:
\begin{align}
& \frac{\prod_{i<j}^N \sh(\sigma_i - \sigma_j)\prod_{i<j}^{\ti N} \sh(\ti \sigma_i - \ti\sigma_j)} {\prod_{i}^N \prod_{j}^{\ti N} \ch(\sigma_i - \ti\sigma_j) }   \no\\
 & \qquad = (-1)^{\Delta \ti N} \sum_{ w \in \scS^N} (-1)^{w} \prod_{j=1}^{\ti N} \frac{e^{-\pi \Delta (\sigma_{w(j)} - \ti\sigma_j)}} {\ch(\sigma_{w(j)} - \ti\sigma_j)} \ \prod_{j=\ti N+1}^{N} e^{2\pi \sigma_{w(j)} \lp \frac{N+\ti N+1}{2}-j \rp } \label{Cauchyformula3} 
\end{align}
This formula was derived from a similar formula in \cite{Matsumoto:2013nya}
\footnote{we use their formula with $x_j = e^{2\pi \ti\sigma_j}$, $y_j = e^{2\pi \sigma_j}$.}

\item Others identities, for $p \in \bN$, $x \in \bC$ :
\begin{align}
& \prod_{j=1}^p 2 \sinh \Big[ \frac{\pi}{p} \lp x + i\,  \frac{p+1}{2} - i j \rp \Big] = \left\{
\begin{array}{c}
2 \cosh(\pi x) \quad , \quad  p \ {\rm even} \\
2 \sinh(\pi x) \quad , \quad  p \ {\rm odd}
\end{array}
\right. \no\\
& \prod_{j=1}^p 2 \cosh \Big[ \frac{\pi}{p} \lp x + i \, \frac{p+1}{2} - i j \rp \Big] = 2 \cosh(\pi x) \label{trigoformulas}\\
& \prod_{1 \le j < k \le p} 2 \sinh \lp \frac{i\pi}{p} (k-j) \rp = i^{\frac{p(p-1)}{2}} \, p^{\frac p 2} \ , \no
\end{align}
and more generally for $A, B \in \bN$ with $p=A+B$ :
\begin{align}
& \prod_{j=1}^A 2 \sinh \Big[ \frac{\pi}{p} \lp x + i\,  \frac{A+1}{2} - i j \rp \Big] \prod_{j=1}^B 2 \cosh \Big[ \frac{\pi}{p} \lp x + i\,  \frac{B+1}{2} - i j \rp \Big] \no\\
& \qquad= \left\{
\begin{array}{c}
2 \cosh(\pi x) \quad , \quad  A \ {\rm even} \\
2 \sinh(\pi x) \quad , \quad  A \ {\rm odd}
\end{array}
\right. \no\\
& (-i)^{\frac{A(A-1)}{2}} p^{-\frac A 2} \prod_{1 \le j < k \le A} 2 \sinh \lp \frac{i\pi}{p} (k-j) \rp = (-i)^{\frac{B(B-1)}{2}} p^{-\frac B 2} \prod_{1 \le j < k \le B} 2 \sinh \lp \frac{i\pi}{p} (k-j) \rp \ .
\label{trigoformulas2}
\end{align}
We tested these formulas with Mathematica for small values of $p$, but did not prove them in general.

\end{itemize}

\section{Computations}
\label{app:computations}

In this appendix we present computations.

\subsection{Local $S$-transformations}
\label{app:Stransfo}

We assume $\Delta = N-\ti N \ge 0$, $p \neq 0$, $q \neq 0$ and we remind $\mu = \frac{N+\ti N}{2}$.
\begin{align*}
& \lp S \pq{p}{q} S^{-1} \rp_{\sigma \, \ti\sigma} = \int d^N\tau d^{\ti N}\ti\tau \ e^{2\pi i \sum_j^N \sigma_j\tau_j} \  \pq{p}{q}_{\tau \, \ti\tau}  \ e^{-2\pi i \sum_j^{\ti N} \ti\tau_j\ti\sigma_j}  \\
& =  \frac{|p|^{-\mu}}{\sqrt{N!\ti N!}} \int d^N\tau d^{\ti N}\ti\tau \ e^{2\pi i \sum_j^N \sigma_j\tau_j} \ e^{-2 \pi i \frac{t}{p} \lp \sum_j^N \tau_j - \sum_j^{\ti N} \ti\tau_j \rp}  \\
& \hspace{4cm} . \ e^{\pi i \, \frac{q}{p} \lp \sum_j^N \tau_j^2 - \sum_j^{\ti N} \ti\tau_j^2 \rp} \
\frac{\prod_{i<j}^N \sh [ \, p^{-1} \tau_{ij}] \ \prod_{i<j}^{\ti N} \sh [\,p^{-1} \ti\tau_{ij}] } {\prod_{i,j}^{N,\ti N} \ch[ \, p^{-1}(\tau_i - \ti\tau_j)] }
 \ e^{-2\pi i \sum_j^{\ti N} \ti\tau_j\ti\sigma_j} \\
& = \frac{|p|^{\mu}}{\sqrt{N!\ti N!}} \int d^N\tau d^{\ti N}\ti\tau \ e^{2\pi i \sum_j^N \tau_j( p \sigma_j - t)} \ 
e^{\pi i \, pq \lp \sum_j^N \tau_j^2 - \sum_j^{\ti N} \ti\tau_j^2 \rp} \\
& \hspace{4cm} . \  
\frac{\prod_{i<j}^N \sh (\tau_{ij}) \ \prod_{i<j}^{\ti N} \sh (\ti\tau_{ij}) } {\prod_{i,j}^{N,\ti N} \ch(\tau_i - \ti\tau_j) }
 \ e^{-2\pi i \sum_j^{\ti N} \ti\tau_j ( p \ti\sigma_j-t)} \\
& = \frac{|p|^{\mu}}{\sqrt{N!\ti N!}} (-1)^{\Delta \ti N} \sum_{ w \in \scS^N} (-1)^{w} 
\int d^N\tau d^{\ti N}\ti\tau \ e^{2\pi i \sum_j^N \tau_j( p \sigma_j - t)} \
e^{\pi i \, pq \lp \sum_j^N \tau_j^2 - \sum_j^{\ti N} \ti\tau_j^2 \rp}  \\
& \hspace{4cm} \ . \ 
 \prod_{j=1}^{\ti N} \frac{e^{- \Delta (\tau_{w(j)} - \ti\tau_j)}}{\ch(\tau_{w(j)} - \ti\tau_j)} \prod_{j= \ti N +1 }^N e^{2\pi \tau_{w(j)} \lp \mu + \half - j \rp}
\ e^{-2\pi i \sum_j^{\ti N} \ti\tau_j ( p \ti\sigma_j-t)}
\end{align*}
where we have rescaled $\tau_j, \ti\tau_j \rightarrow p\tau_j, p \ti\tau_j$ (3rd equality) and used the generalized Cauchy determinant formula~\eqref{Cauchyformula3}.

In each integral we can reshuffle the eigenvalues $\tau_{w(j)} \rightarrow \tau_j$ for $j=1, \cdots, N$ and then shift $\tau_{w(j)} \rightarrow \tau_{w(j)} + \ti\tau_j$ for $j=1, \cdots, \ti N$ to obtain
\begin{align*}
& = (-1)^{\Delta \ti N}  \frac{|p|^{\mu}}{\sqrt{N!\ti N!}}  \sum_{ w \in \scS^N} (-1)^{w} 
\int d^N\tau d^{\ti N}\ti\tau \ e^{2\pi i \sum_j^N \tau_j( p \sigma_{w(j)} - t)} \
e^{\pi i \, pq \sum_j^N \tau_j^2}  \\
& \hspace{4cm} \ . \ 
 \prod_{j=1}^{\ti N} \frac{ e^{- \Delta \tau_j }}{ \ch(\tau_j) } \prod_{j= \ti N +1 }^N e^{2\pi \tau_j \lp \mu + \half - j \rp}
\ e^{2\pi i \, p \sum_j^{\ti N} \ti\tau_j (\sigma_{w(j)} + q \tau_j - \ti\sigma_j )} 
\end{align*}
Integrating over the $\ti\tau_j$ yields $\delta[p(\sigma_{w(j)} + q \tau_j - \ti\sigma_j)]$ that can be used to integrate over the $\tau_j$ for $j=1,\cdots, \ti N$ (since $pq \neq 0$). We get
\begin{align*}
& = (-1)^{\Delta \ti N} \frac{|q|^{-\ti N} |p|^{\frac{\Delta}{2}}}{\sqrt{N!\ti N!}}  \sum_{ w \in \scS^N} (-1)^{w} 
\int d^{\Delta}\tau \ e^{2\pi i \sum_{j=\ti N+1}^N \tau_j( p \sigma_{w(j)} - t)} \ e^{ \frac{2\pi i}{q} \sum_j^{\ti N} (\ti\sigma_j-\sigma_{w(j)})( p \sigma_{w(j)} - t) }  \\
& \hspace{3cm} \ . \
e^{\pi i \, \frac p q \sum_j^{\ti N} (\ti\sigma_j-\sigma_{w(j)})^2} e^{\pi i \, pq \sum_{j=\ti N+1}^N \tau_j^2}  \ 
 \prod_{j=1}^{\ti N} \frac{ e^{- \frac{\Delta}{q} (\ti\sigma_j-\sigma_{w(j)}) }}{ \ch[q^{-1}(\ti\sigma_j-\sigma_{w(j)})] } \prod_{j= \ti N +1 }^N e^{2\pi \tau_j \lp \mu + \half - j \rp}  \\
& = (-1)^{\Delta \ti N} \frac{|q|^{-\ti N}|p|^{\frac{\Delta}{2}}}{\sqrt{N!\ti N!}}  \sum_{ w \in \scS^N} (-1)^{w} 
e^{\pi i \, \frac p q \sum_j^{\ti N} (\ti\sigma_j-\sigma_{w(j)})^2}
\ e^{2\pi i \frac p q \sum_j^{\ti N} (\ti\sigma_j-\sigma_{w(j)})( \sigma_{w(j)} - \frac t p) }
 \prod_{j=1}^{\ti N} \frac{ e^{- \frac{\Delta}{q} (\ti\sigma_j-\sigma_{w(j)}) }}{ \ch[q^{-1}(\ti\sigma_j-\sigma_{w(j)})] }  \\
 & \hspace{2cm} . \ \prod_{j= \ti N +1 }^N 
\int d\tau \  e^{\pi i \, pq \, \tau^2}  \ e^{2\pi i \tau \lp p \sigma_{w(j)} - t -i ( \mu + \half - j ) \rp}   \\
& = (-1)^{\Delta \ti N} \frac{|q|^{-\ti N}|p|^{\frac{\Delta}{2}}}{\sqrt{N!\ti N!}}  \sum_{ w \in \scS^N} (-1)^{w} 
e^{\pi i \, \frac p q \sum_j^{\ti N} (\ti\sigma_j-\sigma_{w(j)})^2}
\ e^{2\pi i \frac p q \sum_j^{\ti N} (\ti\sigma_j-\sigma_{w(j)})( \sigma_{w(j)} - \frac t p) }
 \prod_{j=1}^{\ti N} \frac{ e^{- \frac{\Delta}{q} (\ti\sigma_j-\sigma_{w(j)}) }}{ \ch[q^{-1}(\ti\sigma_j-\sigma_{w(j)})] }  \\
 & \hspace{2cm} . \  e^{\frac{\pi i}{12\, pq}\Delta(\Delta^2-1)} \, e^{-  \frac{\pi i}{pq} \sum_{j=\ti N+1}^{N} ( p \sigma_{w(j)} - t)^2}  |pq|^{-\frac{\Delta}{2}} \prod_{j= \ti N +1 }^N e^{-\frac{2\pi}{q} \sigma_{w(j)} (\mu +\half-j)} \ .
\end{align*}
It is now possible to rearrange the factors and then to use the Cauchy formula~\eqref{Cauchyformula3} backward to obtain
\begin{align}
 \lp S \pq{p}{q} S^{-1} \rp_{\sigma \, \ti\sigma}  
 & =  e^{-\frac{i \pi \Delta t^2}{pq}} \,  \frac{|q|^{-\mu}}{\sqrt{N!\ti N!}} \, e^{\frac{\pi i}{12\, pq}\Delta(\Delta^2-1)} e^{-\pi i \frac p q \lp \sum_j^N \sigma_j^2 - \sum_j^{\ti N} \ti\sigma_j^2 \rp } e^{ 2\pi i \frac t q \lp \sum_j^N \sigma_j - \sum_j^{\ti N}\ti\sigma_j \rp }  \no\\
 & \qquad . \ \frac{\prod_{i<j}^N \sh [ - q^{-1} \sigma_{ij}] \ \prod_{i<j}^{\ti N} \sh [-q^{-1} \ti\sigma_{ij}] } {\prod_{i,j}^{N,\ti N} \ch[- q^{-1}(\sigma_i - \ti\sigma_j)]} \no\\
&= e^{\frac{\pi i}{12\, pq}\Delta(\Delta^2-1)}  \  e^{-\frac{i \pi \Delta t^2}{pq}} \  \pq{-q}{p}_{\sigma \, \ti\sigma} \ .
\end{align}

When $N < \ti N$, the computation amounts to changing $N \leftrightarrow \ti N$, $\sigma \leftrightarrow -\ti\sigma$, $t \rightarrow -t$ and $q \rightarrow -q$. 
The above result is invariant under these exchanges, so it holds also for $N < \ti N$.

\subsection{HW-move identity}
\label{app:HWmove}

Here we give the details of the computations proving the identity \eqref{HWidentity1}. 

We consider the case $N_1 + p \le N_2 \le N_3$ with $p>0$.
This implies $N_1 +p \le \ti N_2 \le N_3$.
The part of the matrix model corresponding to the graph on the left of figure (\ref{HWmove}-a) is given by the product of the two 5-brane factors:
\begin{align}
& \lp \pq{0}{1}{}_{(N_2)} \pq{p}{q} \rp_{\sigma \ti\sigma}  \no\\
& \  = \int d^{N_2}\tau \, 
\frac{(N_1! N_3!)^{-\half}}{\prod_{j}^{N_1} \ch \Big[ \sigma_j - m + i \frac{\Delta_{21}}{2} \Big]} \  \prod_{j=1}^{N_1} \delta \lp \tau_j-\sigma_j \rp  \ \prod_{j=1}^{\Delta_{21}}  \hat\delta \lp \tau_{j+N_1} - \bar m^{(\Delta_{21})}_{j} \rp   \no\\
&  \quad  . \ |p|^{-\mu_{23}} \ e^{\pi i \, \frac{q}{p} \lp \sum_j^{N_2} \tau_j^2 - \sum_j^{N_3} \ti\sigma_j^2 \rp} \ e^{-2\pi i \frac t p (\sum_j^{N_2} \tau_j - \sum_j^{N_3} \ti\sigma_j) } \ \frac{\prod_{i<j}^{N_2} \sh(p^{-1}\tau_{ij})\prod_{i<j}^{N_3} \sh(p^{-1}\ti\sigma_{ij})} { \prod_{i,j}^{N_2,N_3} \ch[p^{-1}(\tau_i - \ti\sigma_j)] }  \no\\
 & \  = \sum_{w \in \scS_{N_3}} \frac{(-1)^{w}}{\sqrt{N_1! N_3!}} \int d^{N_2}\tau \, 
\frac{ \prod_{j=1}^{N_1} \delta \lp \tau_j-\sigma_j \rp }{\prod_{j}^{N_1} \ch \Big[ \sigma_j - m + i \frac{\Delta_{21}}{2} \Big]}  \ \prod_{j=1}^{\Delta_{21}}  \hat\delta \lp \tau_{j+N_1} - \bar m^{(\Delta_{21})}_{j} \rp \  
e^{\pi i \, \frac{q}{p} \lp \sum_j^{N_2} \tau_j^2 - \sum_j^{N_3} \ti\sigma_j^2 \rp} \no\\
&  \quad  . \ (-1)^{\Delta_{32}N_2} |p|^{-\mu_{23}}  \ e^{-2\pi i \frac t p (\sum_j^{N_2} \tau_j - \sum_j^{N_3} \ti\sigma_j) } \ 
\prod_{j=1}^{N_2} \frac{ e^{-\frac{\Delta_{32}}{p}(\ti\sigma_{w(j)} - \tau_j)} }{\ch[p^{-1}(\ti\sigma_{w(j)} - \tau_j)]} \prod_{N_2 < i < j \le N_3} e^{\frac{2\pi}{p}\ti\sigma_{w(j)} \lp \mu_{23} + \half - j \rp } \no\\
& \  = \sum_{w \in \scS_{N_3}} (-1)^{w} \frac{(-1)^{\Delta_{32}N_2} |p|^{-\mu_{23}}}{\sqrt{N_1! N_3!}} \,
\frac{e^{2\pi i \frac t p (\sum_j^{N_3} \ti\sigma_j- \sum_j^{N_1} \sigma_j)} 
e^{\pi i \, \frac{q}{p} \lp \sum_j^{N_1} \sigma_j^2 - \sum_j^{N_3} \ti\sigma_j^2 \rp} }{\prod_{j}^{N_1} \ch \Big[ \sigma_j - m + i \frac{\Delta_{21}}{2} \Big]} \ 
 \prod_{j=1}^{N_1} \frac{e^{-\frac{\Delta_{32}}{p}(\ti\sigma_{w(j)} - \sigma_j)}}{\ch[p^{-1}(\ti\sigma_{w(j)} - \sigma_j)]} \no\\
&  \quad . \prod_{N_2 < i < j \le N_3}  e^{\frac{2\pi}{p}\ti\sigma_{w(j)} \lp \mu_{23} + \half - j \rp } \ \prod_{j=1}^{\Delta_{21}} \Big[ \int d\tau  \ e^{-2\pi i \frac t p \tau} \, e^{\pi i \frac q p \tau^2}
\ \frac{e^{-\frac{\Delta_{32}}{p}(\ti\sigma_{w(j+N_1)} - \tau)}}{\ch[p^{-1}(\ti\sigma_{w(j+N_1)} - \tau)]} \
\hat\delta \lp \tau - \bar m^{(\Delta_{21})}_{j} \rp  \Big] 
\\
& \  = \sum_{w \in \scS_{N_3}} (-1)^{w} \frac{(-1)^{\Delta_{32}N_2} |p|^{-\mu_{23}}}{\sqrt{N_1! N_3!}} \,
\frac{e^{2\pi i \frac t p (\sum_j^{N_3} \ti\sigma_j- \sum_j^{N_1} \sigma_j)} 
e^{\pi i \, \frac{q}{p} \lp \sum_j^{N_1} \sigma_j^2 - \sum_j^{N_3} \ti\sigma_j^2 \rp} }{\prod_{j}^{N_1} \ch \Big[ \sigma_j - m + i \frac{\Delta_{21}}{2} \Big]} \ 
 \prod_{j=1}^{N_1} \frac{e^{-\frac{\Delta_{32}}{p}(\ti\sigma_{w(j)} - \sigma_j)}}{\ch[p^{-1}(\ti\sigma_{w(j)} - \sigma_j)]} \no\\
&  \quad .  \quad 
 e^{-2\pi i \frac t p \Delta_{21} m} \  
 e^{\pi i \frac q p  \Delta_{21} m^2 } \
 e^{\frac{\pi i}{12} \frac{q}{p} \Delta_{21}(\Delta_{21}^2-1)} 
 \prod_{N_2 < i < j \le N_3}  e^{\frac{2\pi}{p}\ti\sigma_{w(j)} \lp \mu_{23} + \half - j \rp } \no\\
&  \quad .  \quad
\prod_{j=1}^{\Delta_{21}}  \int d\tau \, e^{\frac{\Delta_{32}}{p}(\tau-\ti\sigma_{w(j+N_1)})} \, 
\ \frac{\hat\delta \lp \tau - \bar m^{(\Delta_{21})}_{j} \rp }{\ch[p^{-1}(\tau - \ti\sigma_{w(j+N_1)})]} 
\label{eqnTot1}
\end{align}
At the first line we have de-anti-symmetrized the D5-factor, using the fact that the $(p,q)$-factor is already antisymmetric under permutation of the $\tau_j$ eigenvalues. At the second line we have replaced the NS5-factor using (\ref{Cauchyformula3}). At the third line we have integrated over $\tau_j$ for $j=1,\cdots, N_1$, using the delta functions. 
At the fourth line we have replaced $e^{-2\pi i \frac t p \tau} \, e^{\pi i \frac q p \tau^2} \rightarrow e^{-2\pi i \frac t p \bar m_{j}} \, e^{\pi i \frac q p \bar m_{j}{}^2}$ using the property \eqref{NoPoleFunction}.
The next step is to evaluate the remaining $\Delta_{21}$ integrals using (\ref{defdelta}).

We introduce $j^{\ast} \equiv \frac{\Delta_{21}+1}{2}$ and first assume $j^{\ast} \in \bN$ corresponding to $\Delta_{21}$ odd. Then we translate the labels $j \rightarrow j + j^{\ast}$, so that $\prod_{j=1}^{\Delta_{21}} \rightarrow \prod_{j=- j^{\ast}+1}^{ j^{\ast}-1}$ , $ \bar m^{(\Delta_{21})}_{j} \rightarrow \bar m^{(\Delta_{21})}_{j+ j^{\ast}} = m + i j$ , $\ti\sigma_{w(j+N_1)} \rightarrow \ti\sigma_{w(j+\mu +1/2)} \equiv \ti\tau_j$ and we 
focus on a single integral :
\begin{align}
\label{eq00}
\scI(\ti\tau_j) &= \int d\tau  \  e^{\frac{\Delta_{32}}{p}(\tau - \ti\tau_j)} \  \frac{\hat\delta (\tau -m - i j)}{\ch[p^{-1}(\tau - \ti\tau_j)]}  \ .
\end{align}
We also assume for the moment that $p$ is odd and positive, so that $\frac p 2 \in \bN + \half$.
Then for $-\frac{p}{2} < j < \frac{p}{2}$ the integrand has no pole in the region $0 \le |$Im$(\tau)| \le |j|$, so the evaluation of the $\hat\delta$ is simply
\begin{align}
\scI(\ti\tau_j) &= \ \frac{ e^{\frac{\Delta_{32}}{p}(m + ij - \ti\tau_j)}}{\ch[p^{-1}(\ti\tau_j - m -ij)]}
\label{eq01}
\end{align}
For $ \frac p 2 < |j| < \frac{3p}{2}$, the pole at $\tau = \ti\tau_j \pm i \frac{p}{2}$ contributes to the evaluation of the $\hat\delta$ giving
\begin{align}
\scI(\ti\tau_j) &= \ \frac{ e^{\frac{\Delta_{32}}{p}(m + ij - \ti\tau_j)}}{\ch[p^{-1}(\ti\tau_j - m -ij)]} 
 \ + \  p \  e^{\pm i \frac{\Delta_{32}}{2}} \,  \hat\delta \Big[ \ti\tau_j-m - i\lp j \mp \frac{p}{2} \rp  \Big] 
\label{eq02}
\end{align}
where $\mp$ is the sign of $-j$.

\noindent More generally for $\lp k- \half \rp p < |j| < \lp k + \half \rp p$ with $k \in \bN$, the poles $\tau = \ti\tau_j \pm i \lp n + \half \rp p$ with $n=0,1,\cdots,k-1$ contribute to the evaluation of the integral with the $\hat\delta$ giving:
\begin{align}
& \scI(\ti\tau_j) \ = \ \frac{ e^{\frac{\Delta_{32}}{p}(m + ij - \ti\tau_j)}}{\ch[p^{-1}(\ti\tau_j - m -ij)]} 
 \ + \  \sum_{n=0}^{k-1} p \
 (-1)^{n} \  e^{\pm i \Delta_{32}\lp n + \half \rp} \, \hat\delta \Big[ \ti\tau_j-m - i\lp j \mp p \lp n + \half \rp \rp  \Big] 
\label{eq03}
\end{align}
where $\pm$ is the sign of $j$.

We have now to gather these results and consider $\prod_{j=- j^{\ast}+1}^{ j^{\ast}-1}\scI(\ti\tau_j)$, which is a product of sums. However the final result in~\eqref{eqnTot1} is anti-symmetrized over permutations of the $\ti\tau_j$. This means that 
when we expand the product $\prod_{j=- j^{\ast}+1}^{ j^{\ast}-1}\scI(\ti\tau_j)$, all the terms symmetric in $\ti\tau_{j_1}, \ti\tau_{j_2}$ (=invariant under $\ti\tau_{j_1}\leftrightarrow \ti\tau_{j_2}$), for any $j_1 \neq j_2$, will not contribute to the final result and can be dropped.
For $0 \le |j| < \frac{p}{2}$, $\scI(\ti\tau_j)$ contains a single term~\eqref{eq01} that will contribute to the final result.
 For $ \frac p 2 < |j| < \frac{3p}{2}$, $\scI(\ti\tau_j)$ is a sum of two terms~\eqref{eq02}, but the first term in~\eqref{eq02} can be dropped because of the anti-symmetrization with the previous terms $0 \le |j| < \frac{p}{2}$. Then only the second term in~\eqref{eq02} with a $\hat\delta$ will contribute. Similarly for $\lp k- \half \rp p < |j| < \lp k + \half \rp p$, $\scI(\ti\tau_j)$ is a sum of $1+(k-1)$ terms~\eqref{eq03}, but $1+(k-2)$ terms can be dropped because of the anti-symmetrization with the previous terms $0 \le |j| < \lp k- \half \rp p$. Only the term with a $\hat\delta$ for $n=0$ in~\eqref{eq03} will contribute.
 
In total we get
\begin{align}
 \prod_{j=- j^{\ast}+1}^{ j^{\ast}-1}\scI(\ti\tau_j) & = \prod_{j=-(p-1)/2}^{(p-1)/2} \frac{ e^{\frac{\Delta_{32}}{p}(m + ij - \ti\tau_j)}}{\ch[p^{-1}(\ti\tau_j - m -ij)]} \no\\
& \ . \ \prod_{j=(p+1)/2}^{j^{\ast}-1} p \  e^{ i \frac{\Delta_{32}}{2}} \, \hat\delta \Big[ \ti\tau_j-m - i\lp j - \frac{p}{2} \rp  \Big] \
  \prod_{j=-j^{\ast}+1}^{-(p+1)/2} p \  e^{- i \frac{\Delta_{32}}{2}} \, \hat\delta \Big[ \ti\tau_j-m - i\lp j + \frac{p}{2} \rp  \Big] \no\\
  &  \ + \quad {\rm sym} 
\label{eqn04} 
\end{align}
where ``sym" denotes terms symmetric in $\ti\tau_{j_1}, \ti\tau_{j_2}$ for some $j_1 \neq j_2$, that drop from the computation.

The eigenvalues $\ti\sigma_{w(j+N_1)}=\ti\tau_{j-j^{\ast}}$ can be permuted $\ti\sigma_{w(j)} \rightarrow \ti\sigma_{w(w'((j))}$, with some permutation $w'$, to rearrange the result as:
\begin{align}
 \prod_{j=1}^{ \Delta_{21}}\scI(\ti\sigma_{w(j+N_1)}) &=
  \ p^{\Delta_{3\ti 2}} \ 
 \prod_{j=1}^{p} \frac{ e^{-\frac{\Delta_{32}}{p}(\ti\sigma_{w(j+N_1)} - m^{(p)}_j)} }{ \ch[p^{-1}(\ti\sigma_{w(j+N_1)} - m^{(p)}_j )]}  
 \ \prod_{j=1}^{\Delta_{3\ti 2}} \hat\delta\lp \ti\sigma_{w(j+N_1+p)} - \bar m_j \rp
 \label{eqn05}
\end{align}
where $\Delta_{3 \ti 2} = \Delta_{21}-p = N_3 - \ti N_2$, $m^{(p)}_j =  m + i(\frac{p+1}{2}-j)$ and $\bar m_j = \bar m^{(\Delta_{3 \ti 2})}_j = m - i(\frac{\Delta_{3 \ti 2}+1}{2}-j)$.
The permutation of the $\ti\sigma_{w(j)}$ eigenvalues affects the total result~\eqref{eqnTot1} only by a sign (which is the signature of the permutation) that we will not keep track of.\\
\noindent Let us define $\sigma_{j+N_1} = m^{(p)}_j$, for $j=1, \cdots, p$, as $p$ ``frozen eigenvalues".
Plugging~\eqref{eqn05} back in~\eqref{eqnTot1} and permuting again the $\ti\sigma_{w(j)}$ eigenvalues conveniently yields (up to a sign):
\begin{align}
& \lp \pq{0}{1}{}_{(N_2)} \pq{p}{q} \rp_{\sigma \ti\sigma}  \no\\
& \qquad = \sum_{w \in \scS_{N_3}} (-1)^{w} \frac{ p^{-\mu_{1\ti 2} -\frac p 2}}{\sqrt{N_1! N_3!}} \,
e^{-2\pi i t m} \,
 e^{\pi i q m^2 } \,
 e^{\frac{\pi i}{12} \frac q p \Phi} \
\frac{e^{2\pi i \frac t p (\sum_j^{\ti N_2} \ti\sigma_{w(j)}- \sum_j^{N_1} \sigma_j)} 
e^{\pi i \, \frac{q}{p} \lp \sum_j^{N_1} \sigma_j^2 - \sum_j^{\ti N_2} \ti\sigma_j^2 \rp} }{\prod_{j}^{N_1} \ch \Big[ \sigma_j - m + i \frac{\Delta_{21}}{2} \Big]} \no\\
& \qquad \quad . 
 \prod_{j=1}^{N_1+p} \frac{ e^{-\frac{\Delta_{32}}{p}(\ti\sigma_{w(j)} - \sigma_j)} }{ \ch[p^{-1}(\ti\sigma_{w(j)} - \sigma_j)] } \ 
 \prod_{j=N_1+p+1}^{\ti N_2}  e^{\frac{2\pi}{p}\ti\sigma_{w(j)} \lp \mu_{23} + \half - j \rp }
  \ \prod_{j=1}^{\Delta_{3\ti 2}} \hat\delta\lp \ti\sigma_{w(j+\ti N_2)} - \bar m^{(\Delta_{3 \ti 2})}_j \rp  \ ,
\end{align}
with $\mu_{1\ti 2}= \frac{N_1 + \ti N_2}{2}$, $\Phi = \Delta_{21}(\Delta_{21}^2-1) - \Delta_{3\ti 2}(\Delta_{3\ti 2}^2-1)$ and we have used the property \eqref{NoPoleFunction} again to modify some exponential terms.\\
Transforming
$\sum_{w \in \scS_{N_3}} (-1)^w f[\ti\sigma_{w(j)}]= \sum_{w \in \scS_{N_3}} \frac{1}{\ti N_2!} \sum_{w' \in \scS_{\ti N_2}} (-1)^{w+w'} f[\ti\sigma_{w(w'(j))}]$ and using the Cauchy formula~\eqref{Cauchyformula3} backwards yields:
\begin{align}
& \lp \pq{0}{1}{}_{(N_2)} \pq{p}{q} \rp_{\sigma \ti\sigma}  \no\\
& \qquad = \sum_{w \in \scS_{N_3}} (-1)^{w} \frac{p^{-\mu_{1\ti 2} -\frac p 2}}{|W|} \,
e^{-2\pi i t m} \,
 e^{\pi i q m^2 } \,
 e^{\frac{\pi i}{12} \frac q p \Phi}  \,
\frac{e^{2\pi i \frac t p (\sum_j^{\ti N_2} \ti\sigma_{w(j)}- \sum_j^{N_1} \sigma_j)} 
e^{\pi i \, \frac{q}{p} \lp \sum_j^{N_1} \sigma_j^2 - \sum_j^{\ti N_2} \ti\sigma_j^2 \rp} }{\prod_{j}^{N_1} \ch \Big[ \sigma_j - m + i \frac{\Delta_{21}}{2} \Big]} \no\\
& \qquad \quad . \
\frac{\prod_{i<j}^{N_1+p} \sh(p^{-1}\sigma_{ij})\prod_{i<j}^{\ti N_2} \sh(p^{-1}\ti\sigma_{w(ij)})} {\prod_{i,j}^{N_1+p, \ti N_2} \ch[p^{-1}(\sigma_i - \ti\sigma_{w(j)})] }
\ \prod_{j=1}^{\Delta_{3\ti 2}} \hat\delta\lp \ti\sigma_{w(j+\ti N_2)} - \bar m^{(\Delta_{3 \ti 2})}_j \rp \no\\
\end{align}
where $ |W| \equiv \ti N_2! \sqrt{N_1! N_3!}$ \, .
Remembering that $\sigma_{j+N_1} = m - i(\frac{p+1}{2}-j)$ and using the trigonometric formulas~\eqref{trigoformulas}, the above simplifies to:
\begin{align}
& \lp \pq{0}{1}{}_{(N_2)} \pq{p}{q} \rp_{\sigma \ti\sigma}  \no\\
& \qquad = \sum_{w \in \scS_{N_3}} (-1)^{w} \frac{p^{-\mu_{1\ti 2}}}{|W|} \,
e^{-2\pi i t m} \,
 e^{\pi i q m^2 } \,
 e^{\frac{\pi i}{12} \frac q p \Phi}  \,
e^{2\pi i \frac t p \lp \sum_j^{\ti N_2} \ti\sigma_{w(j)}- \sum_j^{N_1} \sigma_j \rp} 
e^{\pi i \, \frac{q}{p} \lp \sum_j^{N_1} \sigma_j^2 - \sum_j^{\ti N_2} \ti\sigma_j^2 \rp} \no\\
& \qquad \quad . \
\frac{\prod_{i<j}^{N_1} \sh(p^{-1}\sigma_{ij})\prod_{i<j}^{\ti N_2} \sh(p^{-1}\ti\sigma_{w(ij)})} {\prod_{i,j}^{N_1, \ti N_2} \ch[p^{-1}(\sigma_i - \ti\sigma_{w(j)})] }
\ \frac{\prod_{j=1}^{\Delta_{3\ti 2}} \hat\delta\lp \ti\sigma_{w(j+\ti N_2)} - \bar m^{(\Delta_{3 \ti 2})}_j \rp } {\prod_{j}^{\ti N_2} \ch(\ti\sigma_{w(j)}-m)} \quad , \quad \Big[ \, \Delta_{3 \ti 2} \ {\rm even}  \Big]
\label{finaleqn1}
\end{align}
up to factors of $i$ that we do not keep track of.

We have treated the case when $\Delta_{21}$ and $p$ are both odd. When $\Delta_{21}$ and $p$ are both even, the analysis proceeds similarly and yields the same result~\eqref{finaleqn1}. Together these two cases correspond to $\Delta_{3 \ti 2}$ being even. When $\Delta_{3 \ti 2}$ is odd, corresponding to $\Delta_{21}$ and $p$ with different parities, there is an extra complication in the evaluation of the $\hat\delta$ due to the presence of half pole contributions (see app.\ref{app:deltahat}). However the computations can still be performed following the same steps. The ``frozen eigenvalues" are $\sigma_{j+N_1} = m - i(\frac{p}{2}-j)$ in this case and the result is given, up to factors of $i$, by: 
\begin{align}
& \lp \pq{0}{1}{}_{(N_2)} \pq{p}{q} \rp_{\sigma \ti\sigma}  \no\\
& \qquad = \sum_{w \in \scS_{N_3}} (-1)^{w} \frac{p^{-\mu_{1\ti 2}}}{|W|} \,
e^{-2\pi i t m} \,
 e^{\pi i q m^2 } \,
 e^{\frac{\pi i}{12} \frac q p \Phi}  \,
e^{2\pi i \frac t p \lp \sum_j^{\ti N_2} \ti\sigma_{w(j)}- \sum_j^{N_1} \sigma_j \rp} 
e^{\pi i \, \frac{q}{p} \lp \sum_j^{N_1} \sigma_j^2 - \sum_j^{\ti N_2} \ti\sigma_j^2 \rp} \no\\
& \qquad \quad . \
\frac{\prod_{i<j}^{N_1} \sh(p^{-1}\sigma_{ij})\prod_{i<j}^{\ti N_2} \sh(p^{-1}\ti\sigma_{w(ij)})} {\prod_{i,j}^{N_1, \ti N_2} \ch[p^{-1}(\sigma_i - \ti\sigma_{w(j)})] }
\ \frac{\prod_{j=1}^{\Delta_{3\ti 2}} \hat\delta\lp \ti\sigma_{w(j+\ti N_2)} - \bar m^{(\Delta_{3 \ti 2})}_j \rp } {\prod_{j}^{\ti N_2} \sh(\ti\sigma_{w(j)}-m)} \quad , \quad \Big[ \, \Delta_{3 \ti 2} \ {\rm odd}  \Big]
\label{finaleqn1bis}
\end{align}
which differs from~\eqref{finaleqn1} only by the ``$\sh$" in the denominator of the last factor.

\bigskip

We may now consider the other combination of 5-brane factors involved in the HW identity:
\begin{align}
& \lp \pq{p}{q}{}_{(\ti N_2)} \pq{0}{1} \rp_{\sigma \ti\sigma}  \no\\
& = \int d^{\ti N_2}\tau \, \frac{|p|^{-\mu_{1\ti 2}}}{|W|} \, \ e^{\pi i \, \frac{q}{p} \lp \sum_j^{N_1} \sigma_j^2 - \sum_j^{\ti N_2} \tau_j^2 \rp} \ e^{-2\pi i \frac t p (\sum_j^{N_1} \sigma_j - \sum_j^{\ti N_2} \tau_j) } \ \frac{\prod_{i<j}^{N_1} \sh(p^{-1}\sigma_{ij})\prod_{i<j}^{\ti N_2} \sh(p^{-1}\tau_{ij})} {\prod_{i,j}^{N_1,\ti N_2} \ch[p^{-1}(\sigma_i - \tau_j)] } 
  \no\\
& \qquad \qquad  .  \Big[ \sum_{w^ \in \scS_{N_3}} (-1)^{w} \frac{ \prod_{j=1}^{\ti N_2} \delta \lp \ti\sigma_{w(j)}- \tau_j \rp }{ \prod_{j}^{\ti N_2} \ch \Big[ \tau_j - m + i \frac{\Delta_{3\ti 2}}{2} \Big]}  \ \prod_{j=1}^{\Delta_{3\ti 2}}  \hat\delta \lp \ti\sigma_{w(j+\ti N_2)} - \bar m^{(\Delta_{3\ti 2})}_{j} \rp  \Big] \no\\
&= \sum_{w^ \in \scS_{N_3}} (-1)^{w} \, 
\frac{|p|^{-\mu_{1\ti 2}}}{|W|} \, \ e^{\pi i \, \frac{q}{p} \lp \sum_j^{N_1} \sigma_j^2 - \sum_j^{\ti N_2} \ti\sigma_{w(j)}^2 \rp} \ e^{-2\pi i \frac t p \lp \sum_j^{N_1} \sigma_j - \sum_j^{\ti N_2} \ti\sigma_{w(j)} \rp } \no\\
& \qquad \qquad  . \frac{\prod_{i<j}^{N_1} \sh(p^{-1}\sigma_{ij})\prod_{i<j}^{\ti N_2} \sh(p^{-1}\ti\sigma_{w(ij)} )} {\prod_{i,j}^{N_1,\ti N_2} \ch[p^{-1}(\sigma_i - \ti\sigma_{w(j)} )] }    \
 \frac{ \prod_{j=1}^{\Delta_{3\ti 2}}  \hat\delta \lp \ti\sigma_{w(j+\ti N_2)} - \bar m^{(\Delta_{3\ti 2})}_{j} \rp }{ \prod_{j}^{\ti N_2} \ch \Big[ \ti\sigma_{w(j)} - m + i \frac{\Delta_{3\ti 2}}{2} \Big]}  \ , 
 \label{finaleqn2}
\end{align}
where we have simply integrated over the $\tau_j$ with the $\dd$-functions.

From~\eqref{finaleqn1}, ~\eqref{finaleqn1bis} and ~\eqref{finaleqn2}, we observe the relation (up to factors of $i$) :
\begin{align}
\lp \pq{0}{1}{}_{(N_2)} \pq{p}{q} \rp_{\sigma \ti\sigma} = e^{- 2\pi i t m} \  e^{ \pi i q m^2 } \  e^{\frac{\pi i}{12} \frac q p \Phi} \ 
\lp \pq{p}{q}{}_{(\ti N_2)} \pq{0}{1} \rp_{\sigma \ti\sigma} \ .
\label{finalmap}
\end{align}

There are various other cases to consider with different orderings between $N_2$ and $N_1, N_1+p, N_3$, plus cases with negative $p$. In each of these cases the computations  involve the same tricks as in the case described above. It is also necessary to use the more general relations~\eqref{trigoformulas2}. These computations are long and tedious, so we do not reproduce them here. Also we mention that we were not able to complete the computations for the extremal cases $ 0 < N_2 < |p|$ and $N_1+N_3 < N_2 < N_1 + N_3 + |p|$ due to some additional complications appearing. In practice we will only consider $p=\pm 1$ for which these cases do not exist. 

The final result is the relation:
\begin{align}
\lp \pq{0}{1}{}_{(N_2)} \pq{p}{q} \rp_{\sigma \ti\sigma} = e^{\mp 2\pi i t m} \  e^{\pm \pi i q m^2 } \  e^{\frac{\pi i}{12} \frac q p \Phi} \ 
\lp \pq{p}{q}{}_{(\ti N_2)} \pq{0}{1} \rp_{\sigma \ti\sigma} \ ,
\label{finalrelation}
\end{align}
where $\pm$ is the sign of $p$ (and $\mp$ its opposite) and $\Phi = |\Delta_{21}|(\Delta_{21}^2-1) - |\Delta_{3\ti 2}|(\Delta_{3\ti 2}^2-1)$.

\subsection{Matrix model of a separated graph}
\label{app:Check}

\begin{figure}[h]
\centering
\includegraphics[scale=0.8]{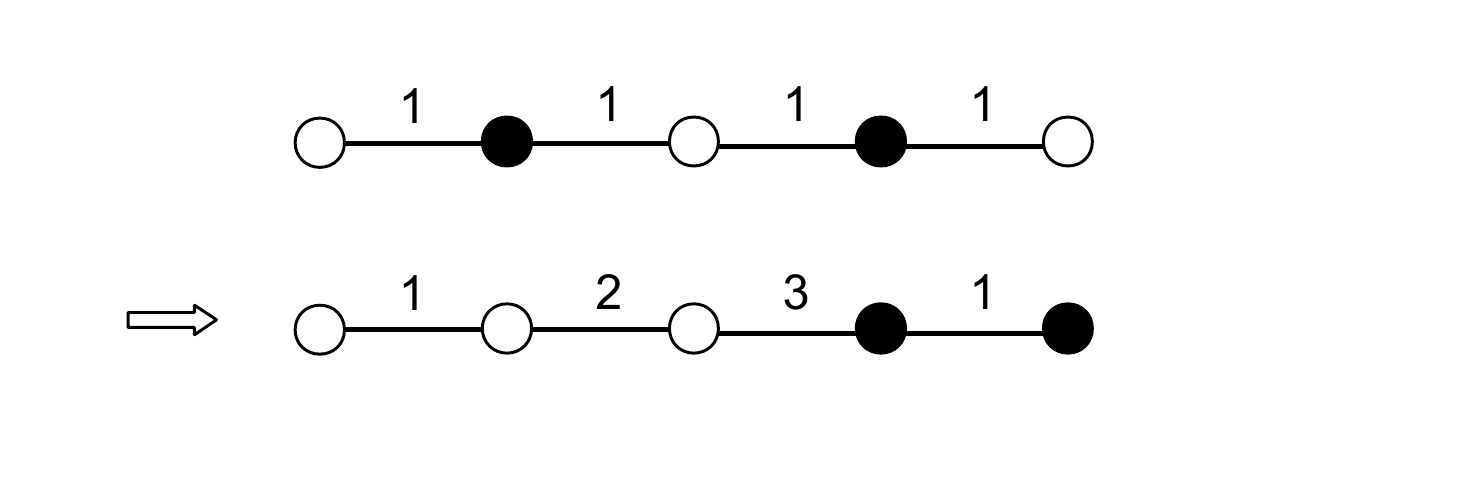}
\vskip -10mm
\caption{\footnotesize  Graph of the $T^{(12)}_{(111)}(SU(3))$ quiver SCFT and its separated graph (black dots pushed to the right). }
\label{checkgraph}
\end{figure}

Here we show explicitly how the matrix model of the separated graph is mapped to the matrix model of the quiver theory in an explicit example.
We consider the theory with $U(1)\times U(1)$ gauge group, one bifundamental hypermultiplet and one fundamental hypermultiplet in each $U(1)$ node. The corresponding SCFT deformed by mass and FI parameters is $T^{(12)}_{(111)}(SU(3))$ and its graph is shown in figure \ref{checkgraph}.
After moving the black dots to the right of the graph, one obtains the separated graph of figure \ref{checkgraph}. The matrix model associated to this graph is:
\begin{align*}
Z_{\rm separ} &= \int \frac{d\sigma d^2\ti\sigma d^3\ti\lambda d\lambda}{2.3!} \, \frac{e^{2\pi i (t_1-t_2)\sigma}}{\prod_j^2 \ch(\sigma-\ti\sigma_j)} 
\frac{e^{2\pi i (t_2-t_3)(\ti\sigma_1 + \ti\sigma_2)}}{\prod_{j,k}^{2,3} \ch(\lambda_k-\ti\sigma_j)} \sh(\ti\sigma_{12})^2 e^{2\pi i t_3(\sum_k^3 \ti\lambda_k)} \prod_{j<k}\sh(\ti\lambda_{jk}) \\
 & \qquad . \ \lp \sum_{w \in S_3} (-1)^w \frac{\delta(\ti\lambda_{w(1)}- \lambda)}{\ch(\lambda-m_2+i)} \hat\delta[ \ti\lambda_{w(2)} -m_2 +\frac i 2 ]  \hat\delta[\ti\lambda_{w(3)} -m_2 -\frac i 2 ] \rp \delta(\lambda - m_1)
\end{align*}
After a few simplifications the matrix model can be expressed as
\begin{align*}
 &= \frac{e^{2\pi i t_3 m_1}}{2 \ch(m_1-m_2)}\int d\sigma d^2\ti\sigma d^2\ti\lambda \, \frac{e^{2\pi i (t_1-t_2)\sigma}}{\prod_j^2 \ch(\sigma-\ti\sigma_j)} 
\frac{e^{2\pi i (t_2-t_3)(\ti\sigma_1 + \ti\sigma_2)}}{\prod_{j,k}^{2,2} \ch(\lambda_k-\ti\sigma_j)} 
\frac{\sh(\ti\sigma_{12})^2 e^{2\pi i t_3( \ti\lambda_1 +\ti\lambda_2)}}{\prod_{j}^2 \ch(m_1 - \ti\sigma_j)}  \\
 & \qquad . \ \sh(\ti\lambda_1 - \ti\lambda_2)  \sh(m_1 - \ti\lambda_1) \sh(m_1 - \ti\lambda_2)  
 \hat\delta[ \ti\lambda_2 -m_2 +\frac i 2 ]  \hat\delta[\ti\lambda_1 -m_2 -\frac i 2 ] \ . 
\end{align*}
Integrating over $\ti\lambda_1$ with the $\hat\delta$ distribution yields after simplifications (contributions appear from the poles at $\ti\lambda_1 = \sigma_j + \frac i 2$):
\begin{align*}
 &= \frac{e^{2\pi i t_3 (m_1+m_2+\frac i 2)}}{2}\int d\sigma d^2\ti\sigma d\ti\lambda_2 \, \frac{e^{2\pi i (t_1-t_2)\sigma}}{\prod_j^2 \ch(\sigma-\ti\sigma_j)} 
\frac{e^{2\pi i (t_2-t_3)(\ti\sigma_1 + \ti\sigma_2)}}{\prod_{j}^{2} \ch(\lambda_2-\ti\sigma_j)} 
\frac{\sh(\ti\sigma_{12})^2 e^{2\pi i t_3\ti\lambda_2}}{\prod_{j}^2 \ch(m_1 - \ti\sigma_j)} \sh(m_1 - \ti\lambda_2)  \\
 & \quad . \ \ch(m_2 - \ti\lambda_2)   \hat\delta[ \ti\lambda_2 -m_2 +\frac i 2 ] \  \lp \frac{1}{\sh(\ti\sigma_1-m_2)\sh(\ti\sigma_2-m_2)} + \frac{i}{2} \sum_{j=1,2} (-1)^j \frac{\delta(\ti\sigma_j-m_2)}{\sh(\ti\sigma_{12})} \rp \ .
\end{align*}
Then the integration over $\ti\lambda_2$ can be performed using the remaining $\hat\delta$. Four terms appear out of which two vanishes and the two others are equal. In total we obtain
\begin{align*}
&= i \, e^{2\pi i t_3 (m_1+2m_2)} \int d\sigma d^2\ti\sigma \, \frac{e^{2\pi i (t_1-t_2)\sigma}}{\prod_j^2 \ch(\sigma-\ti\sigma_j)} 
e^{2\pi i (t_2-t_3)(\ti\sigma_1 + \ti\sigma_2)} \frac{\sh(\ti\sigma_{12})}{\ch(\ti\sigma_1-m_1) \sh(\ti\sigma_1-m_2)}  
\delta(\ti\sigma_2 -m_2) \\
&= i \, e^{2\pi i [t_2 m_2 + t_3 (m_1+m_2)]} \int d\sigma d\ti\sigma \, \frac{e^{2\pi i (t_1-t_2)\sigma}}{\ch(\sigma-m_2)\ch(\sigma-\ti\sigma)} 
\frac{e^{2\pi i (t_2-t_3)\ti\sigma}}{\ch(\ti\sigma-m_1)} \\
&= i \, e^{2\pi i [t_2 m_2 + t_3 (m_1+m_2)]} \ Z_{\rm quiver} \ .
\end{align*}
As expected, we recover the matrix model for the partition function of the quiver theory we started with. The extra phase is also as expected from \eqref{PhaseGap} (as usual we ignore overall factors of $i$).


\bibliographystyle{JHEP}
\bibliography{MMbib}

\end{document}